\documentclass[11pt,a4paper]{article}

\usepackage{longtable}
\usepackage{multirow}
\usepackage{jheppub}
\usepackage{orcidlink}
\usepackage{amsmath, amsthm}
\usepackage{amsfonts}
\usepackage{amssymb}
\usepackage{comment}
\usepackage{graphicx}
\usepackage{graphics}
\usepackage{amscd, amstext, hyperref}
\usepackage{pictexwd}
\usepackage{dcpic}
\usepackage[mathscr]{eucal}
\usepackage{multicol}
\usepackage{multirow}
\usepackage{color}
\usepackage{bm}
\usepackage[dvipsnames]{xcolor}
\usepackage[section]{placeins} 
\usepackage{float}
\usepackage{tikz}
\usepackage{booktabs}
\usepackage[vcentermath]{youngtab}
\usepackage{mathtools}

\newtheorem{theorem}{Theorem}[section]

\newtheorem{proposition}[theorem]{Proposition}

\newtheorem*{maintheorem*}{Main Theorem}
\newtheorem*{corollary*}{Corollary}
\newtheorem*{lemma*}{Lemma}
\newtheorem*{keylemma*}{Key Lemma}

\theoremstyle{definition}
\newtheorem{definition}[theorem]{Definition}
\newtheorem{notation}[theorem]{Notation}
\newtheorem{example}[theorem]{Example}

\theoremstyle{remark}
\newtheorem{remark}[theorem]{Remark}

\numberwithin{equation}{section}
\numberwithin{table}{section}

\newcommand{\widesim}[2][1.5]{
  \mathrel{\underset{#2}{\scalebox{#1}[1]{$\sim$}}}
}

\usepackage{pgfplots}

\usepackage{mathrsfs}
\usetikzlibrary{automata,arrows,positioning,calc}
\usepackage{tikz}
\usepackage{tikz-cd}
\usetikzlibrary {arrows.meta}
\usetikzlibrary{cd}

\usepackage{youngtab}

\def\fracm#1#2{\hbox{\large{${\frac{{#1}}{{#2}}}$}}}
\def\e{\epsilon}
\def\a{\alpha}

\title{Topological Characterization of Adinkra Heights}

\author[a]{Amanda E. Francis\, \orcidlink{0009-0006-7368-4388}}\emailAdd{aefr@umich.edu}
\author[b]{S. James Gates Jr.\, \orcidlink{2}}\emailAdd{gatess@umd.edu}
\author[c, d]{Digvijay Roy Varier\,\orcidlink{0000-0002-4695-7428}}\emailAdd{digvijayroyvarier@berkeley.edu}
\author[a]{Ursula A. Whitcher\,\orcidlink{0009-0002-3764-294X}}\emailAdd{uaw@umich.edu}

\affiliation[a]{Mathematical Reviews, American Mathematical Society, Ann Arbor, MI 48103, USA}
\affiliation[b]{University of Maryland, Department of Physics, College Park, MD 20742, USA}
\affiliation[c]{Leinweber Institute for Theoretical Physics, University of California, Berkeley, CA 94720, USA}
\affiliation[d]{Ernest Orlando Lawrence Berkeley National Laboratory, Berkeley, CA 94720, USA}
\affiliation[e]{}

\abstract{Adinkras are highly structured bipartite graphs that encapsulate 1D supersymmetry representations. We classify hypercube Adinkra heights with one, two, three or four edge colors using a graph-theoretic notion of equivalence. For the 990 distinct height functions in the case of four edge colors, we show that these classes match modified twisted character classes bijectively and provide a finer classification than spectral Height Yielding Matrix Numbers (HYMNs). We also relate our classification to the physics of the two-dimensional chiral and twisted chiral supermultiplets.}

\begin{document}
\maketitle
\section{Introduction}
Adinkras are certain decorated bipartite graphs that can be used to represent supersymmetric algebras and their representation theory. Traditionally, the off-shell representation theory of space-time supersymmetry (SUSY) is analyzed via the superspace formalism, where supermultiplets are realized through the $\theta$-expansion of superfields such as the four-dimensional $\mathcal{N}=1$ real scalar superfield $V(x,\theta)$. Dimensional reduction of these space-time manifolds down to a single temporal dimension simplifies the governing differential constraints. In this 1D worldline setting, the component fields and their SUSY transformation laws are modeled entirely by Adinkras, which can be viewed as suitably decorated quotients of the $N$-dimensional hypercube by binary linear codes. 

The problem of classifying Adinkras is of interest to both mathematicians and physicists.  Mathematically speaking, an Adinkra consists of a \emph{chromotopology} equipped with two additional structures: an \emph{odd dashing} and a \emph{height function}. Here, a chromotopology is an $N$-regular bipartite graph whose edges are equipped with an $N$-coloring satisfying a constraint on the subgraphs given by pairs of edge colors. An odd dashing adds decorations to the edges and a height function assigns an integer height to each vertex (see \S~\ref{sec:comb} for more details). A great deal of progress on the combinatorial classification of Adinkras has already been made. Chromotopologies were classified in \cite{codes} using binary linear codes. In \cite{YXZhang}, Zhang enumerated odd dashings by applying the chromotopology classification together with work of Douglas, Gates, Segler, and Wang in \cite{DGSW}. 

Physically, the choice of a height function on an Adinkra corresponds to a consistent choice of engineering dimension for each field. There are many possible height functions. For example, up to an overall shift by a constant,  the $N=4$ hypercube chromotopology admits 990 distinct height functions and the $N=5$ chromotopology admits 395094 height functions. The problem of enumerating all height functions on an arbitrary chromotopology is $NP$-complete. Though the sheer number of height functions may seem daunting, they are related by a rich internal structure. We can move from one height function to another by raising or lowering fermion or boson nodes. In \cite{YXZhang}, Zhang describes an algorithm for enumerating hypercube height functions by bootstrapping from lower dimensions. 

Naive counts of height functions often ignore important commonalities. For example, the $N=2$ hypercube chromotopology admits four distinct fully extended (diamond) height functions, one with each possible fermion or boson as a base. Classification schemes that can detect meaningful physical and mathematical invariants of height functions are sorely needed.   Historically, three distinct frameworks have been proposed to track and classify all valid heights on Adinkras. Twisted Characters are trace-based algebraic invariants that have been used to classify flat valise configurations \cite{Char09,Char13}. HYMNs (Height Yielding Matrix Numbers) are spectral fingerprints of Banchoff matrices that have been shown to change deterministically as nodes are lifted or lowered \cite{HYMN12,HYMN19,HYMN20}. More recently, FWI (Francis–Whitcher Index) classes were proposed in the context of an algebraic-geometric framework related to embedding the Adinkra as a ribbon graph onto a compact Riemann surface, and thence mapping vertex heights to divisors on a Jacobian surface \cite{FW}. 

We begin by discussing the superfields represented by 4-color Adinkras, demonstrating a projection from the 4-dimensional spacetime manifold to two distinct 3-manifolds, related by an outer automorphism. This creates a clean partition of the three spatial dimensions, independent of the temporal dimension, and demonstrates that the associated Lorentz group changes. Repeating this dimensional reduction two more times takes us to the Adinkra graphs of the 2D chiral and twisted chiral supermultiplets, which feature later in the discussion of the origin of twisted characters (one of the frameworks we will use in this paper to classify height functions). We illustrate an example of this projection via Adinkras for a 4D real scalar superfield. As an interesting aside, we illustrate the importance of the outer automorphism between the chiral and twisted chiral supermultiplets by drawing an algebraic analogy to the representation theory of $\mathfrak{su}(3)$.

We then recall the notion of FWI-equivalence of height functions, first introduced in \cite{FW}, and present an explicit classification of 1-, 2-, 3-, and 4-color hypercube Adinkra heights under FWI-equivalence. The $N=4$ case is particularly important for physical reasons explained in Section \ref{sec:hist}. For this case, we  introduce {\it modified twisted characters} on Adinkras with raised heights by applying ideas from \cite{Char13}, and compare the associated notion of equivalence with that of HYMNs and FWI-equivalence. We discuss how modified twisted characters and HYMNs vary with vertex lowering and raising, an interesting topic for future explorations. We  show that the  FWI classes are in bijective correspondence with the equivalence classes generated by modified twisted characters, and provide a slightly finer classification than that of HYMNs.

The specific content of each section is as follows. In Section \ref{sec:hist}, we discuss the relevant physical motivation  and historical background on $N=4$ Adinkras. In Section \ref{sec:comb} we review the combinatorics of Adinkras and FWI-equivalence and classify FWI-equivalence for the $N \leq 4$ cases. For $N=3$ and 4, additional information  is given in Tables~\ref{tab:hts3} and \ref{tab:hts4}. Color adjacency matrices for non-valise Adinkras are described in Section \ref{sec:mat}. In Section \ref{sec:char} we review the original twisted characters, define  modified twisted characters, and prove that the equivalence classes they produce correspond bijectively to the FWI classes. Section \ref{sec:HYMNs} defines  HYMNs and studies the coincidence between HYMN equivalence classes and FWI classes. Finally, we give our concluding remarks and avenues for future work.

\section{Historical and physical background}\label{sec:hist}

\subsection{4D $\mathcal{N} = 1$ chiral and vector supermultiplets}

In 1971, Golfand and Likhtman \cite{Golf-Likt} introduced the concepts of
\emph{space-time supersymmetry algebras}.
Following in 1974, Salam and Strathdee introduced the concepts of \emph{superfields} and \emph{superspace} \cite{S&S-S&S1} in order to create 
sets of fields defined over four-dimensional Minkowski space
where these sets provide representations of the space-time
supersymmetry algebra. The key to their construction of superfields is the $\theta$-expansion; an example appears below.

The $\theta$-expansion of the general real scalar superfield $V(x,\theta)$ can be extracted from equations 
(6) and (10) of their work, thus leading to the result
\begin{align}
V(x,\theta) &= {\phi}(x) + {\phi}^{\alpha}(x)\,\theta_\alpha
+ \fracm{1}{2}\, {\phi}^{[\alpha\beta]}(x)\,\theta_\beta \theta_\alpha 
+ \frac{1}{6}\, {\phi}^{[\alpha\beta\rho]}(x)\,\theta_\rho \theta_\beta \theta_\alpha
 \nonumber  \\
&\qquad 
+ \frac{1}{24}\, {\phi}^{[\alpha\beta\rho\delta]}(x)\,\theta_\delta \theta_\rho \theta_\beta \theta_\alpha,
\label{EQ1}
\end{align}
with the component fields being solely functions of $x$, defined by
\begin{align}
{\phi}(x) &= A(x),  \nonumber\\
{\phi}^{\alpha}(x) &= {\psi}^{\alpha}(x),    \nonumber\\
{\phi}^{[\alpha\beta]}(x) &= (C^{-1})^{\alpha\beta}F(x) + (C^{-1}\gamma_5)^{\alpha\beta}G(x) \nonumber
\\
&\quad\ +(C^{-1}\gamma^\mu\gamma_5)^{\alpha\beta}\left[A_\mu(x) + \frac{1}{2}\partial_\mu B(x)\right], &  \label{EQ2}\\
{\phi}^{[\alpha\beta\rho]}(x) &= \epsilon^{\alpha\beta\rho\delta}\left[\lambda_\delta(x) + 
\frac{1}{2}(\gamma^\mu)_\delta{}^{\e}\,\partial_\mu \psi_{\e}(x)\right],  \nonumber\\
{\phi}^{[\alpha\beta\rho\delta]}(x) &= \epsilon^{\alpha\beta\rho\delta}\left[D(x) - 
\frac{1}{4}\partial^2 A(x)\right], \nonumber
\end{align}
and where we have made some minor notational modifications\footnote{Note that while $B(x)$ is a new bosonic field that does not appear at any lower level of $V(x, \theta)$, it enters the $\theta_\beta\theta_\alpha$ level solely through its spacetime derivative $\partial_\mu B(x)$. It merely shifts the longitudinal mode of the 4-component vector combination $A_\mu + \frac{1}{2}\partial_\mu B$ rather than adding an independent directional component. Under an appropriate gauge transformation, $B(x)$ can shift precisely to cancel this longitudinal component, which allows us to interpret $A_\mu$ as a gauge field with only 3 physical degrees of freedom. Consequently, the off-shell count of the antisymmetric tensor slice $\phi^{[\alpha\beta]}(x)$ remains strictly limited to $\binom{4}{2} = 6$ (comprising $F$, $G$, $B$ and the 3 transverse components of gauge field $A_{\mu}$), consistent with the 6 central bosonic nodes of the Adinkra in Figure ~\ref{fig:A1}. Also note that $A_{\mu}$ comprises 3 of the 4 bosonic degrees of freedom in the vector supermultiplet presented in Eq. (2.8), whereas $\partial_\mu B(x)$ is part of the chiral supermultiplet in Eq. (2.7).}. Here, as usual, $\epsilon^{\alpha \beta \rho \delta}$ is the totally antisymmetric 4-tensor, $\gamma^{\mu}$ are the $4 \times 4$ Dirac matrices representing the Clifford Algebra, $\gamma_5 = i\gamma^0 \gamma^1 \gamma^2 \gamma^3$, and \\
\begin{equation*}
C_{\alpha \beta} = \begin{pmatrix}
    0&-1&0&0 \\
    1&0&0&0\\
    0&0&0&1\\
    0&0&-1&0
\end{pmatrix} 
\end{equation*}

Following are the transformation laws of the supercovariant derivation, $D_{\alpha}$, acting on each component of the real scalar superfield:
\begin{align}
    D_{\alpha}A &= \psi_{\alpha} \nonumber \\
    D_{\alpha}B &= i(\gamma_5)^{\beta}_{\alpha}\psi_{\beta} \nonumber \\
    D_{\alpha} \psi_{\beta} &= i(\gamma^{\mu})_{\alpha \beta}\partial_{\mu}A - (\gamma_5\gamma^{\mu})_{\alpha \beta}\partial_{\mu}B - i C_{\alpha \beta}F + (\gamma_5)_{\alpha \beta}G \nonumber \\
    D_{\alpha}F &= (\gamma^{\mu})_{\alpha}^{\beta}\partial_{\mu}\psi_{\beta} \nonumber \\
    D_{\alpha}G &= i(\gamma_5 \gamma^{\mu})_{\alpha}^{\beta}\partial_{\mu}\psi_{\beta} \nonumber \\
    D_{\alpha}A_{\mu} &= (\gamma^{\mu})_{\alpha}^{\beta}\lambda_{\beta} \nonumber \\
    D_{\alpha}\lambda_{\beta} &= -i\frac{1}{4}(\left[ \gamma^{\mu}, \gamma^{\nu}\right])_{\alpha \beta}(\partial_{\mu}A_{\nu}-\partial_{\nu}A_{\mu}) + (\gamma_5)_{\alpha \beta}D \nonumber \\
    D_{\alpha}D &= i(\gamma_5 \gamma^{\mu})_{\alpha}^{\beta}\partial_{\mu}\lambda_{\beta} 
    \label{transrel}
\end{align}

Without detailed explanation, it is noteworthy that the superfield $V(x,\theta)$ is a collection of {\it {two distinct}}
 irreducible representations of supersymmetry,  i.e. $V(x,\theta) = V_{{\it CM}}(x,\theta) + V_{{\it VM}}(x,\theta)$.  
These two distinct collections can be seen by setting some of the fields above to zero.  One choice leads to
\begin{align}
V_{{\it CM}}(x,\theta) &= A(x) + {\psi}^{\alpha}(x)\,\theta_\alpha + \fracm{1}{2}\, [\, (C^{-1})^{\alpha\beta}F(x) + 
(C^{-1}\gamma_5)^{\alpha\beta}G(x) \, ] \,\theta_\beta \theta_\alpha 
 \nonumber 
 \\
&\quad \  + \frac{1}{4} (C^{-1}\gamma^\mu\gamma_5)^{\alpha\beta}\left[ \partial_\mu B(x)\right] \,\theta_\beta 
\theta_\alpha + \frac{1}{12}\,  \epsilon^{\alpha\beta\rho\delta}\left[ (\gamma^\mu)_\delta{}^{\e}\,\partial_\mu 
\psi_{\e}(x)\right] \,\theta_\rho \theta_\beta \theta_\alpha  \nonumber
  \\
&\quad \
- \frac{1}{96}\, \epsilon^{\alpha\beta\rho\delta} \left[ \frac{1}{4}\partial^2 A(x)\right]
\,\theta_\delta \theta_\rho \theta_\beta \theta_\alpha \, .
\end{align}
and with a different choice for the vanishing fields we find
\begin{align}
V_{{\it VM}}(x,\theta) &= 
\fracm{1}{2}\, (C^{-1}\gamma_\mu\gamma_5)^{\alpha\beta}\left[A_\mu(x) \right]
\theta_\beta \theta_\alpha 
+ \frac{1}{6}\,  \epsilon^{\alpha\beta\rho\delta}\, \lambda_\delta(x) \,\theta_\rho 
\theta_\beta \theta_\alpha
  \\ \nonumber
&\quad 
+ \frac{1}{24}\, {\epsilon}^{[\alpha\beta\rho\delta]} D(x) \,\theta_\delta \theta_\rho \theta_\beta \theta_\alpha \, .
\end{align}
The first collection was given the name of \emph{the chiral supermultiplet} and the second indicated by the name \emph{the vector supermultiplet}.  Often the simpler terms the \emph{chiral multiplet} and \emph{vector multiplet}, respectively, are used.

The two collections of component fields  are thus 
\begin{equation}\label{eq:comp_fd}
    \left( \, A, \, B, \,  \psi_{\a}, \, F, \, G  \right) \quad \text{and}\quad \left( \, A_{\mu}, \,  \lambda_{\a}, \, D \, \right).
\end{equation}  
Each field is regarded as a function of four space-time coordinates ($x^0$, $x^1$, $x^2$, $x^3$) for points lying on a 
Minkowski-signatured 4-manifold endowed with an invariance with respect to a set of $\mathfrak{so}(1,3)$ Lorentz transformations.  The 
independent scalar functions $ A, \, B, \,  F, \, G ,  \, D$ map any point on the four-manifold to the real numbers.  The 
four functions $A_{\mu}$ are components of a 1-form on the manifold, and thus map any point on the manifold to its tangent space.  However, this 1-form is a gauge connection on a $\frak{u(1)}$  bundle and it is defined only up to the exterior derivative of a 0-form. Finally the quantities $\psi_{\a}$ and $\lambda_{\a}$ are independent sections of the spin bundle associated with the $\mathfrak{so}(1,3)$ Lorentz transformations.

\subsection{Projection to lower dimensions}

We now begin a process of implementing projections of the manifold to submanifolds with the 
first projection being $ (x^0,x^1,x^2,x^3) \to (x^0,x^1,x^2)$.  A first implication of this transformation is that the Lorentz group associated 
with the discussion changes.  Using more precise language we consider:
\begin{itemize}
    \item restriction of a representation,
    \item decomposition under the subalgebra, and
    \item branching of \(\mathfrak{so}(1,3)\) representations to \(\mathfrak{so}(1,2)\).
\end{itemize}

In the latter, the process is:
$
\mathfrak{so}(1,3)\longrightarrow \mathfrak{so}(1,2)
$
and a representation \(V\) of \(\mathfrak{so}(1,3)\) becomes, when restricted,
$
V\downarrow_{\mathfrak{so}(1,2)},
$
which in general may decompose into irreducible \(\mathfrak{so}(1,2)\) representations.
Examples of this include
\begin{enumerate}    
\item[(a)] Lorentz scalar remains a scalar under \(\mathfrak{so}(1,2)\);
\item[(b)] 4-vector of \(\mathfrak{so}(1,3)\) decomposes as
  $
  \mathbf{4}\to \mathbf{3}\oplus \mathbf{1},
  $
  and this further means
  \begin{enumerate}
  \item[(b1)] the components \(x^0,x^1,x^2\) transform as a 3-vector of \(\mathfrak{so}(1,2)\),
  \item[(b2)]  the component \(x^3\) becomes a scalar under the surviving \(\mathfrak{so}(1,2)\),
   \end{enumerate}
  \item[(c)]  the consideration applies to components of the four-vector $A_{\mu}$,
   \begin{enumerate}
  \item[(c1)] the components \(A_0,A_1,A_2\) transform as a 3-vector of \(\mathfrak{so}(1,2)\),
  \item[(c2)]  the component \(A_3\) becomes a scalar under the surviving \(\mathfrak{so}(1,2)\),
  \end{enumerate}
  \item[(d)] similarly, tensors and spinors decompose into lower-dimensional \(\mathfrak{so}(1,2)\)  representations.

\end{enumerate}
The matter concerning point (b) $ \mathbf{4}\to \mathbf{3}\oplus \mathbf{1},$ can be notationally reflected by 
using a convention where $A_{\mu} \to Y_{\mu} \oplus J$,  where on the right hand side $Y_{\mu}$ and $J$ 
correspond to the  \(\mathfrak{so}(1,2)\) representations of dimensions $\mathbf{3}$ and $\mathbf{1}$,  respectively.
Thus, the projective process leads to the two independent collections of functions 
 \[
 \left( \, A, \, B, \,  \psi_{\a}, \, F, \, G  \right)\quad \text{and}\quad \left( \, J, \,Y_{\mu}, \,  \lambda_{\a}, \, D \, \right)
 \]
 providing
irreducible representations of supersymmetry in three dimensions.  The three functions $Y_\mu$ 
are the components of a 1-form and gauge connection on a $\frak{u}(1)$ bundle, and it is defined only up to the exterior 
derivative of a 0-form. 

\subsubsection{2D chiral and twisted chiral supermultiplets}

Next we consider the process of implementing the
projections $ (x^0,x^1,x^2,x^3) \to (x^0,x^1)$.  After this projection, the two collections of component fields in \eqref{eq:comp_fd} take the form  
\[
\left( \, A, \, B, \,  \psi_{\a}, \, F, \, G  \right)
\quad \text{and}\quad \left( \, J, \, L,  \lambda_{\a}, \, W, \,D\right).
\]
which are a pair of 2D $\mathcal{N} = (2, 2)$ supermultiplets. In
the latter collection, $L$ emerges as the singlet portion of the projection of the \(\mathfrak{so}(1,2)\) $\to$  
\(\mathfrak{so}(1,1)\) applied to $Y_{\mu}$ and $W$ is simply the field strength of the doublet portion of
$Y_{\mu}$. The first collection of fields is commonly called the 2D chiral supermultiplet. This is the 2D projection (dimensional reduction) of the 4D chiral supermultiplet. The second collection of fields, isomorphic to the first collection in terms of its \(\mathfrak{so}(1,1)\) representations, 
was introduced into the physics literature in the works of \cite{Sigma1,Sigma2} and the second of these references 
gave it the name of the 2D \emph{twisted chiral multiplet}\footnote{The two works \cite{Sigma1,Sigma2} mark the historical 
inauguration of applying the projection \(\mathfrak{so}(1,3)\) $\to$  \(\mathfrak{so}(1,1)\) in these contexts.}. Relying simply on the \(\mathfrak{so}(1,1)\) spectra of the two collections, one might be tempted to think they are the same representation.  They are not. This is the 2D projection of the 4D vector multiplet. We will be frequently referring to this chiral and twisted chiral pair in the forthcoming sections. \\

\subsubsection{\textbf{Chiral and twisted chiral 4-4 Adinkras}}
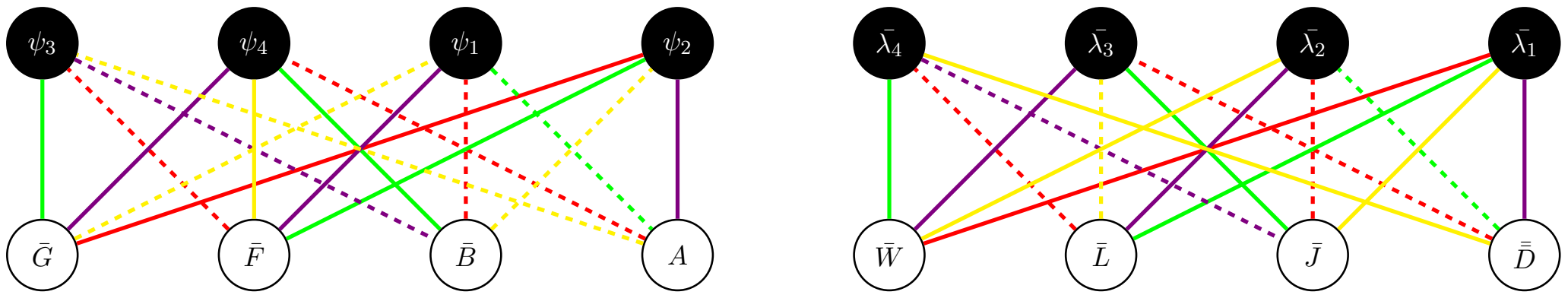
\begin{figure}[h!]
\centering
\resizebox{0.9\textwidth}{!}{
\begin{tikzpicture}[
    boson/.style={circle, draw, thick, fill=white, minimum size=10mm},
    fermion/.style={circle, draw, thick, fill=black, text=white, minimum size=10mm},
    edge1/.style={ultra thick, red},
    edge11/.style={ultra thick, red, dashed},
    edge2/.style={ultra thick, green},
    edge22/.style={ultra thick, green, dashed},
    edge3/.style={ultra thick, violet},
    edge33/.style={ultra thick, violet, dashed},
    edge4/.style={ultra thick, yellow},
    edge44/.style={ultra thick, yellow, dashed},
]
\node[boson] (phi1) at (-10.5, 0){$\Bar{G}$};
\node[boson] (phi2) at (10.5, 0){$\Bar{\Bar{D}}$};

\node[fermion] (psi1) at (-4.5, 3){$\psi_1$};
\node[fermion] (psi2) at (-7.5, 3){$\psi_4$};
\node[fermion] (psi3) at (-10.5, 3){$\psi_3$};
\node[fermion] (psi4) at (-1.5, 3){$\psi_2$};

\node[boson] (phi3) at (1.5, 0){$\Bar{W}$};
\node[boson] (phi4) at (-7.5, 0){$\Bar{F}$};
\node[boson] (phi5) at (4.5, 0){$\Bar{L}$};
\node[boson] (phi6) at (-4.5, 0){$\Bar{B}$};
\node[boson] (phi7) at (7.5, 0){$\Bar{J}$};
\node[boson] (phi8) at (-1.5, 0){$A$};

\node[fermion] (psi5) at (7.5, 3){$\Bar{\lambda_2}$};
\node[fermion] (psi6) at (4.5, 3){$\Bar{\lambda_3}$};
\node[fermion] (psi7) at (1.5, 3){$\Bar{\lambda_4}$};
\node[fermion] (psi8) at (10.5, 3){$\Bar{\lambda_1}$};

\draw[edge1] (phi1) -- (psi4);
\draw[edge11] (phi4) -- (psi3);
\draw[edge11] (phi6) -- (psi1);
\draw[edge11]  (phi8) -- (psi2);
\draw[edge1] (phi3) -- (psi8);
\draw[edge11] (phi5) -- (psi7);
\draw[edge11] (phi7) -- (psi5);
\draw[edge11] (phi2) -- (psi6);

\draw[edge2] (phi1) -- (psi3);
\draw[edge2] (phi3) -- (psi7);
\draw[edge22] (psi1) -- (phi8);
\draw[edge2] (psi4) -- (phi4);
\draw[edge2] (psi2) -- (phi6);
\draw[edge2] (phi5) -- (psi8);
\draw[edge2] (phi7) -- (psi6);
\draw[edge22] (phi2) -- (psi5);

\draw[edge3] (phi1) -- (psi2);
\draw[edge33] (psi3) -- (phi6);
\draw[edge33] (psi7) -- (phi7);
\draw[edge3] (psi6) -- (phi3);
\draw[edge3] (phi4) -- (psi1);
\draw[edge3] (phi5) -- (psi5);
\draw[edge3] (psi8) -- (phi2);
\draw[edge3] (psi4) -- (phi8);

\draw[edge44] (phi1) -- (psi1);
\draw[edge4] (phi3) -- (psi5);
\draw[edge4] (phi4) -- (psi2);
\draw[edge44] (phi5) -- (psi6);
\draw[edge44] (phi6) -- (psi4);
\draw[edge4] (phi7) -- (psi8);
\draw[edge44] (phi8) -- (psi3);
\draw[edge4] (phi2) -- (psi7);

\end{tikzpicture}}
\caption{Adinkra graphs for the chiral supermultiplet (to the left) and the twisted chiral supermultiplet (to the right). Each bar over a field indicates one time integral, i.e. $\Bar{\lambda} \equiv \int dx^0 \lambda$ and $\Bar{\Bar{D}} \equiv \iint dx^0 dx^{0'} D$.} \label{fig:CM_TCM}
\end{figure}

The final projection $ (x^0,x^1,x^2,x^3) \to (x^0)$ of the 2D chiral and twisted chiral supermultiplets allows the two collections of fields\footnote{Since the fields $\psi_{\a}$
and $\lambda_{\a}$ are sections of the original \(\mathfrak{so}(1,3)\) spin-bundle, their explicit `components' can be indicated
by $\psi_{1}, \,  \psi_{2}, \,  \psi_{3}, \,  \psi_{4},$ and $ \lambda_{1}, \, \lambda_{2}, \, \lambda_{3}, \, \lambda_{4}$, respectively.} 
\[
\left( \, A, \, B, \,  \psi_{1}, \,  \psi_{2}, \,  \psi_{3}, \,  \psi_{4}, \, F, \, 
G  \right) \quad \text{and} \quad 
\left( \, J, \, L,  \lambda_{1}, \, \lambda_{2}, \, \lambda_{3}, \, \lambda_{4}, \, W, \, D \,\right)
\]
to enter into the domain of Adinkra graphs. Adinkras encode 1-dimensional supersymmetry algebra representations in bipartite graphs with specific additional combinatorial structure. As seen in Figure \ref{fig:CM_TCM}, the fields are represented by nodes in graphs 
and the orbits of the fields under the generators of supersymmetry are represented as the links between nodes. (See \S~\ref{sec:comb} for a description of the combinatorics of Adinkras that will be used in this paper, and \cite{YXZhang} for a more extensive discussion.)

Ignoring the names of the nodes in each graph reveals something about the relation between the two: one is 
obtained from the other by replacing solid yellow lines by dashed yellow lines and vice-versa.  
The `flipping' between dashed and solid lines is emblematic of the change of a sign $\pm$ in a field's 
transformation laws.  In other words, there is an outer automorphism that distinguishes the transformation
of the chiral supermultiplet from the twisted chiral supermultiplet.

\subsection{Projecting Adinkras into 4-4 components}\label{ssec:projecting}
Generally speaking, 4D, $\mathcal{N} = 1$ off-shell multiplets will have $(4k|4k)$ bosonic and
fermionic degrees of freedom \cite{Char13}, where $k$ is the number of bosonic (or fermionic) off-shell degrees of freedom divided by four. The number $k$ satisfies the relation 
\begin{align*}
    k = n_c + n_t
\end{align*}
where the ``isomer numbers" $n_c$ and $n_t$ are the number of chiral and  twisted chiral Adinkras composing the representation, respectively. At first glance it is not at all obvious that it is possible to decompose the Adinkra corresponding to a 4D, $\mathcal{N} = 1$ representation in
terms of a uniquely specified number of $n_c$ chiral and $n_t$  twisted chiral Adinkras. Remarkably, in studies of specific multiplets, a decomposition in terms of chiral and twisted chiral Adinkras consistent with the structure of the original multiplet has always been found \cite{Char13}. We shall provide here an example of such a decomposition with $n_c = n_t = 1$ (more examples can be found in \cite{G-2}).

\begin{figure}[h!]
\centering
\resizebox{0.7\textwidth}{!}{
\begin{tikzpicture}[
    boson/.style={circle, draw, thick, fill=white, minimum size=10mm},
    fermion/.style={circle, draw, thick, fill=black, text=white, minimum size=10mm},
    edge1/.style={ultra thick, red},
    edge11/.style={ultra thick, red, dashed},
    edge2/.style={ultra thick, green},
    edge22/.style={ultra thick, green, dashed},
    edge3/.style={ultra thick, violet},
    edge33/.style={ultra thick, violet, dashed},
    edge4/.style={ultra thick, yellow},
    edge44/.style={ultra thick, yellow, dashed},
]
\node[boson] (phi1) at (0, -6){$A$};
\node[boson] (phi2) at (0, 6){$D$};

\node[fermion] (psi1) at (-4.5, -3){$\psi_3$};
\node[fermion] (psi2) at (-1.5, -3){$\psi_2$};
\node[fermion] (psi3) at (1.5, -3){$\psi_1$};
\node[fermion] (psi4) at (4.5, -3){$\psi_4$};

\node[boson] (phi3) at (-7.5, 0){$B - J$};
\node[boson] (phi4) at (-4.5, 0){$W - G$};
\node[boson] (phi5) at (-1.5, 0){$F + L$};
\node[boson] (phi6) at (1.5, 0){$F - L$};
\node[boson] (phi7) at (4.5, 0){$W + G$};
\node[boson] (phi8) at (7.5, 0){$B + J$};

\node[fermion] (psi5) at (-4.5, 3){$\lambda_3$};
\node[fermion] (psi6) at (-1.5, 3){$\lambda_2$};
\node[fermion] (psi7) at (1.5, 3){$\lambda_1$};
\node[fermion] (psi8) at (4.5, 3){$\lambda_4$};

\draw[edge1] (phi1) -- (psi4);
\draw[edge11] (psi3) -- (phi8);
\draw[edge11] (psi1) -- (phi5);
\draw[edge11] (phi3) -- (psi6);
\draw[edge1] (psi2) -- (phi7);
\draw[edge1] (phi4) -- (psi7);
\draw[edge1] (phi6) -- (psi8);
\draw[edge11] (psi5) -- (phi2);

\draw[edge2] (phi1) -- (psi3);
\draw[edge22] (phi2) -- (psi6);
\draw[edge2] (psi4) -- (phi8);
\draw[edge2] (psi1) -- (phi4);
\draw[edge22] (psi2) -- (phi6);
\draw[edge2] (phi3) -- (psi5);
\draw[edge2] (phi5) -- (psi7);
\draw[edge2] (phi7) -- (psi8);

\draw[edge3] (phi1) -- (psi2);
\draw[edge3] (psi3) -- (phi6);
\draw[edge33] (psi4) -- (phi7);
\draw[edge33] (psi1) -- (phi3);
\draw[edge3] (phi4) -- (psi5);
\draw[edge3] (phi5) -- (psi6);
\draw[edge3] (psi7) -- (phi2);
\draw[edge3] (psi8) -- (phi8);

\draw[edge4] (phi1) -- (psi1);
\draw[edge4] (phi3) -- (psi2);
\draw[edge44] (phi4) -- (psi3);
\draw[edge4] (phi5) -- (psi4);
\draw[edge4] (phi6) -- (psi5);
\draw[edge4] (phi7) -- (psi6);
\draw[edge44] (phi8) -- (psi7);
\draw[edge4] (phi2) -- (psi8);

\end{tikzpicture}}
\caption{A fully extended 8-8 $N=4$ Adinkra.}\label{fig:A1}

\end{figure}
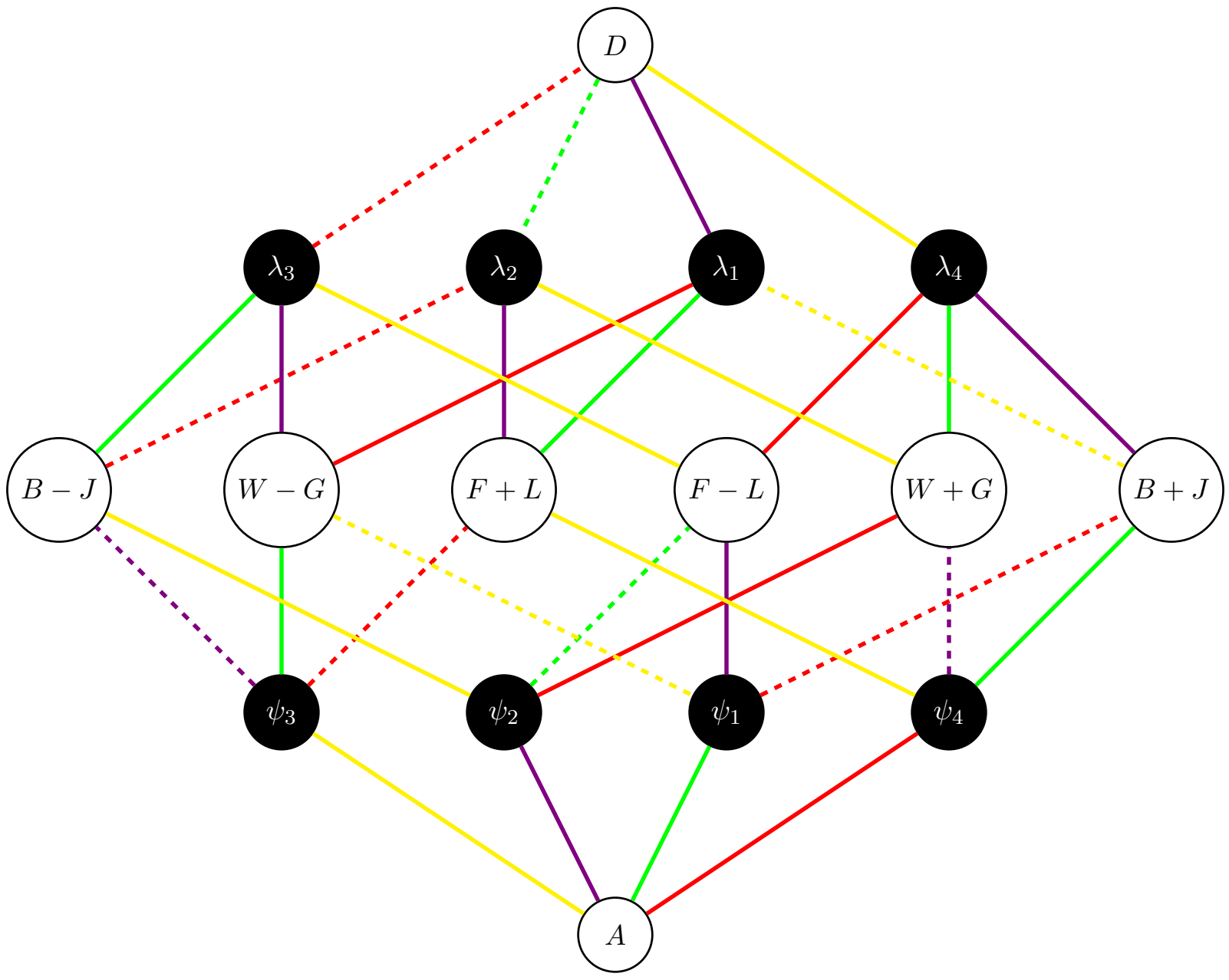

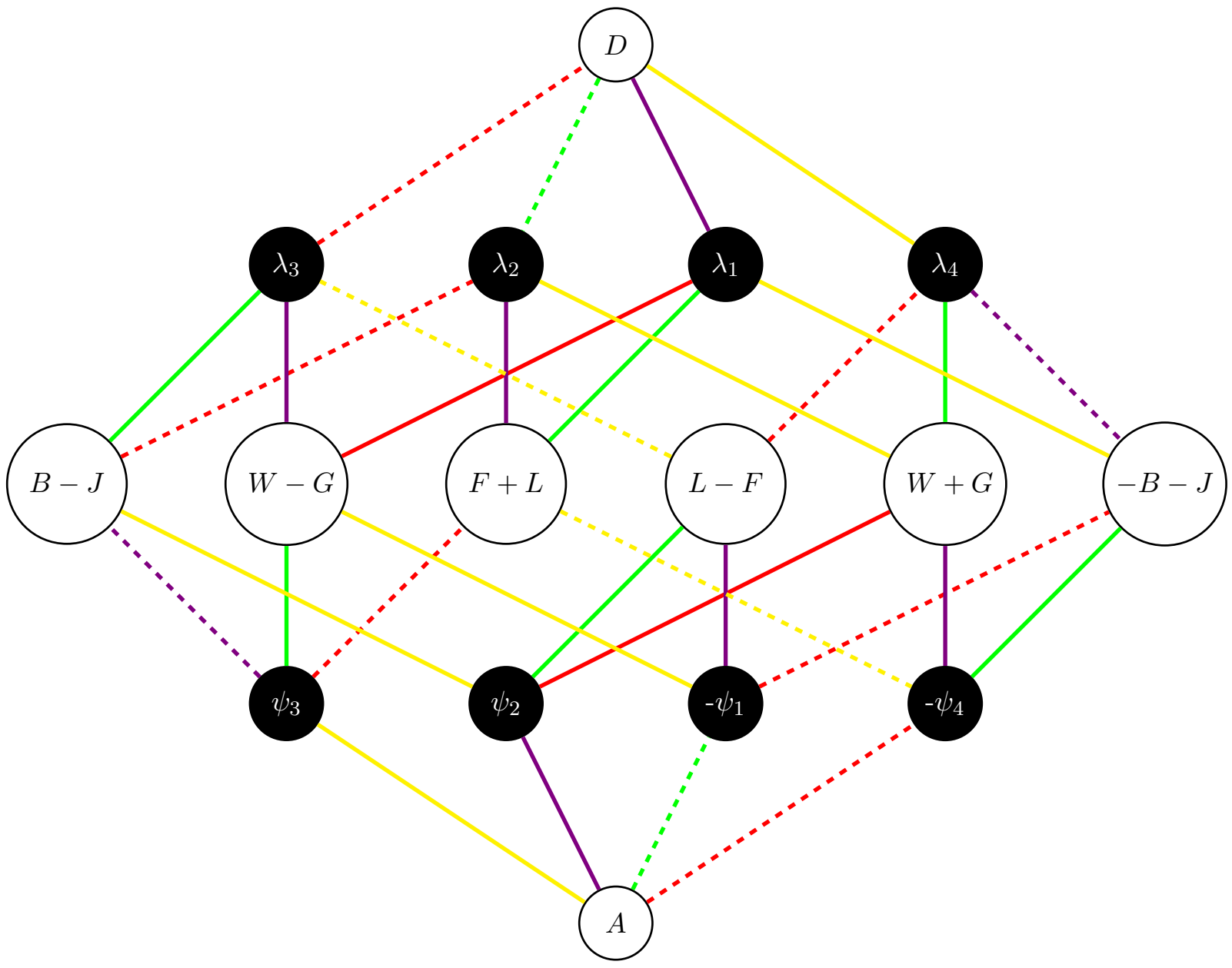
\begin{figure}[h!]
\centering
\resizebox{0.7\textwidth}{!}{
\begin{tikzpicture}[
    boson/.style={circle, draw, thick, fill=white, minimum size=10mm},
    fermion/.style={circle, draw, thick, fill=black, text=white, minimum size=10mm},
    edge1/.style={ultra thick, red},
    edge11/.style={ultra thick, red, dashed},
    edge2/.style={ultra thick, green},
    edge22/.style={ultra thick, green, dashed},
    edge3/.style={ultra thick, violet},
    edge33/.style={ultra thick, violet, dashed},
    edge4/.style={ultra thick, yellow},
    edge44/.style={ultra thick, yellow, dashed},
]
\node[boson] (phi1) at (0, -6){$A$};
\node[boson] (phi2) at (0, 6){$D$};

\node[fermion] (psi1) at (-4.5, -3){$\psi_3$};
\node[fermion] (psi2) at (-1.5, -3){$\psi_2$};
\node[fermion] (psi3) at (1.5, -3){-$\psi_1$};
\node[fermion] (psi4) at (4.5, -3){-$\psi_4$};

\node[boson] (phi3) at (-7.5, 0){$\ B - J\ $};
\node[boson] (phi4) at (-4.5, 0){$\ W - G$};
\node[boson] (phi5) at (-1.5, 0){$\ F + L\ $};
\node[boson] (phi6) at (1.5, 0){$\ L - F\ $};
\node[boson] (phi7) at (4.5, 0){$\ W + G$};
\node[boson] (phi8) at (7.5, 0){$-B - J$};

\node[fermion] (psi5) at (-4.5, 3){$\lambda_3$};
\node[fermion] (psi6) at (-1.5, 3){$\lambda_2$};
\node[fermion] (psi7) at (1.5, 3){$\lambda_1$};
\node[fermion] (psi8) at (4.5, 3){$\lambda_4$};

\draw[edge11] (phi1) -- (psi4);
\draw[edge11] (psi3) -- (phi8);
\draw[edge11] (psi1) -- (phi5);
\draw[edge11] (phi3) -- (psi6);
\draw[edge1] (psi2) -- (phi7);
\draw[edge1] (phi4) -- (psi7);
\draw[edge11] (phi6) -- (psi8);
\draw[edge11] (psi5) -- (phi2);

\draw[edge22] (phi1) -- (psi3);
\draw[edge22] (phi2) -- (psi6);
\draw[edge2] (psi4) -- (phi8);
\draw[edge2] (psi1) -- (phi4);
\draw[edge2] (psi2) -- (phi6);
\draw[edge2] (phi3) -- (psi5);
\draw[edge2] (phi5) -- (psi7);
\draw[edge2] (phi7) -- (psi8);

\draw[edge3] (phi1) -- (psi2);
\draw[edge3] (psi3) -- (phi6);
\draw[edge3] (psi4) -- (phi7);
\draw[edge33] (psi1) -- (phi3);
\draw[edge3] (phi4) -- (psi5);
\draw[edge3] (phi5) -- (psi6);
\draw[edge3] (psi7) -- (phi2);
\draw[edge33] (psi8) -- (phi8);

\draw[edge4] (phi1) -- (psi1);
\draw[edge4] (phi3) -- (psi2);
\draw[edge4] (phi4) -- (psi3);
\draw[edge44] (phi5) -- (psi4);
\draw[edge44] (phi6) -- (psi5);
\draw[edge4] (phi7) -- (psi6);
\draw[edge4] (phi8) -- (psi7);
\draw[edge4] (phi2) -- (psi8);

\end{tikzpicture}}
\caption{Another consistent way of drawing the fully extended 8-8 $N=4$ Adinkra, but such that the edges with same color and dashing get arranged more symmetrically (to be precise, taking successive reflections of any edge across the vertical and horizontal bisectors of the graph lands it on an edge with the same color and dashing as itself).}\label{fig:A2}
\end{figure}

The components of the Adinkra in  in Figure~\ref{fig:A1} comprise a real scalar superfield. This particular organization of the graph is adopted from \cite{G-2}, but there are more than one ways of the drawing the graph such that the nodes, links and dashings are in accordance with the supersymmetry relations \eqref{transrel}. We shall now flip the signs of the two bosonic fields ($F-L$, $B+J$) and the two fermionic fields ($\psi_1$, $\psi_4$), and simultaneously invert the dashing of some of their surrounding edges to keep the supersymmetry relations \eqref{transrel} between the fields unchanged. we do this so that it becomes easier to visualize pairs of nodes that have identical configurations within the Adinkra, namely ($D, A$), ($\lambda_3, -\psi_4$), ($\lambda_2, -\psi_1$), ($\lambda_1, \psi_2$), ($ \psi_3, \lambda_4$), ($B-J, -B-J$), ($W-G, W+G$), and ($F+L, L-F$).   This is shown in Figure~\ref{fig:A2}. \\

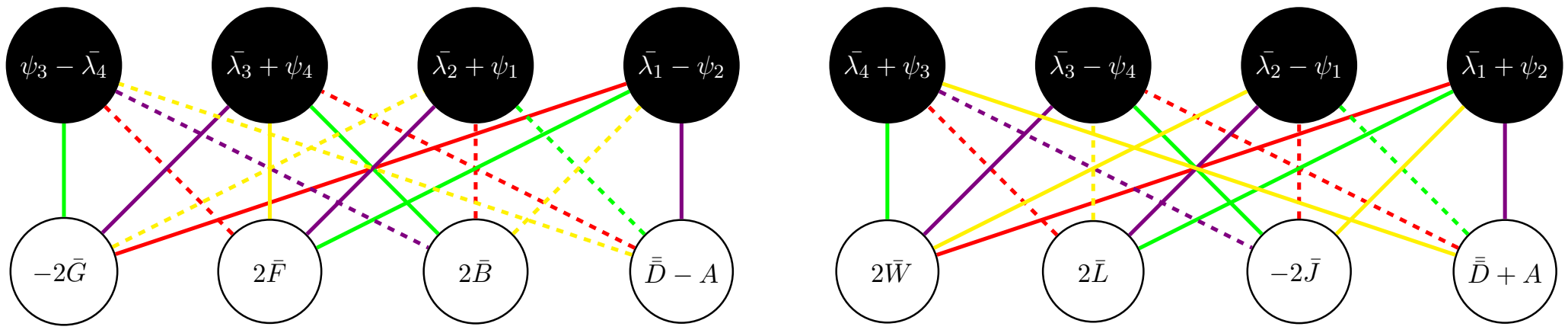
\begin{figure}[h!]
\centering
\resizebox{0.9\textwidth}{!}{
\begin{tikzpicture}[
    boson/.style={circle, draw, thick, fill=white, minimum size=10mm},
    fermion/.style={circle, draw, thick, fill=black, text=white, minimum size=10mm},
    edge1/.style={ultra thick, red},
    edge11/.style={ultra thick, red, dashed},
    edge2/.style={ultra thick, green},
    edge22/.style={ultra thick, green, dashed},
    edge3/.style={ultra thick, violet},
    edge33/.style={ultra thick, violet, dashed},
    edge4/.style={ultra thick, yellow},
    edge44/.style={ultra thick, yellow, dashed},
]
\node[boson] (phi1) at (-10.5, 0){$\, \, -2\Bar{G}\ \ $};
\node[boson] (phi2) at (10.5, 0){$\Bar{\Bar{D}}+A$};

\node[fermion] (psi1) at (-4.5, 3){$\Bar{\lambda_2}+\psi_1$};
\node[fermion] (psi2) at (-7.5, 3){$\Bar{\lambda_3}+\psi_4$};
\node[fermion] (psi3) at (-10.5, 3){$\psi_3 - \Bar{\lambda_4}$};
\node[fermion] (psi4) at (-1.5, 3){$\Bar{\lambda_1}-\psi_2$};

\node[boson] (phi3) at (1.5, 0){$\,\ \ 2\Bar{W}\,\ $ };
\node[boson] (phi4) at (-7.5, 0){$\,\ \ 2\Bar{F}\,\ \ $};
\node[boson] (phi5) at (4.5, 0){$\,\ \ 2\Bar{L}\,\ \ $\  };
\node[boson] (phi6) at (-4.5, 0){$\,\ \ 2 \Bar{B}\,\ \ $};
\node[boson] (phi7) at (7.5, 0){$\, \, -2\Bar{J}\ \ $};
\node[boson] (phi8) at (-1.5, 0){$\Bar{\Bar{D}}-A$};

\node[fermion] (psi5) at (7.5, 3){$\Bar{\lambda_2}-\psi_1$};
\node[fermion] (psi6) at (4.5, 3){$\Bar{\lambda_3}-\psi_4$};
\node[fermion] (psi7) at (1.5, 3){$\Bar{\lambda_4}+\psi_3$};
\node[fermion] (psi8) at (10.5, 3){$\Bar{\lambda_1}+\psi_2$};

\draw[edge1] (phi1) -- (psi4);
\draw[edge11] (phi4) -- (psi3);
\draw[edge11] (phi6) -- (psi1);
\draw[edge11]  (phi8) -- (psi2);
\draw[edge1] (phi3) -- (psi8);
\draw[edge11] (phi5) -- (psi7);
\draw[edge11] (phi7) -- (psi5);
\draw[edge11] (phi2) -- (psi6);

\draw[edge2] (phi1) -- (psi3);
\draw[edge2] (phi3) -- (psi7);
\draw[edge22] (psi1) -- (phi8);
\draw[edge2] (psi4) -- (phi4);
\draw[edge2] (psi2) -- (phi6);
\draw[edge2] (phi5) -- (psi8);
\draw[edge2] (phi7) -- (psi6);
\draw[edge22] (phi2) -- (psi5);

\draw[edge3] (phi1) -- (psi2);
\draw[edge33] (psi3) -- (phi6);
\draw[edge33] (psi7) -- (phi7);
\draw[edge3] (psi6) -- (phi3);
\draw[edge3] (phi4) -- (psi1);
\draw[edge3] (phi5) -- (psi5);
\draw[edge3] (psi8) -- (phi2);
\draw[edge3] (psi4) -- (phi8);

\draw[edge44] (phi1) -- (psi1);
\draw[edge4] (phi3) -- (psi5);
\draw[edge4] (phi4) -- (psi2);
\draw[edge44] (phi5) -- (psi6);
\draw[edge44] (phi6) -- (psi4);
\draw[edge4] (phi7) -- (psi8);
\draw[edge44] (phi8) -- (psi3);
\draw[edge4] (phi2) -- (psi7);

\end{tikzpicture}}
\caption{The chiral (left) and twisted-chiral (right) 4-4 Adinkras that result from projecting the 8-8 Adinkra by following the procedure described in the main text.}\label{fig:44}
\end{figure}

Each of these pairs of nodes is identified by either adding or subtracting their constituent fields, resulting in the 8-8 Adinkra getting projected into a 4-4 Adinkra. If all  pairs are identified via addition (after adjusting the engineering dimensions by performing time integrals), the resulting Adinkra (seen on the right in Figure~\ref{fig:44}) describes a twisted-chiral supermultiplet with the fields 
\begin{equation*}
     \Bar{\Bar{D}}+A, \Bar{\lambda_3}-\psi_4, \Bar{\lambda_2}-\psi_1, \Bar{\lambda_1}+\psi_2, \Bar{\lambda_4}+\psi_3, -2\Bar{J}, 2 \Bar{W}, 2\Bar{L}
\end{equation*}
where, again, each bar over a field indicates one time integral, i.e. $\Bar{\lambda} \equiv \int dx^0 \lambda$ and $\Bar{\Bar{D}} \equiv \iint dx^0 dx^{0'} D$.
If all the pairs are identified by subtraction, the resulting Adinkra (seen on the left in Figure~\ref{fig:44}) describes a chiral supermultiplet with the fields 
\begin{equation*}
     \Bar{\Bar{D}}-A, \Bar{\lambda_3}+\psi_4, \Bar{\lambda_2}+\psi_1, \Bar{\lambda_1}-\psi_2, \psi_3 - \Bar{\lambda_4}, 2\Bar{B}, -2\Bar{G}, 2\Bar{F}
\end{equation*}

Note that this second projection is equivalent to first flipping the sign of the second field in each pair, and then identifying all the pairs via addition. After changing the signs of the appropriate fields in this step, the dashings of the yellow edges in the above Adinkra must also be flipped, in order to preserve the signs in the equations \eqref{EQ1}-\eqref{transrel}. This is what causes the two different projected 4-4 Adinkras to have opposite dashings for all the yellow edges (only), as was observed in Figure \ref{fig:CM_TCM}. To make the chiral and twisted-chiral Adinkras look like those in that figure, we simply perform a redefinition of the above component fields as follows: 
\begin{align*}
\Bar{\Bar{D}}+A&\rightarrow \Bar{\Bar{D}},& \quad \Bar{\Bar{D}}-A&\rightarrow A, &\quad \Bar{\lambda_3}-\psi_4 &\rightarrow \Bar{\lambda_3}, &\quad\Bar{\lambda_3}+\psi_4 &\rightarrow \psi_4,\\
\quad\Bar{\lambda_2}-\psi_1 &\rightarrow \Bar{\lambda_2},
&\quad \Bar{\lambda_2}+\psi_1 &\rightarrow \psi_1,&\quad \Bar{\lambda_1}+\psi_2 &\rightarrow \Bar{\lambda_1}, &\quad\Bar{\lambda_1}-\psi_2 &\rightarrow \psi_2,\\
\quad \Bar{\lambda_4}+\psi_3 &\rightarrow \Bar{\lambda_4}, &\quad\psi_3 - \Bar{\lambda_4} &\rightarrow \psi_3,&\quad
-2\Bar{J} &\rightarrow \Bar{J},&\quad 2\Bar{B} &\rightarrow \Bar{B},\\
\quad 2\Bar{W} &\rightarrow \Bar{W},&\quad -2\Bar{G} &\rightarrow \Bar{G},&\quad 2\Bar{L} &\rightarrow \Bar{L}, &\quad 2\Bar{F} &\rightarrow \Bar{F}. 
\end{align*}

This was just one example of how any 4D, $\mathcal{N} = 1$ representation can be decomposed in terms of $n_c$ ``chiral'' and $n_t$  ``twisted chiral'' Adinkras. These isomer numbers $n_c$ and $n_t$ will be of relevance when we discuss twisted characters as Adinkra height classifiers in section 6.

\subsection{Outer automorphisms in Lie algebras and sigma models}

We have observed that the chiral and twisted chiral Adinkras are dual to each other via an outer automorphism that manifests combinatorially as a change in dashings along certain edges. Though individually the chiral and twisted chiral Adinkras appear to contain the same information, in combination they demonstrate remarkably rich structure. In this section, we discuss two other examples of objects related by outer automorphisms that generate systems with rich physical and mathematical behavior. We first treat the Lie algebra $\mathfrak{su}(3)$, which offers an algebraic model familiar to many physicists. We then discuss the role played by the chiral and twisted chiral multiplets in the context of nonlinear $\sigma$-models. Study of the geometrical structures associated to these dual multiplets drove the development of generalized K\"ahler geometry.

\subsubsection{The $\mathfrak{su}(3)$ analogy}
A familiar context for the appearance of an outer automorphism appearing between distinct representations 
of a Lie algebra occurs in the context of the Lie algebras, with the case of the $\mathfrak{su}(3)$ Lie algebra 
being the simplest example.  It has two minimal-dimension representations that can be represented by the Young 
Tableaux shown in Figure \ref{fig:YT}.

\begin{figure}[h]
$$
\mathcal{A} \;=\; {\young(\ )}    ~~~,~~~ {\mathcal B} \;=\; {\young(\ ,\ )}
$$
\vskip0.14in

\caption{Young ~ Tableaux ~for ~the ~$\mathfrak{su}(3)$ ~quark~ representation~ (to ~the ~left) and ~the ~ $\mathfrak{su}(3)$ ~anti-quark ~representation ~ (to ~the ~right).}
\label{fig:YT}
\end{figure}

There exist eight Hermitian traceless 3$\times$3 matrices that can be used to represent the $\mathfrak{su}(3)$ generators 
acting on the $\mathcal A$ representation and we can express these matrices in the form of $T_{\rm I}$, with $\rm I$ = 1, 
$\dots$, 8.   In this case, the $\mathfrak{su}(3)$ generators acting on the $\mathcal B$ representation are expressed by
matrices  $(-\, T_{\rm I})^*$ where $*$ represents the complex conjugation operator (the additional minus sign here causes 
the sign change $\pm$ in transformation laws).  This set of matrices is thus related by an outer automorphism to the first 
set of matrices.  Using the symbol ($\mathcal {OA})_{\rm I}{}^{\rm J}$ we can write the equation
$$
(-\, T_{\rm I})^* = ({\mathcal {OA}})_{\rm I}{}^{\rm J} \, (T_{\rm J}) ~,
$$
to describe the action of the outer automorphism acting on the $\mathfrak{su}(3)$ generators.

To expand upon this point and its realization on the $T_{\rm I}$ matrices, the fact that they
are hermitian can be expressed in the following:
$$
( T_{\rm I})^{\dag} ~=~ [( T_{\rm I})^*]^{t} ~=~ ( T_{\rm I}) ~~~,
$$
where the superscript $t$ denotes matrix transposition.  The transposition operation allows a partitioning of the $T_{\rm I}$ matrices.   Two of the eight matrices satisfy
$$
( T_{\rm I})^{t} ~=~ ( T_{\rm I}) ~~~,
$$
and these are real. The remaining six matrices do not satisfy this
condition and are purely imaginary.  Thus, the outer automorphism is realized on the eight su(3) generators according to: 
\begin{align*}
({\mathcal {OA}})_{\rm I}{}^{\rm J} \, (T_{\rm J})  &{~~~~}=  {~~~}( T_{\rm I}) 
{~\,\,~\,~} , {~\,~} {\rm {if}}{~~}( T_{\rm I}){~~} {\rm {is}}{~~} {\rm {imaginary}} {,}
 \nonumber  \\
&{~~~~} =~
-\, ( T_{\rm I}) {~~\,~} ,{~\,~} {\rm {if}}{~~}( T_{\rm I}){~~} {\rm {is}}{~~} {\rm {real}}
.
\label{EQx}
\end{align*}
Crucially, the eigenvalues associated with the Cartan–Weyl root-space decomposition are determined by these latter matrices and signify the distinction (in physics) between the quark representation from the anti-quark representation.  Similarly, the switching of the yellow lines only between the two Adinkra diagrams is symptomatic of the action of an outer automorphism acting on only one of the four supersymmetry generators.

Using row lengths $(r_1,r_2,r_3)$ and Dynkin labels
$$
(p,q)=(r_1-r_2,\; r_2-r_3),
\qquad
\dim(p,q)=\frac{(p+1)(q+1)(p+q+2)}{2},
$$
we obtain via the hook-content formula or Weyl's dimension formula

$$
{\mathcal A}=(1)\; \Rightarrow\; (p,q)=(1,0), \qquad \dim({\mathcal A})=3,\qquad
$$

$$
{\mathcal B}=(1,1)\; \Rightarrow\; (p,q)=(0,1), \qquad \dim({\mathcal B})=3,\qquad
$$

Using the Littlewood--Richardson rule applied to $\mathcal A$ and $\mathcal B$, the three products are:

$$
{\mathcal A}  \otimes {\mathcal A} ={\yng(2)} \oplus   {\yng(1,1)} = \{ 6 \} \oplus  {\mathcal B}
$$

$$
{\mathcal B} \otimes {\mathcal B} ={\yng(2,2)}\oplus{\yng(1)}  = \{ {\bar 6} \} \oplus  {\mathcal A}
$$

$$
{\mathcal A} \otimes {\mathcal B} ={\yng(2,1)}\oplus ~{\bm {\cdot}}
$$

Calculations of the dimension of each of the equations yields
\begin{align*}   
\dim( {\mathcal A}  \otimes {\mathcal A} ) &=  6 \oplus  3
\\
\dim(  {\mathcal B} \otimes {\mathcal B} ) &=  6 \oplus  3
\\
\dim( {\mathcal A} \otimes {\mathcal B}) &= 8 \oplus  1.
\end{align*}

Note that $
\dim( {\mathcal A}  \otimes {\mathcal A} ) = 
\dim( {\mathcal B}  \otimes {\mathcal B} )$, and it is possible to show that this result generalizes to
any monomial power solely of $\mathcal A$ on one side
and the same monomial power solely of $\mathcal B$ on the other side, i.e.
$
\dim( {\mathcal A}  \otimes {\mathcal A} \otimes \cdots \otimes {\mathcal A} ) = 
\dim( {\mathcal B}  \otimes {\mathcal B} \otimes  \cdots {\mathcal B})$.  On the other hand, the multiplication
of ${\mathcal A}$ times ${\mathcal B}$ yields an entirely different result. For example, the three-fold products are
\begin{align*}
A\otimes A\otimes A &= 10 \oplus 2\cdot 8 \oplus 1,\\
B\otimes B\otimes B &= \overline{10} \oplus 2\cdot 8 \oplus 1,
\end{align*}
while the mixed three-fold products are
\begin{align*}
A\otimes A\otimes B &= 15 \oplus \bar{6} \oplus 2\cdot 3,\\
A\otimes B\otimes B &= \overline{15} \oplus 6 \oplus 2\cdot\bar{3}.
\end{align*}
With four factors, the purely chiral and purely anti-chiral products give the same dimensional content,
\begin{equation*}
A^{\otimes 4} = 15' \oplus 3\cdot 15 \oplus 2\cdot \bar{6} \oplus 3\cdot 3, \qquad B^{\otimes 4} = \overline{15}{}' \oplus 3\cdot\overline{15} \oplus 2\cdot 6 \oplus 3\cdot\bar{3},
\end{equation*}
again with $\dim(A^{\otimes 4})=\dim(B^{\otimes 4})$. The mixed four-fold products, by contrast, are
\begin{align*}
A^{\otimes 3}\otimes B & = 24 \oplus 3\cdot 6 \oplus 2\cdot \overline{15} \oplus 3\cdot \bar{3},\\
A^{\otimes 2}\otimes B^{\otimes 2} &= 27 \oplus 10 \oplus \overline{10} \oplus 4\cdot 8 \oplus 2\cdot 1,\\
A\otimes B^{\otimes 3} &= \overline{24} \oplus 3\cdot\bar{6} \oplus 2\cdot 15 \oplus 3\cdot 3.
\end{align*}
These computations confirm the pattern observed at the level of two factors: purely chiral or purely anti-chiral products of the same degree are related by complex conjugation (the outer automorphism), and hence have equal dimensions, while products that mix both $A$ and $B$ generically produce different representation content. This is the $\mathfrak{su}(3)$ manifestation of the same outer automorphism that distinguishes the chiral from the twisted chiral Adinkra. 

If we substitute 
\begin{align*}{\mathcal A} ~=~ \left( \, A, \, B, \,  \psi_{1}, \,  \psi_{2}, \,  \psi_{3}, \,  \psi_{4}, \, F, \, 
G  \right) \\
{\mathcal B} ~=~
\left( \, J, \, L,  \lambda_{1}, \, \lambda_{2}, \, \lambda_{3}, \, \lambda_{4}, \, W, \, D \,\right)\end{align*}
and use these two supermultiplets to construct nonlinear $\sigma$-models, the same general behavior occurs, namely, products ($\sigma$-models) built from a single type of supermultiplet yield equivalent theories related by the outer automorphism, while mixing both types of supermultiplets produces qualitatively new and richer geometric structure. 

We now make this parallel precise by describing the nonlinear $\sigma$-model side of the duality in detail.  This shows concretely how the outer automorphism between chiral and twisted chiral supermultiplets, visible combinatorially in the Adinkra dashings, has profound implications for the geometry of the target space.
 
\medskip
 
 
\subsubsection{Sigma models}
The first two-dimensional supersymmetric nonlinear $\sigma$-models with extended supersymmetry in the literature \cite{Zumino1979} were constructed from the starting point of a collection of chiral supermultiplets (denoted by $\Phi^i$ where the index runs from $1,\ldots,d$ and the $d$-parameter is the dimension of the space of the $\sigma$-model). These superfields are defined by $\theta$-expansions which express how the component fields $X^i(\sigma)$, $\psi^i_+(\sigma)$, $\psi^i_-(\sigma)$, and $F^i(\sigma)$ are embedded within a superfield expansion making use of the Grassmann variables $\theta^+, \theta^-, \bar\theta^+, \bar\theta^-$. Chiral superfields satisfy conditions that involve two of the linear super-differential operators $D_+$, $D_-$, $\bar D_+$ and $\bar D_-$ defined in this superspace:
\begin{equation}
\bar{D}_+\Phi^i = 0 \,, \qquad \bar{D}_-\Phi^i = 0 \,. 
\end{equation}
These superfields are complex-valued and thereby possess complex conjugate equations. An action for these superfields can be defined from
\begin{equation}
S^{(c)} = \int d^2\sigma \, d^4\theta \; K^{(c)}\!\left(\Phi^i,\, \Phi^{\bar\jmath}\right), \qquad d^4\theta \equiv d\theta^+d\theta^-d\bar\theta^+d\bar\theta^- \,, 
\end{equation}
where $K^{(c)}$ is the chiral supermultiplet K\"{a}hler potential and the associated K\"{a}hler metric is
\begin{equation}
g_{i\bar\jmath} = \frac{\partial^2 K^{(c)}}{\partial\Phi^i \,\partial\Phi^{\bar\jmath}} \,. 
\end{equation}
 
Later, after the work of \cite{Zumino1979}, two papers introduced the concept of the \emph{twisted chiral supermultiplet} \cite{Sigma1} and their associated nonlinear $\sigma$-models \cite{Sigma2}. These superfields are also defined by $\theta$-expansions which express how the component fields $Y^i(\sigma)$, $\bar\lambda^i_+(\sigma)$, $\lambda^i_-(\sigma)$, and the real fields $D^i(\sigma)$ and $W^i(\sigma)$ are assembled. Twisted chiral superfields also satisfy conditions that involve two of the linear super-differential operators $D_+$, $D_-$, $\bar D_+$ and $\bar D_-$:
\begin{equation}
D_+\Sigma^i = 0 \,, \qquad \bar{D}_-\Sigma^i = 0 \,. 
\end{equation}
An action for these superfields can be defined from
\begin{equation}
S^{(tc)} = \int d^2\sigma \, d^4\theta \left[-K^{(tc)}\!\left(\Sigma^i,\, \Sigma^{\bar\jmath}\right)\right], \qquad d^4\theta \equiv d\theta^+d\theta^-d\bar\theta^+d\bar\theta^- \,, 
\end{equation}
where $K^{(tc)}$ is the twisted chiral supermultiplet K\"{a}hler potential and the associated K\"{a}hler metric is
\begin{equation}
g_{i\bar\jmath} = -\frac{\partial^2 K^{(tc)}}{\partial\Sigma^i \,\partial\Sigma^{\bar\jmath}} \,.
\end{equation}
 
Looking at the relation of the metric to the K\"{a}hler potential for the chiral theories in comparison to the relation of the metric to the K\"{a}hler potential for the twisted chiral theories, one would first suspect a problem: there is an obvious difference in sign. In the work of \cite{Sigma1} a calculation was undertaken to dispel this concern. The simplest $\sigma$-models are those with flat metrics, i.e.\ $g_{ij} = \delta_{ij}$, and thus one needs to calculate
\begin{equation}
S^{(c)} = \int d^2\sigma \, d^4\theta \; \delta_{ij} \left[ \Phi^i \Phi^{\bar\jmath} \right], 
\end{equation}
and
\begin{equation}
S^{(tc)} = \int d^2\sigma \, d^4\theta \left(-\delta_{ij}\right) \left[ \Sigma^i \Sigma^{\bar\jmath} \right]. 
\end{equation}
Expansion of these respective actions in component fields leads to
\begin{equation}
S^{(c)} = \int d^2\sigma \; \delta_{ij}\!\left[ (\partial_+ X^i)(\partial_- \bar X^{\bar\jmath}) + i\psi^i_- \partial_+ \bar\psi^{\bar\jmath}_- + i\psi^i_+ \partial_- \bar\psi^{\bar\jmath}_+ + F^i \bar F^{\bar\jmath} \right],
\end{equation}
\begin{equation}
S^{(tc)} = \int d^2\sigma \left(-\delta_{ij}\right)\!\left[ (-1)(\partial_- Y^i)(\partial_+ \bar Y^{\bar\jmath}) + i\lambda^i_- \partial_+ \bar\lambda^{\bar\jmath}_- + \bar\lambda^i_+ \partial_- \lambda^{\bar\jmath}_+ + \mathcal{F}^i \mathcal{F}^i \right],
\end{equation}
respectively, and where $\mathcal{F}^i = (D^i - iW^i)$ and $\bar{\mathcal{F}}^i = (D^i + iW^i)$. It is easily seen that the two factors of $(-1)$ in the latter case cancel one another so that the resulting actions are identical in form one to the other. Note that each of the component fields is defined by the `component-by-projection' method:
\begin{equation}
\begin{aligned}
\Phi^i\big|_{\theta=0} &= X^i, &\;
D_+\Phi^i\big|_{\theta=0} &= \sqrt{2}\,\psi^i_+, &\;
D_-\Phi^i\big|_{\theta=0} &= \sqrt{2}\,\psi^i_-, &\;
D_+D_-\Phi^i\big|_{\theta=0} &= 2i F^i, \\[4pt]
\Sigma^i\big|_{\theta=0} &= Y^i, &\;
D_+\Sigma^i\big|_{\theta=0} &= \sqrt{2}\,\bar\lambda^i_+, &\;
\bar D_-\Sigma^i\big|_{\theta=0} &= \sqrt{2}\,\lambda^i_-, &\;
D_+\bar D_-\Sigma^i\big|_{\theta=0} &= 2i\mathcal{F}^i.
\end{aligned}
\end{equation}
 
This is not just the case for the simple actions that depend solely on $\Phi^i$ and $\bar\Phi^{\bar\imath}$ in one case, versus $\Sigma^i$ and $\bar\Sigma^{\bar\imath}$ in the other. Similar calculations reveal that the full nonlinear actions in \cite{Sigma2, HowST, HowP2} yield
\begin{align}
S^{(c)} &= \int d^2\sigma \; K^{(c)}_{i\bar\jmath}\!\left[(\partial_+ X^i)(\partial_- \bar X^{\bar\jmath}) + i\psi^i_-\partial_+\bar\psi^{\bar\jmath}_- + i\psi^i_+\partial_-\bar\psi^{\bar\jmath}_+\right] \notag\\
&\quad + i K^{(c)}_{i\bar\jmath\bar k}\!\left[\psi^i_-(\partial_+\bar X^{\bar k})\bar\psi^{\bar\jmath}_- + \psi^i_+(\partial_-\bar X^{\bar k})\bar\psi^{\bar\jmath}_+\right] \notag\\
&\quad - \left[K^{(c)}_{ij\bar k\bar l} - K^{(c)}_{ij\bar m}(K^{(c)})^{n\bar m}K^{(c)}_{n\bar k\bar l}\right]\psi^i_+\psi^j_-\bar\psi^{\bar k}_+\bar\psi^{\bar l}_- \,, \label{eqn:fullnonlinear}
\end{align}
where the quantities $K^{(c)}_{i\bar\jmath\bar k}$ and $K^{(c)}_{ij\bar k\bar l}$ (etc.) denote partial derivatives of $K^{(c)}$ with respect to the chiral superfield or its conjugate after the elimination of the auxiliary fields $F^i$, and
\begin{align}
S^{(tc)} &= \int d^2\sigma \; K^{(tc)}_{i\bar\jmath}\!\left[(\partial_- Y^i)(\partial_+ \bar Y^{\bar\jmath}) + i\lambda^i_-\partial_+\bar\lambda^{\bar\jmath}_- + i\bar\lambda^i_+\partial_-\lambda^{\bar\jmath}_+\right] \notag\\
&\quad + i K^{(tc)}_{i\bar\jmath\bar k}\!\left[\lambda^i_-(\partial_+\bar Y^{\bar\jmath})\bar\lambda^{\bar k}_- + \bar\lambda^i_+(\partial_-\bar Y^{\bar k})\lambda^{\bar\jmath}_+\right] \notag\\
&\quad - \left[K^{(tc)}_{ij\bar k\bar l} - K^{(tc)}_{ij\bar m}(K^{(tc)})^{n\bar m}K^{(tc)}_{n\bar k\bar l}\right]\bar\lambda^i_+\lambda^j_-\bar\lambda^{\bar k}_-\lambda^{\bar l}_+ \,, \label{eqn:fullnonlineartwist}
\end{align}
where the quantities $K^{(tc)}_{i\bar\jmath\bar k}$ and $K^{(tc)}_{ij\bar k\bar l}$ (etc.) denote partial derivatives of $K^{(tc)}$ with respect to the twisted chiral superfield or its conjugate, respectively, after the elimination of the auxiliary field $\mathcal{F}^i$. The quartic coefficient of the fermionic terms describes the Riemann tensor associated with the metrics in Eqs.~\ref{eqn:fullnonlinear} and~\ref{eqn:fullnonlineartwist}. 
 
A careful examination of the two results in Eqs.~\ref{eqn:fullnonlinear} and~\ref{eqn:fullnonlineartwist} shows them to be identical in spite of the sign differences seen in Eq.~(2.8) compared to Eq.~(2.11), if we replace the component fields according to the rules
\begin{equation}
X^i \longleftrightarrow Y^i, \qquad \psi^i_+ \longleftrightarrow \bar\lambda^i_+, \qquad \psi^i_- \longleftrightarrow \lambda^i_-, \qquad F^i \longleftrightarrow \mathcal{F}^i. 
\end{equation}
This is due to the differences between the differential equations in Eqs.~(2.7) and~(2.10). Thus, in the circumstance of considering nonlinear actions that are dependent only on chiral supermultiplets or only on twisted chiral supermultiplets, the two theories are physically equivalent. 
 
At this point, one might ask the question: ``Are the chiral and the twisted chiral supermultiplets the same?'' This was addressed in the work of \cite{Sigma2} where definitively the answer is shown to be ``No.'' The proof of this inequivalence begins by modifying the nonlinear action to include both chiral and twisted chiral superfields simultaneously,
\begin{equation}
S^{(c|tc)} = \int d^2\sigma \, d^4\theta \; K^{(c|tc)}\!\left(\Phi^i,\,\Phi^{\bar\jmath},\,\Sigma^a,\,\Sigma^{\bar b}\right), \qquad d^4\theta \equiv d\theta^+d\theta^-d\bar\theta^+d\bar\theta^- \,, \label{eqn:simultaneous}
\end{equation}
where $K^{(c|tc)}$ is a potential but it is \emph{not} a K\"{a}hler potential. Before revealing the results that follow from Eq.~\ref{eqn:simultaneous}, an augmentation of notation is in order. We define
\begin{equation}
Z^M = \left(X^i,\, \bar X^{\bar\imath},\, Y^a,\, \bar Y^{\bar a}\right), 
\end{equation}
and define
\begin{equation}
\Psi^M_+ = \left(\psi^i_+,\, \bar\psi^{\bar\imath}_+,\, \bar\lambda^a_+,\, \lambda^{\bar a}_+\right), \qquad \Psi^M_- = \left(\psi^i_-,\, \bar\psi^{\bar\imath}_-,\, \lambda^a_-,\, \bar\lambda^{\bar a}_-\right), 
\end{equation}
which allows the following expression to be derived:
\begin{equation} \label{eqn:simPsi}
\begin{split}
S^{(c|tc)} = \int d^2\sigma \Bigl[\, \tfrac{1}{2}(g+B)_{MN}\,\partial_+Z^M\partial_-Z^N + \tfrac{i}{2}\,g_{MN}\,\Psi^M_-\nabla^{(-)}_+\Psi^N_-  \\
+ \tfrac{i}{2}\,g_{MN}\,\Psi^M_+\nabla^{(+)}_-\Psi^N_+ + \tfrac{1}{4}\,R^{(-)}_{MNPQ}\,\Psi^M_+\Psi^N_+\Psi^P_-\Psi^Q_- \Bigr] 
\end{split}
\end{equation}
 
This compact expression results by taking advantage of all of the following definitions. All derivatives of $K^{(c|tc)}$ mean derivatives of
\begin{equation}
K^{(c|tc)}(X,\,\bar X,\,Y,\,\bar Y)\,. 
\end{equation}
The nonzero metric components are
\begin{equation}
g_{i\bar\jmath} = K^{(c|tc)}_{i\bar\jmath}\,, \qquad g_{a\bar b} = -K^{(c|tc)}_{a\bar b}\,, 
\end{equation}
together with their conjugates, while the mixed metric components vanish:
\begin{equation}
g_{i\bar b} = g_{a\bar\jmath} = 0\,. 
\end{equation}
The mixed second derivatives instead generate the antisymmetric $B$-field:
\begin{equation}
B_{i\bar b} = -K^{(c|tc)}_{i\bar b}\,, \qquad B_{a\bar\jmath} = +K^{(c|tc)}_{a\bar\jmath}\,, 
\end{equation}
with
\begin{equation}
B_{\bar b i} = +K^{(c|tc)}_{i\bar b}\,, \qquad B_{\bar\jmath a} = -K^{(c|tc)}_{a\bar\jmath}\,. 
\end{equation}
The connections appearing in Eq.~(2.24) are
\begin{equation}\label{eqn:bihermitian}
\Gamma^{(\pm)P}{}_{MN} = \Gamma^{(g)P}{}_{MN} \pm \tfrac{1}{2}\,g^{PQ}H_{QMN}\,, \qquad H = dB\,, 
\end{equation}
and $R^{(-)}$ is the curvature of $\Gamma^{(-)}$. 
  
At a high level, the differences between the three actions Eqs.~\ref{eqn:fullnonlinear}, \ref{eqn:fullnonlineartwist}, and \ref{eqn:simPsi} are summarized in Table~\ref{tab:geomstructures}.\footnote{We note that $\omega = i\,g_{i\bar\jmath}\,dX^i\wedge d\bar X^{\bar\jmath}$.}
 
\begin{table}[ht]
\renewcommand{\arraystretch}{1.35}
{
\begin{tabular}{@{}l|c|c|c@{}}
\toprule
& \textbf{Chiral only} & \textbf{Twisted chiral only} & \textbf{Chiral $+$ twisted chiral} \\
\midrule
Complex structures & $J_+,\,J_-$ & $J_+,\,J_-$ & $J_+,\,J_-$ \\
Relation & $J_+=J_-$ & $J_+=-J_-$ & $[J_+,J_-]=0$ \\
Metric & K\"{a}hler & K\"{a}hler & bi-Hermitian \\
$B$-field & locally removable & locally removable & generally nonzero \\
$H=dB$ & $0$ & $0$ & generally nonzero \\
$\nabla J$ & $0$ & $0$ & $\nabla^{(\pm)}J_{\pm}=0$ \\
$d\omega$ & $0$ & $0$ & $d\omega_\pm\neq 0$ in general \\
\bottomrule
\end{tabular}
}
\vspace{12pt}
\caption{In this table, the first column lists properties of geometric structures associated with the geometries defined by the sigma-models in Eqs.~(2.18), (2.19), and (2.24). This table makes explicit the structural analogy with the $\mathfrak{su}(3)$ representation theory discussed above: using only $A$ (chiral) or only $B$ (twisted chiral) yields equivalent geometric structures related by the outer automorphism ($J_+=J_-$ versus $J_+=-J_-$), while combining $A$ and $B$ produces the genuinely new bi-Hermitian geometry with torsion.}
\label{tab:geomstructures}
\end{table}

The connection described in Equation~\ref{eqn:bihermitian} demonstrates that K\"{a}hler geometry must be replaced with bi-Hermitian/ torsionful geometry \cite{Sigma2}. At the time that the work \cite{Sigma2} was written, there was no recognition that manifolds of this nature existed. However, soon after it appeared, a number of other researchers (some independently \cite{HowST, HowP2, StrM} and others building on it \cite{HitGCY, GCY, GGP}) greatly extended the recognition of the existence and utility of such constructions. The work of the second author in \cite{Sigma1}, which delineated in superspace the two representations (the chiral and twisted chiral supermultiplets) of the supersymmetry algebra discussed above, led directly to that in \cite{Sigma2} 
where the second author and collaborators showed that these different representations lead to distinct $\sigma$-models and described their dynamics. Independently, other researchers
\cite{HowST,HowP1,HowP2} probed these same structures.  The results in \cite{Sigma2} dramatically 
demonstrated that simultaneous use of both chiral and twisted multiplets opened a pathway 
to Hitchin-type \cite{HitGCY} complex geometry which then led to Calabi-Yau manifolds with torsion. 
The latter work in \cite{Sigma2} identified the bi-Hermitian manifold with torsion explicitly, and flowed into the 
torsionful string compactification analysis of Strominger~\cite{StrM} and Gualtieri's identification of generalized 
K\"ahler geometry with the earlier bi-Hermitian sigma-model geometry \cite{Gualtieri}.  This current paper marks the first discussion of the analogy of behaviors in $\frak{su}(3)$ algebras with the behavior found in the $\sigma$-models. 

Implications in string theory such as those in the works of \cite{GCY,Huybrechts-K3-Bfield,Lindstrom-Rocek-vonUnge-Zabzine-GK}
extended the applications of bi-Hermitian and generalized 
K\"ahler geometry. Exchange of chiral and twisted chiral supermultiplets also appears in the context of the mirror correspondence for $\mathcal{N} = (2,2)$ supersymmetry. In this context, mirror symmetry produces an outer automorphism that maps chiral multiplets on one side to twisted chiral multiplets of the mirror, and vice versa (see \cite[\S 12.4]{MirrorSymmetry} and \cite{lookingglass}).

\section{Combinatorics of Adinkra heights}\label{sec:comb}

An {\it Adinkra topology} is an $N$-regular bipartite graph $A$. It is conventional to refer to the two vertex sets in the bipartition as bosons and fermions. We say that an Adinkra topology with a specific choice of which set is the bosons and which set is the fermions is \emph{spin-labeled}.
 A {\it chromotopology} is an Adinkra topology $A$ equipped with an $n$-coloring of the edges of $A$, such that for any two distinct colors $j$ and $k$, $\{e \in E(A) \mid {\rm color}(e) \in \{j,k\}\}$ is comprised of disjoint  4-cycles. If $u$ and $v$ are adjacent along an edge of color $i$, we denote this by $u \widesim[2]{I} v$. Every $N$-colored Adinkra chromotopology can be obtained as a quotient of the hypercube $H^N$ by a \emph{doubly even code}, where $H^N$ denotes the hypercube with $2^N$ vertices in $\{0,1\}^N$ and vertices $v$ and $v'$ are adjacent if $v$ and $v'$ differ in exactly one coordinate. Our convention will be to say that the vertices whose hypercube representations have even parity are bosons, and those with odd parity are fermions. We refer to a choice of cyclic ordering for the $N$ edge colors as a \emph{rainbow}.\\

 A {\it height function} on a bipartite graph $A$ is a map $h: V(A) \to \mathbb Z_{\geq 0}$ whose values on adjacent vertices differ by 1, that is, $|h(v_1) - h(v_2)| = 1 $ if $(v_1,v_2) \in E(A)$. We refer to an Adinkra topology equipped with a height function as an \emph{acrotopology}. (The Greek prefix \emph{acro-} refers to the highest place, as in `acrobat' or `acrophobia'.)  
 We say that two heights $h$ and $h'$ are the same if they differ by a constant (i.e., there exists $a \in \mathbb{Z}$ such that $h(v) = h'(v)+a$ for all $v \in V(A)$). Our convention will be to use the height representative that satisfies $\min_{v \in V(A)} h(v) = 0$. If the vertices of minimal height are bosons, then we call the Adinkra {\it boson-based}, and we call it {\it fermion-based} otherwise. 

 The two extremal examples of heights on an Adinkra $A$ are the valise height (with one part of the bipartition of vertices at height 1 and the other part at height 0) and the fully extended height (with only a single pegged vertex at height ${\rm diam}(A)$), as already seen in Figures~\ref{fig:CM_TCM} and \ref{fig:A2}. Each height has a {\it height signature} $(n_{k} \mid n_{k-1} \mid \ldots \mid n_1 \mid n_0)$, where $n_j$ is the number of vertices with height equal to $j$. Such a height is called a $k$-level  height.  For example, the height signature of the valise is $(2^{N-1}\mid 2^{N-1})$, and it is a 2-level height.  \\

The classification of acrotopologies is an interesting and highly nontrivial combinatorial problem. The number of heights on an Adinkra is one third of the number of three-colorings of the vertices of the underlying graph \cite{Z13}, which is a well-studied problem in combinatorics. Exact values are known for  $N\leq 6$ \cite{S}, and asymptotic properties have been described in \cite{Gal}. An algorithm for enumerating all heights on the $N$-hypercube was given in \cite{Z13}.

\begin{notation}[Height notations]\label{not:heights}

Every height is uniquely determined by its {\it pinned vertices} $P(h)$ and their heights. That is, $v$ is a pinned vertex for $h$ if it is a local maximum, i.e., $h(v) \geq h(v')$ for any $v'$ adjacent to $v$. The heights of all other vertices are determined by their relationship to the pinned vertices: $h(u) = \max\{ h(v)-{\rm dist}_A(u,v) \mid v \in P(h)\}$.

For hypercube Adinkras, another way to represent a height function is by placing the heights of the vertices into a certain vector. To do this, we set an order on the vertices of an Adinkra $A$, via $u<v$ if $u $ is a boson and $v$ is a fermion, or if $u$ and $v$ are both either bosons or fermions and $u<_{\rm lex} v$ (using the hypercube representation of each vertex). 

This representation of a height will be called the {\it elevator vector}.
\end{notation}

\begin{example}
    The vertex ordering for the $N=3$ hypercube  $H^3$ is 
    \[
    [0,0,0],\ [1,1,0],\ [1,0,1],\ [0,1,1],\ [1,0,0],\ [0,1,0],\ [0,0,1],\ [1,1,1] .
    \]
    A fully extended height on $H^3$ then might have  elevator vector $[0,2,2,2,1,1,1,3]$, whereas a valise height might have elevator vector $[0,0,0,0,1,1,1,1]$.
\end{example}

\subsection{Equivalence classes of heights}
\begin{figure}[h]
\centering
\resizebox{0.8\textwidth}{!}{
       \begin{tikzpicture}[> = stealth,  shorten > = 2pt,   auto,   node distance = .5cm,state/.style={},scale=.9, outer sep =0 pt]

        \node[state]  (ha) at (-9,0) {
                    \resizebox{1.2cm}{!}{
                   \begin{tikzpicture}[> = stealth,  shorten > = 2pt,   auto,   node distance = .5cm,boson/.style={circle, draw, thick, fill=white, minimum size=3mm},fermion/.style={circle, draw, thick, fill=black, text=white, minimum size=5mm}, inner sep = 2pt]
                    \node[boson]  (haa) at (0,1) {$A$};
                    \node[fermion]  (hba) at (-1,0) {$B$};
                    \node[fermion]  (hca) at (1,0) {$C$};
                    \node[boson]  (hda) at (0,-1) {$D$};
                   	\path[color = red] (haa)  edge node {} (hba);
                   	\path[color = blue] (haa)  edge node {} (hca);
                   	\path[color = blue] (hda)  edge node {} (hba);
                   	\path[color = red] (hda)  edge node {} (hca);
                  \end{tikzpicture}}};
        \node[state]  (hb) at (-4.5,2) {
                    \resizebox{1cm}{!}{
                   \begin{tikzpicture}[> = stealth,  shorten > = 2pt,   auto,   node distance = .5cm,state/.style={circle, draw, minimum size=.25cm},boson/.style={circle, draw, thick, fill=white, minimum size=3mm},fermion/.style={circle, draw, thick, fill=black, text=white, minimum size=5mm}, inner sep = 2pt]
                    \node[boson]  (hab) at (0,0) {$A$};
                    \node[fermion]  (hbb) at (0,1.5) {$B$};
                    \node[fermion]  (hcb) at (1.5,1.5) {$C$};
                    \node[boson]  (hdb) at (1.5,0) {$D$};
                   	\path[color = red] (hab)  edge node {} (hbb);
                   	\path[color = blue] (hab)  edge node {} (hcb);
                   	\path[color = blue] (hdb)  edge node {} (hbb);
                   	\path[color = red] (hdb)  edge node {} (hcb);
                  \end{tikzpicture}}};

        \node[state]  (hc) at (-6,0) {
                    \resizebox{1.2cm}{!}{
                   \begin{tikzpicture}[> = stealth,  shorten > = 2pt,   auto,   node distance = .5cm,state/.style={circle, draw, minimum size=.25cm},boson/.style={circle, draw, thick, fill=white, minimum size=3mm},fermion/.style={circle, draw, thick, fill=black, text=white, minimum size=5mm}, inner sep = 2pt]
                    \node[fermion]  (hac) at (0,1) {$B$};
                    \node[boson]  (hbc) at (-1,0) {$A$};
                    \node[boson]  (hcc) at (1,0) {$D$};
                    \node[fermion]  (hdc) at (0,-1) {$C$};
                   	\path[color = red] (hac)  edge node {} (hbc);
                   	\path[color = blue] (hac)  edge node {} (hcc);
                   	\path[color = blue] (hdc)  edge node {} (hbc);
                   	\path[color = red] (hdc)  edge node {} (hcc);
                  \end{tikzpicture}}};
        \node[state]  (hd) at (-3,0) {
                    \resizebox{1.2cm}{!}{
                   \begin{tikzpicture}[> = stealth,  shorten > = 2pt,   auto,   node distance = .5cm,state/.style={circle, draw, minimum size=.25cm},boson/.style={circle, draw, thick, fill=white, minimum size=3mm},fermion/.style={circle, draw, thick, fill=black, text=white, minimum size=5mm},inner sep = 2pt]
                    \node[fermion]  (had) at (0,1) {$C$};
                    \node[boson]  (hbd) at (-1,0) {$A$};
                    \node[boson]  (hcd) at (1,0) {$D$};
                    \node[fermion]  (hdd) at (0,-1) {$B$};
                   	\path[color = red] (had)  edge node {} (hbd);
                   	\path[color = blue] (had)  edge node {} (hcd);
                   	\path[color = blue] (hdd)  edge node {} (hbd);
                   	\path[color = red] (hdd)  edge node {} (hcd);
                  \end{tikzpicture}}};
        \node[state]  (he) at (-4.5,-2) {
                    \resizebox{1cm}{!}{
                   \begin{tikzpicture}[state/.style={circle, draw},boson/.style={circle, draw, thick, fill=white, minimum size=3mm},fermion/.style={circle, draw, thick, fill=black, text=white, minimum size=5mm}, inner sep = 2pt]
                    \node[boson]  (hae) at (0,1.5) {$A$};
                    \node[fermion]  (hbe) at (0,0) {$B$};
                    \node[fermion]  (hce) at (1.5,0) {$C$};
                    \node[boson]  (hde) at (1.5,1.5) {$D$};
                   	\path[color = red] (hae)  edge node {} (hbe);
                   	\path[color = blue] (hae)  edge node {} (hce);
                   	\path[color = blue] (hde)  edge node {} (hbe);
                   	\path[color = red] (hde)  edge node {} (hce);
                  \end{tikzpicture}}};
        \node[state]  (hf) at (0,0) {
                    \resizebox{1.2cm}{!}{
                   \begin{tikzpicture}[> = stealth,  shorten > = 2pt,   auto,   node distance = .5cm,state/.style={circle, draw},boson/.style={circle, draw, thick, fill=white, minimum size=3mm},fermion/.style={circle, draw, thick, fill=black, text=white, minimum size=5mm}, inner sep = 2pt]
                    \node[boson]  (haf) at (0,1) {$D$};
                    \node[fermion]  (hbf) at (-1,0) {$B$};
                    \node[fermion]  (hcf) at (1,0) {$C$};
                    \node[boson]  (hdf) at (0,-1) {$A$};
                   	\path[color = red] (haf)  edge node {} (hbf);
                   	\path[color = blue] (haf)  edge node {} (hcf);
                   	\path[color = blue] (hdf)  edge node {} (hbf);
                   	\path[color = red] (hdf)  edge node {} (hcf);
                  \end{tikzpicture}}};
       
		\path[->, violet] (ha)  edge node {} (hb);
		\path[->, violet] (hb)  edge node {} (hc);
		\path[->, violet] (hb)  edge node {} (hd);
		\path[->, violet] (hc)  edge node {} (he);
		\path[->, violet] (hd)  edge node {} (he);
		\path[->, violet] (he)  edge node {} (hf);
		\path[->, violet] (he)  edge node {} (ha);
		\path[->, violet] (hf)  edge node {} (hb);

      \end{tikzpicture}}

\caption{The digraph $\Gamma_2$ whose vertices are the  heights on $H^4$ and whose directed edges come from vertex lowering. }\label{fig:Gamma_2}
\end{figure}
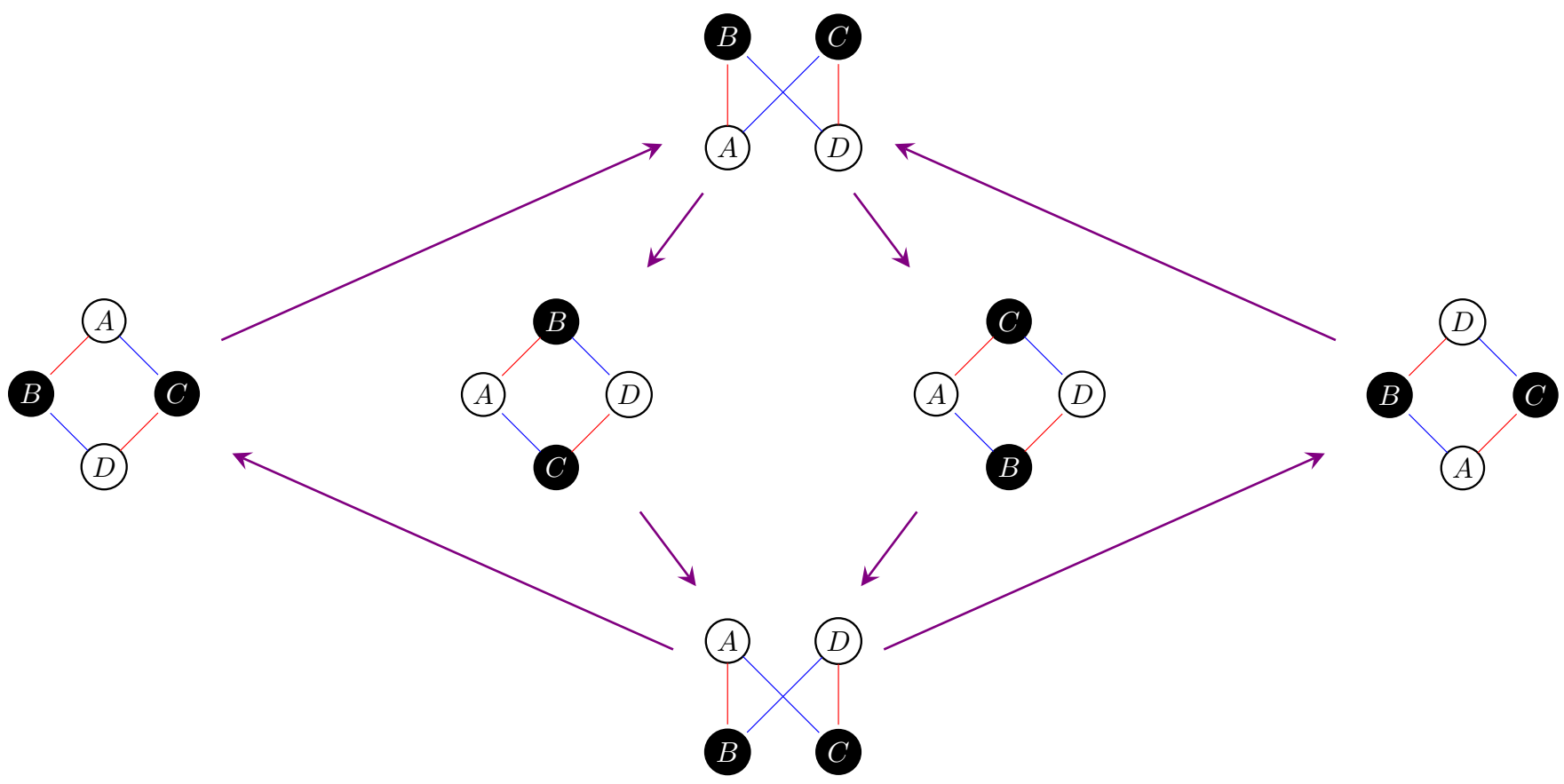

In this section, we classify heights by considering some idea of the shape of their acrotopologies. Consider the following situation for the $N = 2$ case: 
In Figure \ref{fig:Gamma_2}, the six distinct height functions on a 2-cube (a square) are shown, classified by two combinatorial shapes: diamonds and bow-ties. We aim to make this classification into shapes explicit for more complicated Adinkras, by introducing decorated-graph isomorphisms, which preserve the ``shape'' of the acrotopology, while permuting vertices and edges of the graph in an admissible way.

We introduce two types of graph isomorphisms, which produce not only a related graph, but a new Adinkra with related structure. If the Adinkra $A_1$ is the image of another $A_2$ along a composition of the maps described here, vertices and colors may be permuted, but the vertex bipartition and the edge $N$-partition are maintained, as is the adjacency structure of vertices. In other words, if you forgot the names of the vertices and the names of each color, the two underlying acrotopologies would be the same. 

\begin{definition}
    Let $I \in \{1, \ldots, N\}$, $v \in V(H^N)$, and $\sigma \in S_N$, where $H^N$ is the $N$-dimensional hypercube Adinkra. Then we define the acrotopology isomorphisms $N_I, C_{v,\sigma}\colon H^N \to H^N$  in the following way:
    
    \begin{itemize}
        \item $N_I$ sends each vertex to its neighbor along color $I$, $N_I(v) = u$ if $v\widesim[2]{I} u$.  
        \item $C_{v,\sigma}$ fixes a base vertex $v$, and permutes edge colors (i.e., $I \mapsto \sigma(I)$) and  vertices $u\neq v$ according to the  colors in a path between $u$ and $v$, i.e.,   if a minimal path from $v$ to $u$ travels along colors $I_1, \ldots, I_s$, then $C_{v,\sigma}$ sends $u$ to the vertex obtained by traveling from $v$ along colors $\sigma(I_1), \ldots, \sigma(I_s)$. 
    \end{itemize}
\end{definition}

\begin{figure}[h]
\centering
       \begin{tikzpicture}[> = stealth,  shorten > = 2pt,   auto,   node distance = .5cm,state/.style={},scale=.9, outer sep =0 pt]

        \node[state]  (h1) at (-4,2) {
                    \resizebox{.3\textwidth}{!}{
                   \begin{tikzpicture}[> = stealth,  shorten > = 2pt,   auto,   node distance = .5cm,boson/.style={ draw, thick, fill=white, minimum size=3mm},fermion/.style={ draw, thick, fill=black, text=white, minimum size=5mm}, inner sep = 2pt]
                    \node[boson]  (ha1) at (0,1) {$0,0,0$};
                    \node[boson]  (ha2) at (1.5,1) {$1,1,0$};
                    \node[boson]  (ha3) at (3,-1) {$1,0,1$};
                    \node[boson]  (ha4) at (4.5,-1) {$0,1,1$};
                    \node[fermion]  (hb1) at (0,0) {$1,0,0$};
                    \node[fermion]  (hb2) at (1.5,0) {$0,1,0$};
                    \node[fermion]  (hb3) at (3,0) {$0,0,1$};
                    \node[fermion]  (hb4) at (4.5,0) {$1,1,1$};
                   	\path[color = red] (ha1)  edge node {} (hb1);
                   	\path[color = blue] (ha1)  edge node {} (hb2);
                   	\path[color = orange] (ha1)  edge node {} (hb3);
                   	\path[color = red] (ha2)  edge node {} (hb2);
                   	\path[color = blue] (ha2)  edge node {} (hb1);
                   	\path[color = orange] (ha2)  edge node {} (hb4);
                   	\path[color = red] (ha3)  edge node {} (hb3);
                   	\path[color = blue] (ha3)  edge node {} (hb4);
                   	\path[color = orange] (ha3)  edge node {} (hb1);
                   	\path[color = red] (ha4)  edge node {} (hb4);
                   	\path[color = blue] (ha4)  edge node {} (hb3);
                   	\path[color = orange] (ha4)  edge node {} (hb2);
                  \end{tikzpicture}}};
        \node[state]  (h2) at (-4,-2) {
                    \resizebox{.3\textwidth}{!}{
                   \begin{tikzpicture}[> = stealth,  shorten > = 2pt,   auto,   node distance = .5cm,boson/.style={ draw, thick, fill=white, minimum size=3mm},fermion/.style={ draw, thick, fill=black, text=white, minimum size=5mm}, inner sep = 2pt]
                    \node[boson]  (ha1) at (0,1) {$0,1,1$};
                    \node[boson]  (ha2) at (1.5,1) {$1,0,1$};
                    \node[boson]  (ha3) at (3,-1) {$1,1,0$};
                    \node[boson]  (ha4) at (4.5,-1) {$0,0,0$};
                    \node[fermion]  (hb1) at (0,0) {$1,1,1$};
                    \node[fermion]  (hb2) at (1.5,0) {$0,0,1$};
                    \node[fermion]  (hb3) at (3,0) {$0,1,0$};
                    \node[fermion]  (hb4) at (4.5,0) {$0,0,1$};
                   	\path[color = red] (ha1)  edge node {} (hb1);
                   	\path[color = blue] (ha1)  edge node {} (hb2);
                   	\path[color = orange] (ha1)  edge node {} (hb3);
                   	\path[color = red] (ha2)  edge node {} (hb2);
                   	\path[color = blue] (ha2)  edge node {} (hb1);
                   	\path[color = orange] (ha2)  edge node {} (hb4);
                   	\path[color = red] (ha3)  edge node {} (hb3);
                   	\path[color = blue] (ha3)  edge node {} (hb4);
                   	\path[color = orange] (ha3)  edge node {} (hb1);
                   	\path[color = red] (ha4)  edge node {} (hb4);
                   	\path[color = blue] (ha4)  edge node {} (hb3);
                   	\path[color = orange] (ha4)  edge node {} (hb2);
                     \end{tikzpicture}}};

        \node[state]  (h3) at (4,2) {
                    \resizebox{.3\textwidth}{!}{
                   \begin{tikzpicture}[> = stealth,  shorten > = 2pt,   auto,   node distance = .5cm,boson/.style={ draw, thick, fill=white, minimum size=3mm},fermion/.style={ draw, thick, fill=black, text=white, minimum size=5mm}, inner sep = 2pt]
                    \node[boson]  (ha1) at (0,1) {$0,0,0$};
                    \node[boson]  (ha2) at (1.5,1) {$0,1,1$};
                    \node[boson]  (ha3) at (3,-1) {$1,0,1$};
                    \node[boson]  (ha4) at (4.5,-1) {$1,1,0$};
                    \node[fermion]  (hb1) at (0,0) {$0,0,1$};
                    \node[fermion]  (hb2) at (1.5,0) {$0,1,0$};
                    \node[fermion]  (hb3) at (3,0) {$1,0,0$};
                    \node[fermion]  (hb4) at (4.5,0) {$1,1,1$};
                   	\path[color = red] (ha1)  edge node {} (hb1);
                   	\path[color = blue] (ha1)  edge node {} (hb2);
                   	\path[color = orange] (ha1)  edge node {} (hb3);
                   	\path[color = red] (ha2)  edge node {} (hb2);
                   	\path[color = blue] (ha2)  edge node {} (hb1);
                   	\path[color = orange] (ha2)  edge node {} (hb4);
                   	\path[color = red] (ha3)  edge node {} (hb3);
                   	\path[color = blue] (ha3)  edge node {} (hb4);
                   	\path[color = orange] (ha3)  edge node {} (hb1);
                   	\path[color = red] (ha4)  edge node {} (hb4);
                   	\path[color = blue] (ha4)  edge node {} (hb3);
                   	\path[color = orange] (ha4)  edge node {} (hb2);
                     \end{tikzpicture}}};
        \node[state]  (h4) at (4,-2) {
                    \resizebox{.3\textwidth}{!}{
                   \begin{tikzpicture}[> = stealth,  shorten > = 2pt,   auto,   node distance = .5cm,boson/.style={ draw, thick, fill=white, minimum size=3mm},fermion/.style={ draw, thick, fill=black, text=white, minimum size=5mm}, inner sep = 2pt]
                    \node[boson]  (ha1) at (0,1) {$0,1,1$};
                    \node[boson]  (ha2) at (1.5,1) {$0,0,0$};
                    \node[boson]  (ha3) at (3,-1) {$1,1,0$};
                    \node[boson]  (ha4) at (4.5,-1) {$1,0,1$};
                    \node[fermion]  (hb1) at (0,0) {$0,1,0$};
                    \node[fermion]  (hb2) at (1.5,0) {$0,0,1$};
                    \node[fermion]  (hb3) at (3,0) {$1,1,1$};
                    \node[fermion]  (hb4) at (4.5,0) {$1,0,0$};
                   	\path[color = red] (ha1)  edge node {} (hb1);
                   	\path[color = blue] (ha1)  edge node {} (hb2);
                   	\path[color = orange] (ha1)  edge node {} (hb3);
                   	\path[color = red] (ha2)  edge node {} (hb2);
                   	\path[color = blue] (ha2)  edge node {} (hb1);
                   	\path[color = orange] (ha2)  edge node {} (hb4);
                   	\path[color = red] (ha3)  edge node {} (hb3);
                   	\path[color = blue] (ha3)  edge node {} (hb4);
                   	\path[color = orange] (ha3)  edge node {} (hb1);
                   	\path[color = red] (ha4)  edge node {} (hb4);
                   	\path[color = blue] (ha4)  edge node {} (hb3);
                   	\path[color = orange] (ha4)  edge node {} (hb2);
                     \end{tikzpicture}}};
       
		\path[->, violet] (h1)  edge node {} (h2);
		\path[->, violet] (h1)  edge node {} (h3);
		\path[->, violet] (h2)  edge node {} (h4);
		\path[->, violet] (h3)  edge node {} (h4);

        \node[violet] at (0,2.5) {{\small $C_{[0,0,0],(13)}$}};
        \node[violet] at (0,-2.5) {{\small $C_{[0,0,1],(13)}$}};
        \node[violet] at (5,0) {{\small $N_2\circ N_3$}};
        \node[violet] at (-5,0) {{\small $N_2\circ N_3$}};
      \end{tikzpicture}

\caption{The maps $N_2$, $N_3$,  $C_{[0,0,0],(13)}$, and $C_{[1,0,0],(13)}$  permute heights and horizontal ordering of the hypercube Adinkra $H^3$. }\label{fig:N_C}
\end{figure}
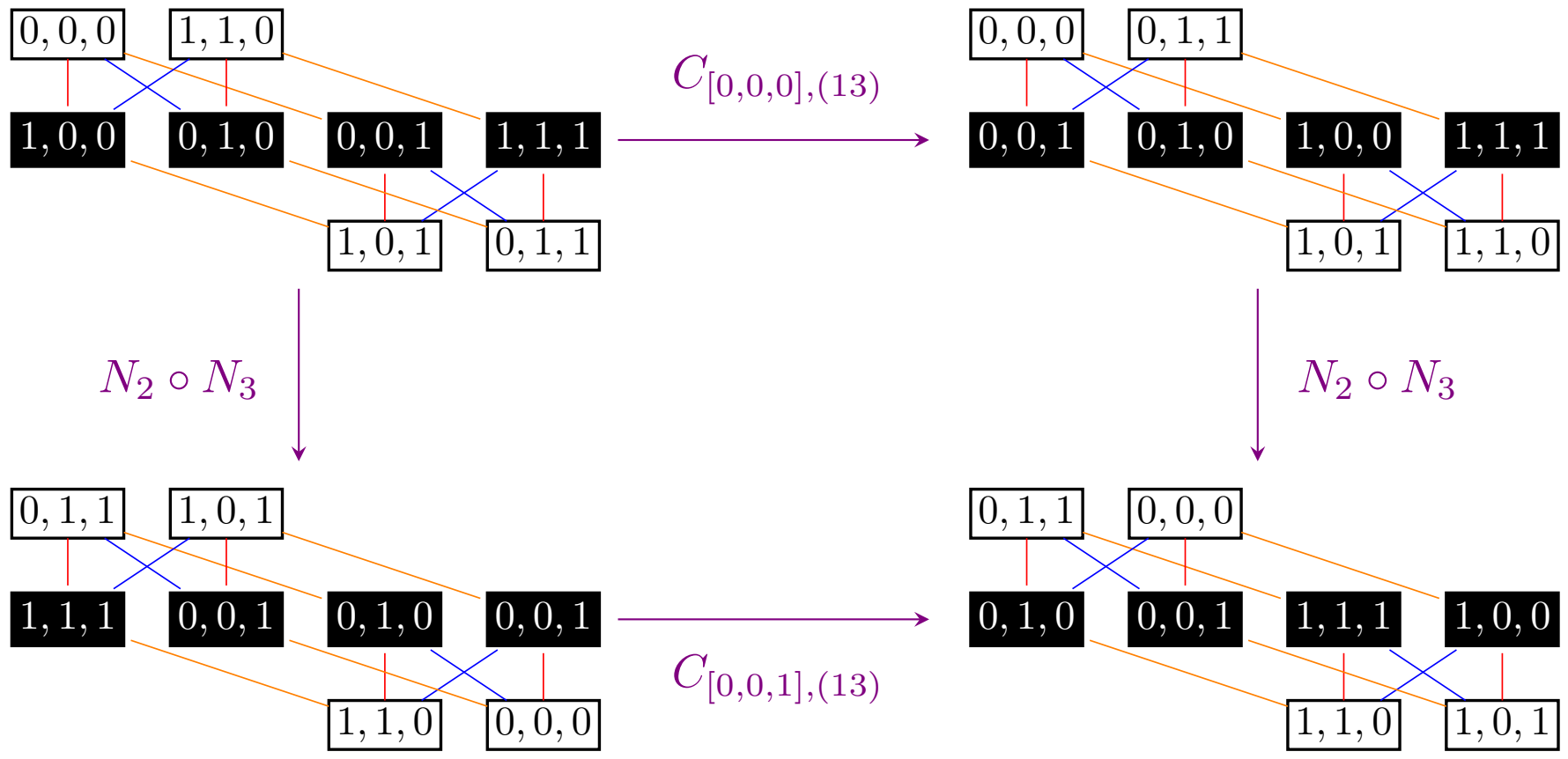

The 4-cycle condition on Adinkras implies that $u \widesim[2]{J} u'$ if and only if $N_I(u) \widesim[2]{J} N_I(u')$ if an only if $C_{v,\sigma}(u) \widesim[2]{\sigma(J)} C_{v,\sigma}(u')$. Thus, $N_I$ and $C_{v,\sigma}$ are indeed graph isomorphisms. 

Note that $N_i$ has order 2 and acts as its own inverse, and $C_{v,\sigma}$ has order $|\sigma|$ and has $C_{v,\sigma^{-1}}$ as its inverse. For example, the maps $N_I$ ($I = 1,2,3$) and $C_{v,\sigma}$ ($v \in H^3$, $\sigma \in S_3$) permute the vertices of the hypercube Adinkra $H^3$. These actions do not, in general, commute, as seen in Figure~\ref{fig:N_C}. In Figure~\ref{fig:N_C}, we also see that both the horizontal and vertical locations (the heights and the ordering of the vertices, respectively) are changed by $N_i$, and $C_{v,\sigma}$.  It will be convenient for us to also consider these maps as acting on the elevator vectors (Notation~\ref{not:heights}), as demonstrated in the following example.

\begin{example}
   For the $N=3$ hypercube  $H^3$, the matrix representations of $N_2$, $N_3$, and $C_{[0,0,0],\sigma_1}$ are given below.
   \[
   N_2 = \left[\begin{array}{cc|cc}
   O &O & J_2 & O\\
   O & O & O & J_2 \\\hline
   J_2 & O & O & O \\
   O & J_2 & O & O 
   \end{array}\right], \quad 
   N_3 = \left[\begin{array}{cc|cc}
   O &O & O & I_2\\
   O & O & I_2 & O \\\hline
   O &I_2 & O & O \\
   I_2 & O & O & O 
   \end{array}\right], \quad 
   C_{v,(1\, 3)} = \left[\begin{array}{cc|cc}
   E_{1\,1} & E_{2\,2} & O & O\\
   E_{2\,2} & E_{1\,1} & O & O \\\hline
   O & O & E_{2\,2} & E_{1\,1} \\
   O & O & E_{1\,1} & E_{2\,2} 
   \end{array}\right], 
   \]

   where $J_2 = \left[\begin{array}{cc}0 & 1 \\ 1 & 0  \end{array}\right]$, $I_2 = \left[\begin{array}{cc}1 & 0 \\ 0 & 1  \end{array}\right]$, $E_{1\, 1} = \left[\begin{array}{cc}1 & 0 \\ 0 & 0  \end{array}\right]$, $E_{2\, 2} = \left[\begin{array}{cc}0 & 0 \\ 0 & 1  \end{array}\right]$, and $O = \left[\begin{array}{cc}0 & 0 \\ 0 & 0  \end{array}\right]$.
   
Consider the elevator vector for the height shown in the upper left corner of Figure~\ref{fig:N_C}: $h = [2,2,0,0,1,1,1,1]$. Then, 
\[
N_2\circ N_3(h) = \left[\begin{array}{cc|cc}
   O &O & J_2 & O\\
   O & O & O & J_2 \\\hline
   J_2 & O & O & O \\
   O & J_2 & O & O 
   \end{array}\right]\cdot
   \left[\begin{array}{cc|cc}
   O &O & O & I_2\\
   O & O & I_2 & O \\\hline
   O &I_2 & O & O \\
   I_2 & O & O & O 
   \end{array}\right] \cdot [ 2\ 2\ 0\ 0\ 1\ 1\ 1\ 1]^T = [ 0\ 0\ 2\ 2\ 1\ 1\ 1\ 1]^T
\]

\[
C_{[0,0,0],(1\, 3)}(h) = \left[\begin{array}{cc|cc}
   E_{1\,1} & E_{2\,2} & O & O\\
   E_{2\,2} & E_{1\,1} & O & O \\\hline
   O & O & E_{2\,2} & E_{1\,1} \\
   O & O & E_{1\,1} & E_{2\,2} 
   \end{array}\right] \cdot [2\ 2\ 0\ 0\ 1\ 1\ 1\ 1]^T = [ 2\ 0\ 0\ 2\ 1\ 1\ 1\ 1]^T
\]

\end{example}

\begin{proposition}\label{prop:CNmatrices}
    \[
    N_J = \left[\begin{array}{c|c}
   0 & N_J' \\ \hline N_J' & 0 \end{array} \right]
    \quad \text{and} \quad C_{v,\sigma} = \left[\begin{array}{c|c}
    C_{v,\sigma}' & 0 \\ \hline 0& C_{v,\sigma}'\cdot N'_{\sigma(1)} \end{array} \right],
    \]
    where $N_J'$ and $C_{v,\sigma}'$ are $8 \times 8 $ permutation matrices. 
\end{proposition}
\begin{proof}
    The form of $N_J$ follows directly from the observations that 
    \[
    (N_J')_{i,j} = \begin{cases}
        1 & b_i \widesim[2]{J} f_j \\ 
        0 & \text{otherwise}
    \end{cases}
    \]
    and $b_i \widesim[2]{J} f_j $ iff $b_j \widesim[2]{J} f_i$ by the 4-cycle property of Adinkras. 
    
    Similarly, the structure of $C_{v,\sigma}$ follows from the observation that for $1\leq i,j\leq 8$, $(C_{v,\sigma})_{i,j} = 1$ iff a path from $v$ to $b_i$ consists of colors $I_1, \ldots, I_\ell$, and the path from $v$ traveling along colors $\sigma(I_1),\ldots, \sigma(I_\ell)$ terminates at $b_j$. In this case, $I_1, \ldots, I_\ell,1 $ gives a path from $v$ to $f_i$, and the path $ \sigma(I_1), \ldots, \sigma(I_\ell), \sigma(1)$ gives a path from $f_i$ to $f_j$. 
\end{proof}

\begin{definition}
    Let $v = (0,0,\ldots 0) \in H^N$, and  let $\sigma_1 = (1 \ 2)$, $\sigma_2 = (1\ 2 \ \ldots \ N)$ be permutations in $S_N$.
    The orbits of the elevator vectors under $\langle N_1, \ldots, N_N, C_{v,\sigma_1}, C_{v,\sigma_2} \rangle $ give equivalence classes called  {\it FWI} classes. The orbits of elevator vectors under $\langle N_1, \ldots, N_N, C_{v,\sigma_2} \rangle $ give equivalence classes called  {\it rainbow} classes.  
\end{definition}

\begin{remark}
    Note that for $h$ and $h'$ in the same FWI-class, it is always possible to find $I_1,\ldots, I_\ell \in [N]$, $v\in H^N$, and $\sigma \in S_N$, such that $h' = C_{v,\sigma} \cdot N_{I_1} \cdots N_{I_\ell} h $.
\end{remark}

Here FWI stands for Francis--Whitcher Index, after unpublished work of the first and last authors. We use the terminology rainbow classes because these classes are obtained by preserving the order of edge colors, that is, by only allowing rotations of the rainbow.

The equivalence classes mentioned above can be used to organize height functions on the $N$-hypercube $H^N$ in a directed graph $\Gamma_N$ in the following way. Given height functions $h$ and $h'$,  we say that $h \rightarrow h'$ if for some vertex $v \in V(H^N)$, 
 $h'|_{V(H^N)\setminus \{v\}}  = h$ and $h'(v) = h(v)-2$ (in this case $v$ is a local max in $h$ and a local min in $h'$). In other words, adjacency in $\Gamma_N$ is given by vertex lowering.
Note also that the ``reduced'' digraph $\widetilde \Gamma_N$ of height equivalence classes $[h]_{\rm c}$ (where, once again, adjacency is given by vertex-lowering) is well-defined. 
The details of the number and digraph structure of these equivalence classes is a nontrivial problem. 

In \cite{YXZhang}, Zhang organized heights on an Adinkra $A$ into symmetric ranked posets $P_v(A)$, which depend on a choice of vertex $v \in A$, and are defined carefully to avoid cycles. He also proved that $P_v(A)$ is a finite distributive lattice.  Here, we allow cycles in both $\Gamma_N$ and $\widetilde \Gamma_N$, to demonstrate the relationships between heights in a different way.  

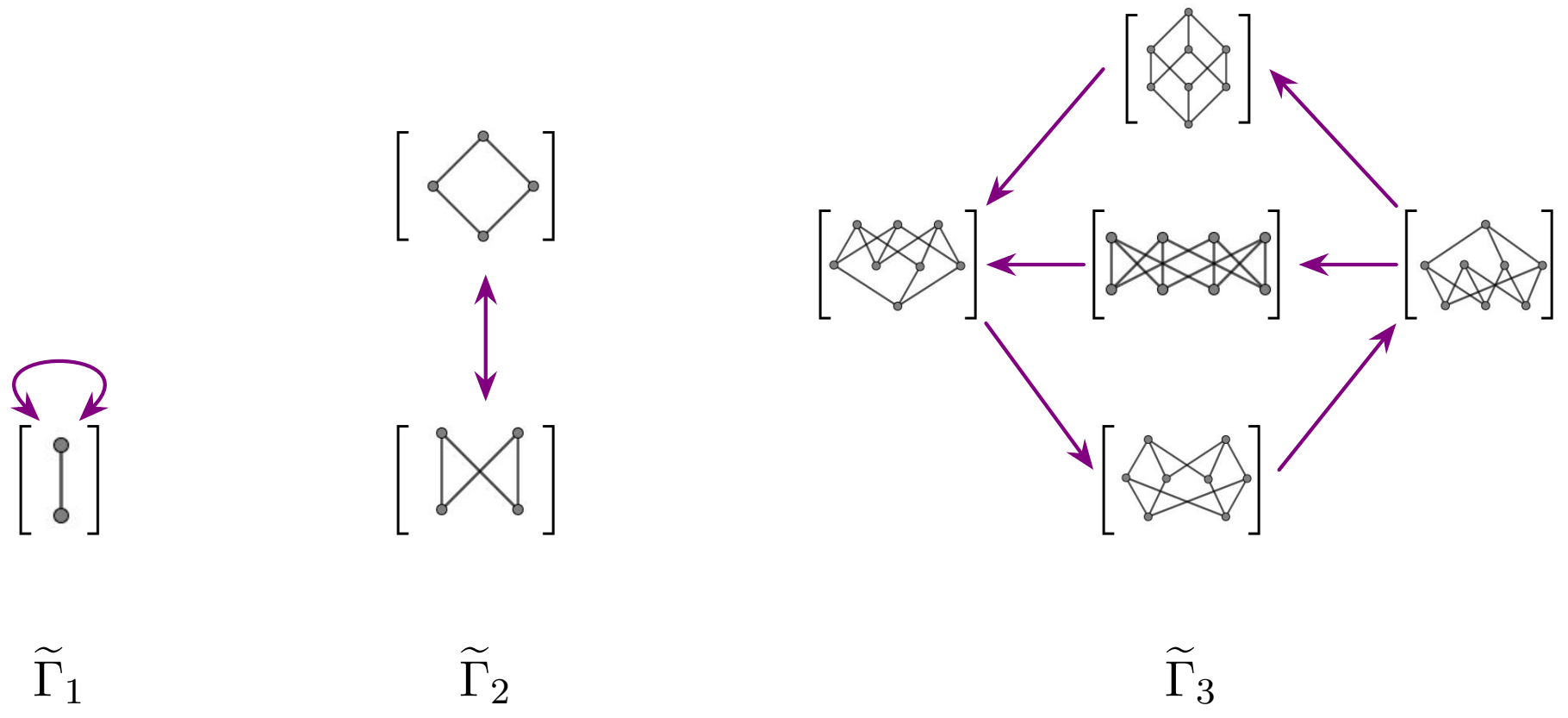
\begin{figure}[H]
\begin{center}
\begin{tikzpicture}[line cap=round,line join=round,>=triangle 45,x=1cm,y=1cm,scale = 1.5,c/.style={insert path={circle[radius=.8pt]}}]

 \draw[Stealth-Stealth, line width = .8pt, violet] (.4,.3) to [out=135,in=45,looseness=7] (.6,.3);

\begin{scriptsize}
\draw[color=black] (.3,0) node {$\Bigg[$};

  \node at (.52,0) {\includegraphics[height = .8cm]{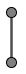}};
  \draw[color=black] (.7,0) node {$\Bigg]$};
\end{scriptsize}

\node at (.5,-1) {$\widetilde \Gamma_1$};
\end{tikzpicture}
\qquad\qquad
\begin{tikzpicture}[line cap=round,line join=round,>=triangle 45,x=1cm,y=1cm,scale = 1.5,c/.style={insert path={circle[radius=.8pt]}}]
\draw [Stealth-Stealth,line width=.8pt, violet] (.75,0.4)-- (.75,1.05);

\begin{scriptsize}
  \node at (.72,.05) {\includegraphics[height = .8cm]{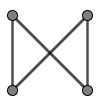}};
\draw[color=black] (.3,0) node {$\Bigg[$};
  \draw[color=black] (1.1,0) node {$\Bigg]$};

\draw[color=black] (.3,1.5) node {$\Bigg[$};
  \node at (.74,1.5) {\includegraphics[height = 1cm]{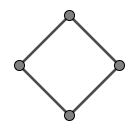}};

  \draw[color=black] (1.1,1.5) node {$\Bigg]$};

\end{scriptsize}
\node at (.75,-1) {$\widetilde \Gamma_2$};

\end{tikzpicture}
\qquad\qquad
\begin{tikzpicture}[line cap=round,line join=round,>=triangle 45,x=1cm,y=1cm,scale = 1.5,c/.style={insert path={circle[radius=.8pt]}}]

\begin{scriptsize}
  \node at (.72,0) {\includegraphics[height = .8cm]{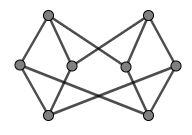}};
\draw[color=black] (.3,0) node {$\Bigg[$};
  \draw[color=black] (1.1,0) node {$\Bigg]$};

  \node at (.74,1.1) {\includegraphics[height = .65cm]{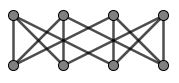}};

\draw[color=black] (.25,1.1) node {$\Bigg[$};
  \draw[color=black] (1.2,1.1) node {$\Bigg]$};

  \node at (.74,2.1) {\includegraphics[height = 1cm]{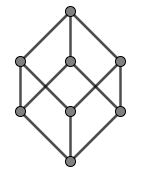}};

\draw[color=black] (.42,2.1) node {$\Bigg[$};
  \draw[color=black] (1.05,2.1) node {$\Bigg]$};

  \node at (-.75,1.1) {\includegraphics[height = .8cm]{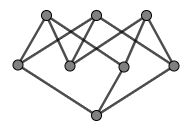}};

\draw[color=black] (-1.15,1.1) node {$\Bigg[$};
  \draw[color=black] (-.35,1.1) node {$\Bigg]$};

  \node at (2.25,1.1) {\includegraphics[height = .85cm]{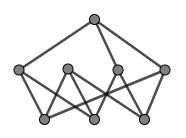}};

\draw[color=black] (1.85,1.1) node {$\Bigg[$};
  \draw[color=black] (2.6,1.1) node {$\Bigg]$};

\end{scriptsize}
\draw [Stealth-,line width=.8pt, violet] (1.8,.8)-- (1.2,.05);
\draw [-Stealth,line width=.8pt, violet] (1.8,1.1)-- (1.3,1.1);
\draw [-Stealth,line width=.8pt, violet] (0.2,1.1)-- (-.3,1.1);
\draw [-Stealth,line width=.8pt, violet] (1.8,1.4)-- (1.15,2.1);
\draw [-Stealth,line width=.8pt, violet] (0.3,2.1)-- (-.3,1.4);
\draw [Stealth-,line width=.8pt, violet] (0.25,.05)-- (-.3,.8);

\node at (.75,-1) {$\widetilde \Gamma_3$};

\end{tikzpicture}

\end{center}
\caption{The digraph $\widetilde \Gamma_N$ of equivalence classes of heights on $H^N$ (where adjacency is determined by vertex lowering) for $N = 1, 2, 3$.}\label{fig:digraphs1}
\end{figure}

For $N = 1,2,3$, there are relatively few heights, which can be worked out in a minute or two; $\widetilde\Gamma_1$, $\widetilde \Gamma_2$, and $\widetilde \Gamma_3$ are displayed in Figure \ref{fig:digraphs1}. 
Notice that for $N = 1$, $V(H^1) = \{0,1\}$. There are two heights: one where 0 is pinned, one where 1 is pinned. These are both in the same equivalence class. For $N=2$, there are 6 heights and two equivalence classes. For $N=3$, there are 38 heights, and 5 equivalence classes. Note, however that two of these equivalence classes exhibit an additional equivalence under vertical reflections. If such reflections are added to our  sense of equivalence, then for $N = 3$, there are 4 equivalence classes. Table~\ref{tab:hts3} lists the 38 heights using their elevator vectors, separated by height signatures. In this case, the height signatures fully classify the FWI and rainbow classes of heights. 

 \begin{table}       
            \begin{tabular}{|c|c|c|c|c|c|}
            \hline
              \text{FWI name} & \text{Picture} & \#[h] & \multicolumn{2}{c|}{\text{elevator vectors}} & \text{height signature}\\\hline
            & & & boson based & fermion based & \\\hline\hline
            $h_{\rm fe}$ &  \multirow{2}{1cm}{\includegraphics[height=.8cm]{fe3.JPG}} & 8 & 
            $\begin{matrix} 
            [0, 2, 2, 2, 1, 1, 1, 3]  \\[.25mm]
            [2, 0, 2, 2, 1, 1, 3, 1]  \\[.25mm]
            [2, 2, 0, 2, 1, 3, 1, 1]  \\[.25mm]
            [2, 2, 2, 0, 3, 1, 1, 1]  \\[.25mm]
            \end{matrix} $ &
            $\begin{matrix} 
             [3,1,1,1, 2,2,2,0 ]\\[.25mm]
            [1,3,1,1, 2,2,0,2 ]\\[.25mm]
             [1,1,3,1, 2,0,2,2 ]\\[.25mm]
             [1,1,1,3, 0,2,2,2 ]\\[.25mm]
            \end{matrix} $ 
             & $(1\mid 3 \mid 3 \mid 1)$ 
            \\ \hline\hline
            $h_1$ &  \includegraphics[height=.8cm]{h1.JPG} & 8 & 
            $\begin{matrix} 
            [0, 2, 2, 2, 1, 1, 1, 1] \\[.25mm]
            [2, 0, 2, 2, 1, 1, 1, 1] \\[.25mm]
            [2, 2, 0, 2, 1, 1, 1, 1]  \\[.25mm]
            [2, 2, 2, 0, 1, 1, 1, 1] \\[.25mm]
            \end{matrix} $ &
            $\begin{matrix} 
             [1,1,1,1, 2,2,2,0 ]\\[.25mm]
            [1,1,1,1, 2,2,0,2 ]\\[.25mm]
            [1,1,1,1, 2,0,2,2 ]\\[.25mm]
            [1,1,1,1, 0,2,2,2 ]\\[.25mm]
            \end{matrix} $ 
            & $( 3 \mid 4 \mid 1)$  
            \\ \hline\hline
            $h_2$ &  \includegraphics[height=.8cm]{h2.JPG} & 12 & 
            $\begin{matrix} 
            [0, 0, 2, 2, 1, 1, 1, 1]  \\ [2,2,0,0, 1,1,1,1 ]\\[.25mm]
            [0, 2, 0, 2, 1, 1, 1, 1]  \\ [2,0,2,0, 1,1,1,1 ]\\[.25mm]
            [0, 2, 2, 0, 1, 1, 1, 1]  \\ [2,0,0,2, 1,1,1,1 ]\\[.25mm]
            \end{matrix} $ &
                        $\begin{matrix} 
            [ 1, 1, 1, 1,0, 0, 2, 2]  \\ [1,1,1,1,2,2,0,0 ]\\[.25mm]
            [ 1, 1, 1, 1,0, 2, 0, 2]  \\ [1,1,1,1,2,0,2,0 ]\\[.25mm]
            [ 1, 1, 1, 1,0, 2, 2, 0]  \\ [1,1,1,1,2,0,0,2 ]\\[.25mm]
            \end{matrix} $& $(2 \mid 4 \mid 2)$ 
            \\ \hline\hline
            $h_{1'}$ &  \includegraphics[height=1cm]{h3.JPG} & 8 & 
             $\begin{matrix} 
            [2,0,0,0, 1, 1, 1, 1]  \\[.25mm]
            [0,2,0,0, 1, 1, 1, 1]  \\[.25mm]
            [0,0,2,0, 1, 1, 1, 1]  \\[.25mm]
            [0,0,0,2, 1, 1, 1, 1]  \\[.25mm]
            \end{matrix} $ & 
                         $\begin{matrix} 
            [1,1,1,1, 2,0,0,0 ]\\[.25mm]
             [1,1,1,1, 0,2,0,0 ]\\[.25mm]
             [1,1,1,1, 0,0,2,0]\\[.25mm]
             [1,1,1,1, 0,0,0,2 ]\\[.25mm]
            \end{matrix} $ 

	    & $( 1 \mid 4 \mid 3)$  
            \\ \hline\hline
            $h_{\rm v}$ &  \includegraphics[height=.8cm]{hv3.JPG} & 2 & 
            $ [0,0,0,0,1,1,1,1]$ & $[1,1,1,1,0,0,0,0] $
            & $(8\mid 8)$             \\ \hline\hline 
            \end{tabular}
            \vspace*{.5cm}
            
\caption{Distribution of the 38 $N=3$ height functions among the 5 FWI equivalence classes.  }\label{tab:hts3}
\end{table}

For $N = 4$, the situation is already much more complicated. There are 990 distinct height functions, which can be grouped into 24 (or 15) equivalence classes. Representatives of these heights (described by their elevator vectors) appear in Table~\ref{tab:hts4}, and a picture of the associated digraph structure appears below in Figure \ref{fig:Gamma_4}. Already in the $N=4$ case, height signatures are no longer sufficient to distinguish between the FWI classes. For example, $h_7$ and $h_8$ both have height signature $(6\mid 8 \mid 2)$, but we can see that there can be no acrotopology isomorphism between them because in $h_7$ the two vertices at height 0 are antipodes in the 4-cube, while in $h_8 $ there is a path of length 2 between them. 

\begin{small}
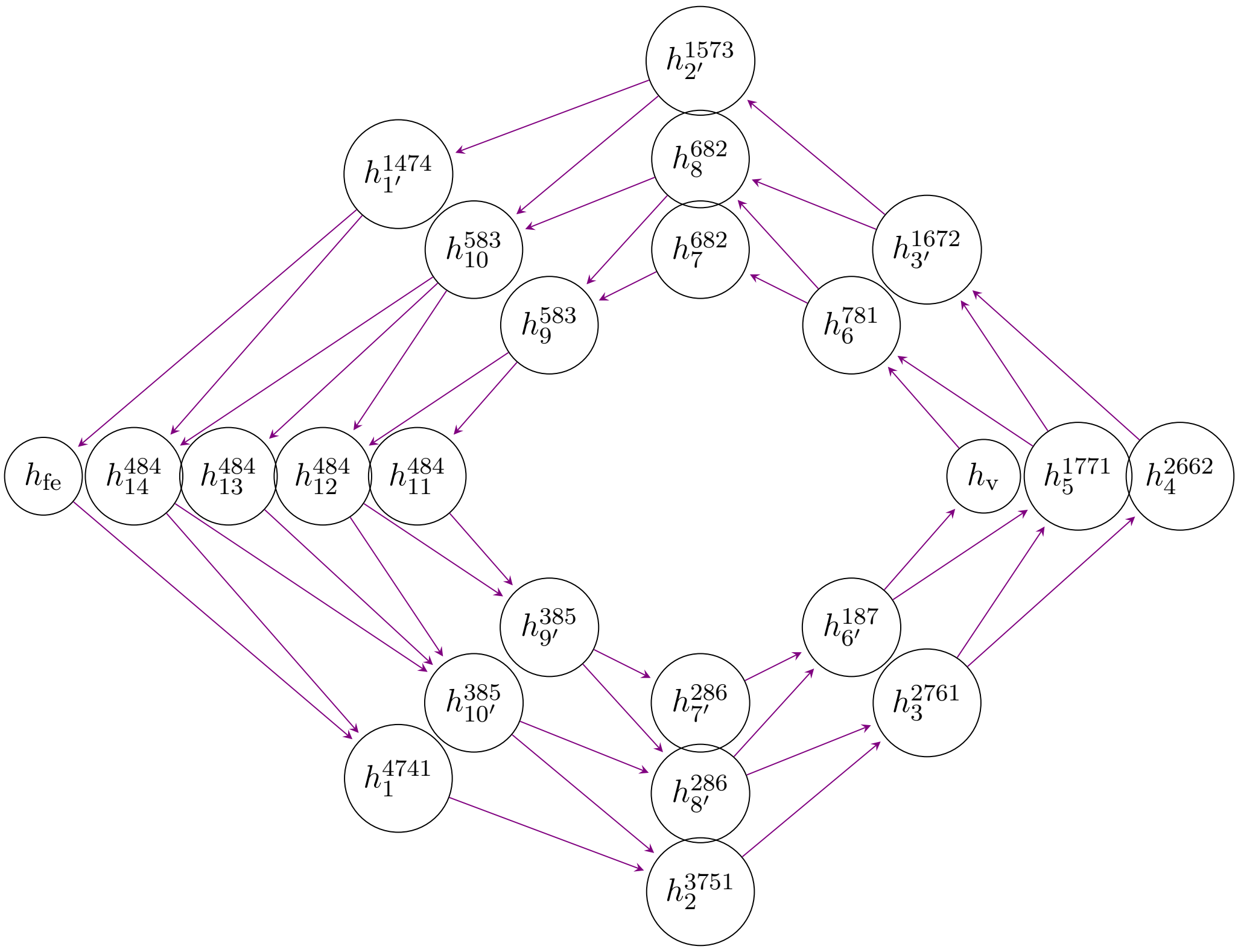
\begin{figure}[h]
\centering
\resizebox{0.9\textwidth}{!}{
       \begin{tikzpicture}[> = stealth,  shorten > = 2pt,   auto,   node distance = 1.5cm,state/.style={circle, draw, minimum size=.5cm},scale=.9]

        \node[state]  (hfe) at (-8.7,0) {$h_{\rm fe}$};
        \node[state]  (h4741) at (-4,-4) {$h_{1}^{4741}$};
        \node[state]  (h1474) at (-4,4) {$h_{1'}^{1474}$};
        \node[state]  (h3751) at (0,-5.5) {$h_{2}^{3751}$};
        \node[state]  (h1573) at (0,5.5) {$h_{2'}^{1573}$};
        \node[state]  (h484a) at (-7.5,0) {$h_{14}^{484}$};
        \node[state]  (h484c)  at (-6.25,0){$h_{13}^{484}$};
        \node[state]  (h484b)  at (-5,0) {$h_{12}^{484}$};
        \node[state]  (h484d)  at (-3.75,0) {$h_{11}^{484}$};
        \node[state]  (h88) at (3.75,0) {$h_{\rm v}$};
        \node[state]  (h1771) at (5,0){$h_{5}^{1771}$};
        \node[state]  (h2761) at(3,-3) {$h_{3}^{2761}$};
        \node[state]  (h1672) at(3,3) {$h_{3'}^{1672}$};
        \node[state]  (h2662)  at (6.35,0){$h_{4}^{2662}$};
        \node[state]  (h781) at (2,2) {$h_{6}^{781}$};
        \node[state]  (h187) at (2,-2) {$h_{6'}^{187}$};
        \node[state]  (h286b) at (0,-3) {$h_{7'}^{286}$};
        \node[state]  (h286a) at (0,-4.2) {$h_{8'}^{286}$};
        \node[state]  (h682b)  at (0,3) {$h_{7}^{682}$};
        \node[state]  (h682a) at (0,4.2) {$h_{8}^{682}$};
        \node[state]  (h385b) at (-2,-2) {$h_{9'}^{385}$};
        \node[state]  (h385a) at (-3,-3) {$h_{10'}^{385}$};
        \node[state]  (h583b) at (-2,2) {$h_{9}^{583}$};
        \node[state]  (h583a)  at (-3,3) {$h_{10}^{583}$};
       
		\path[->, violet] (hfe)  edge node {} (h4741);
		\path[->, violet] (h4741)  edge node {} (h3751);
		\path[->, violet] (h1474)  edge node {} (hfe);
		\path[->, violet] (h1474)  edge node {} (h484a);
		\path[->, violet] (h484a)  edge node {} (h385a);
		\path[->, violet] (h484a)  edge node {} (h4741);
		\path[->, violet] (h385a)  edge node {} (h3751);
		\path[->, violet] (h385a)  edge node {} (h286a);
		\path[->, violet] (h583a)  edge node {} (h484a);
		\path[->, violet] (h583a)  edge node {} (h484b);
		\path[->, violet] (h583a)  edge node {} (h484c);
		\path[->, violet] (h484b)  edge node {} (h385a);
		\path[->, violet] (h484b)  edge node {} (h385b);
		\path[->, violet] (h484c)  edge node {} (h385a);
		\path[->, violet] (h3751)  edge node {} (h2761);
		\path[->, violet] (h1573)  edge node {} (h583a);
		\path[->, violet] (h1573)  edge node {} (h1474);
		\path[->, violet] (h286a)  edge node {} (h187);
		\path[->, violet] (h286a)  edge node {} (h2761);
		\path[->, violet] (h682a)  edge node {} (h583a);
		\path[->, violet] (h682a)  edge node {} (h583b);
		\path[->, violet] (h385b)  edge node {} (h286a);
		\path[->, violet] (h385b)  edge node {} (h286b);
		\path[->, violet] (h583b)  edge node {} (h484b);
		\path[->, violet] (h583b)  edge node {} (h484d);
		\path[->,violet] (h2761)  edge node {} (h1771);
		\path[->, violet] (h2761)  edge node {} (h2662);
		\path[->, violet] (h1672)  edge node {} (h682a);
		\path[->, violet] (h1672)  edge node {} (h1573);
		\path[->, violet] (h1771)  edge node {} (h1672);
		\path[->, violet] (h1771)  edge node {} (h781);
		\path[->, violet] (h187)  edge node {} (h1771);
		\path[->, violet] (h187)  edge node {} (h88);
		\path[->, violet] (h781)  edge node {} (h682a);
		\path[->, violet] (h781)  edge node {} (h682b);
		\path[->, violet] (h286b)  edge node {} (h187);
		\path[->, violet] (h682b)  edge node {} (h583b);
		\path[->, violet] (h484d)  edge node {} (h385b);
		\path[->, violet] (h88)  edge node {} (h781);
		\path[->, violet] (h2662)  edge node {} (h1672);

      \end{tikzpicture}}

\caption{The digraph $\widetilde \Gamma_4$ whose vertices are the equivalence classes of heights on $H^4$ under decorated graph isomorphism. A representative from each height class is described  in Table~\ref{tab:hts4}. The height signature numbers for each class are added as superscripts. }\label{fig:Gamma_4}
\end{figure}

\renewcommand{\arraystretch}{1.3}
 \begin{table}       
            \[
            \begin{array}{|c|l|c|}
            \hline
            \text{height class name} & \text{elevator vector representative} & \#[h]\\\hline \hline
            h_{\rm fe} & [0,2,2,2,2,2,2,4;1,1,1,1,3,3,3,3] & 16\\ \hline\hline
            h_{\rm v} &  [0,0,0,0,0,0,0,0;1,1,1,1,1,1,1,1]& 2\\\hline\hline
            h_{1}^{4741} & [0,2,2,2,2,2,2,2;1,1,1,1,3,3,3,3] & 16 \\\hline 
            h_{1'}^{1474} & [3,1,1,1,1,1,1,1;2,2,2,2,0,0,0,0]  & 16\\\hline\hline 
            h_{2}^{3751} & [0,2,2,2,2,2,2,2;1,1,1,1,1,3,3,3]  & 64\\\hline 
            h_{2'}^{1573} &  [3,1,1,1,1,1,1,1;2,2,2,2,2,0,0,0] & 64\\\hline\hline 
            h_{3}^{2761} & [0,2,2,2,2,2,2,2;1,1,1,1,1,1,3,3]  & 96\\\hline 
            h_{3'}^{1672} &  [3,1,1,1,1,1,1,1;2,2,2,2,2,2,0,0] & 96\\ \hline\hline 
            h_{4}^{2662} &  [0,0,2,2,2,2,2,2;1,1,1,1,1,1,3,3] & 48\\\hline\hline 
            h_{5}^{1771} & [0,2,2,2,2,2,2,2;1,1,1,1,1,1,1,3] & 64 \\\hline\hline 
            h_{6}^{781} & [0,2,2,2,2,2,2,2;1,1,1,1,1,1,1,1]  & 16\\\hline 
            h_{6'}^{187} &[2,0,0,0,0,0,0,0;1,1,1,1,1,1,1,1] & 16\\ \hline\hline 
            h_{7}^{682} & [0,2,2,2,2,2,2,0;1,1,1,1,1,1,1,1]& 8\\\hline
            h_{7'}^{286} &[2,0,0,0,0,0,0,2;1,1,1,1,1,1,1,1] & 8\\\hline\hline 
            h_{8}^{682} & [0,0,2,2,2,2,2,2;1,1,1,1,1,1,1,1]& 48\\\hline 
            h_{8'}^{286} &[2,2,0,0,0,0,0,0;1,1,1,1,1,1,1,1] & 48\\ \hline\hline 
            h_{9}^{583} &  [0,0,2,2,2,2,2,0;1,1,1,1,1,1,1,1] & 48\\\hline 
            h_{9'}^{385} &[2,2,0,0,0,0,0,2,1,1,1,1,1,1,1,1]& 48\\\hline\hline
            h_{10}^{583} & [0,0,0,2,2,2,2,2;1,1,1,1,1,1,1,1] & 64\\\hline 
            h_{10'}^{385} &[2,2,2,0,0,0,0,0;1,1,1,1,1,1,1,1] & 64 \\\hline\hline 
            h_{11}^{484} &  [0,0,2,2,2,2,0,0;1,1,1,1,1,1,1,1] & 12\\ \hline\hline 
            h_{12}^{484} &  [0,0,0,2,2,2,2,0;1,1,1,1,1,1,1,1] & 96 \\\hline\hline 
            h_{13}^{484} &  [0,0,0,2,0,2,2,2;1,1,1,1,1,1,1,1] & 16 \\\hline\hline 
            h_{14}^{484} &  [0,0,0,0,2,2,2,2;1,1,1,1,1,1,1,1] & 16 \\\hline 
            \end{array}
            \]
\caption{ Descriptions of a boson-based representative from each height class.}\label{tab:hts4}
\end{table} 
\renewcommand{\arraystretch}{1.1}

\end{small}
\section{Matrix-theoretic height descriptors}\label{sec:mat}
In this section, we focus on characterizing the $N=4$ heights in another way. First, we define three $8 \times 8$ matrices $L_i^h$, $R_i^h$ and $M_{ij}^h$, which will help describe heights.

\begin{definition}[Color-level adjacency matrices]\label{def:LR}
Let $b_1, \ldots, b_8$, $f_1, \ldots, f_8$ be the  lists of bosons and fermions ordered as described in Notation~\ref{not:heights}. Then, for $1\leq I, J \leq 4$, the ($8 \times 8 $)   $L$- and $R$-matrices $L_I^{\rm v}$ and $R_I^{\rm v}$ corresponding to the (boson-based or fermion-based) valise height ${\rm v}$ are defined by: 
    \begin{gather*}
    L_I^{\rm v} (i,j) =  \begin{cases}
        1 & b_i\widesim[2]{I} f_j \\
        0 & \text{otherwise}, 
    \end{cases} \\[2mm]
    R_I^{\rm v}(i,j) =  \begin{cases}
        1 & f_i\widesim[2]{I} b_j \\
        0 & \text{otherwise}, 
    \end{cases} 
    \end{gather*}
     If $h'$ can be obtained from $h$ by raising the boson $b_i$, then 
    \[
    L_I^{h'} = D_{B_i} L_I^h \quad \text{and} \quad  R_I^{h'} =R_I^h \widehat D_{B_i} 
    \]
    where $D_{B_i}  = \mathrm{diag}(1,1,\ldots,1, m_B,1, \ldots, 1)  $ and $\widehat D_{B_i}  = \mathrm{diag}(1,1,\ldots,1, 1/\mu_B,1, \ldots, 1)  $ (i.e., non-one entries in the $i$th entries). Similarly, if $h'$ can be obtained from $h$ by raising fermion $f_i$, then 
    \[
    L_I^{h'} =  L_I^h D_{F_i} \quad \text{and} \quad  R_I^{h'} =\widehat D_{F_i}  R_I^h 
    \]
    where $D_{F_i}  = \mathrm{diag}(1,1,\ldots,1, m_F,1, \ldots, 1)  $ and $\widehat D_{F_i}  = \mathrm{diag}(1,1,\ldots,1, 1/\mu_F,1, \ldots, 1)  $.

    Any height $h$ can be obtained from the boson-based valise by raising some sequence of bosons and fermions. Suppose that $i_1, \ldots, i_k$ and $j_1, \ldots, j_\ell$ are the indices of these bosons and fermions, respectively. Then, letting 
    \[
    D_B^h = \prod_{s = 1}^k D_{B_{i_s}},\quad \widehat D_B^h = \prod_{s = 1}^k \widehat D_{B_{i_s}},\quad D_F^h = \prod_{s = 1}^\ell D_{F_{j_s}}, \quad \widehat D_F^h = \prod_{s = 1}^\ell \widehat D_{F_{j_s}},
    \]
    we have
    \[
    L^h_I = D_B^h L_I^{\rm v} D_F^h, \quad \text{and} \quad R_I^h = \widehat D_F^h L_I^{\rm v} \widehat D_B^h.
    \]
    A similar sequence of vertices could be used to obtain any height from the fermion-based valise. Note that raising all bosons once takes the boson-based valise to the fermion-based valise (and vice versa for raising all fermions), which corresponds to multiplying each $L_I$ by $m_B$ ($m_F$) and each $R_I$ by $\frac1{\mu_B}$ ($\frac1{\mu_F}$). If $h$ can be obtained from the boson-based valise by a given sequence of vertex-raisings, then it can be obtained from the fermion-based vaise by first raising all fermions and then following the same sequence. Thus, we adopt the convection that, for any $k, \ell \in \mathbb{Z}$,  the collection $\{L_1, \ldots, L_N;R_1,\ldots,R_N\}$ is equivalent to the collection 
    \[
    \left\{m_B^k m_F^\ell L_1, \ldots, m_B^k m_F^\ell L_N;\frac1{\mu_B^k} \frac1{\mu_F^\ell} R_1,\ldots,\frac1{\mu_B^k} \frac1{\mu_F^\ell} R_N\right\}.
    \]
\end{definition}

We think of the rows of the $L_I$ matrices and the columns of the $R_I$ matrices as being indexed by the bosons $b_1, \ldots, b_8$. Similarly, the columns of the $L_I$ matrices and the rows of the $R_I$ matrices are indexed by the fermions $f_1, \ldots, f_8$.

\begin{proposition}\label{prop:FWI_L}
    If $h$ and $h'$ are two heights on a hypercube Adinkra that belong to the same FWI class and they are either both boson-based or both fermion-based, then their color-level adjacency matrices are the same up to permutations of the colors and permutation similarity. If $h$ and $h'$ are two heights on a hypercube Adinkra that belong to the same FWI class and one is  boson-based and one fermion-based, then their color-level adjacency matrices are the same up to permutations of the colors, permutation similarity, and swapping the variables $m_B$ and $\mu_B$ for $m_F$ and $\mu_F$, respectively. 
    
    More precisely, if $h'$ and $h$ are in the same FWI class, then $h' =  C_{v,\sigma}\cdot N_{I_1} \cdots N_{I_\ell}(h)$ for some $\sigma\in S_4$, $v \in H^4$, and $I_1, \ldots, I_\ell \in [4]$. Let $P = C_{v,\sigma}'\cdot  N'_{I_1}\cdot N'_{I_\ell}\cdot $, then if $\ell$ is even, we have
    \[
    P  L^{h}_{J} P^T N_{\sigma(1)} = L^{h'}_{\sigma(J)} \quad \text{and} \quad  N'_{\sigma(1)} P^T  R^{h}_{J} P = R^{h'}_{\sigma(J)}.
    \]

   If $\ell$ is odd, then 
  \[
    P  L^{h}_{J} P^T N_{\sigma(1)} = (L^{h'}_{\sigma(J)})^T \quad \text{and} \quad  N'_{\sigma(1)} P^T  R^{h}_{J} P = (R^{h'}_{\sigma(J)})^T.
    \]

\end{proposition}
\begin{proof}
It is clear from Proposition~\ref{prop:CNmatrices}, that $N'_J L_I^h N'_J$ acts via sending rows (bosons) to columns (bosons), e.g., sending $b_i$ to $f_j$ and $f_j$ to $b_i$ iff $b_i \widesim[2]{J} f_i$. Thus if  $(L_I^h)_{ij} = m_B^{k_i} m_F^{\ell_j}$, then $b_i \widesim[2]{I} f_j$ and $b_i$ and $f_j$ have been raised $k_i$ and $\ell_j$ times, respectively, from their initial locations in the relevant valise. For some $t$ and $s$, we have that $b_i \widesim[2]{J} f_s$ and $f_j \widesim[2]{J} b_t$ and $f_s$ and $b_t$ have been raised $k_i$ and $\ell_j$ times, respectively, from their initial locations in the relevant valise. It must also be true that $f_s \widesim[2]{I} b_t$ by the 4-cycle property of Adinkras. So, we have that $(L_I^{h'})_{st} = m_F^{k_i} m_B^{\ell_j} $, and hence $N'_J L_I^h N'_J = (L_I^{h'})^T$.

Similarly, $C_{v,\sigma}' L_I^h (C_{v,\sigma}'\cdot N'_{\sigma(1)})^T$ acts by permuting the rows via the permutation matrix $C_{v,\sigma}'$ and permuting the columns via the permutation matrix $C_{v,\sigma}'\cdot N'_{\sigma(1)}$. If $(L_I^h )_{ij} = m_B^{k_i} m_F^{\ell_j}$, then $b_i \widesim[2]{I} f_j$ and $b_i$ and $f_j$ have been raised $k_i$ and $\ell_j$ times, respectively, from their initial locations in the relevant valise. For some $I_1 \ldots I_\ell$, we have a path $$v\widesim[2]{I_1}v_1\widesim[2]{I_2} \cdots \widesim[2]{I_\ell} b_i $$ and let $s,t$ be such that 
$$v \widesim[2]{\sigma(I_1)}v_1\widesim[2]{\sigma(I_2)} \cdots \widesim[2]{\sigma(I_\ell)} b_s \widesim[2]{\sigma(I)}f_t.$$ 
Then, in $h'$   $b_s$ and $f_t$ have been raised $k_i$ and $\ell_j$ times, respectively, from their initial locations in the relevant valise.  So, we have that $(L_I^{h'})_{st} = m_F^{k_i} m_B^{\ell_j} $, and hence $C_{v,\sigma}' L_I^h (C_{v,\sigma}'\cdot N'_{\sigma(1)})^T = L_{\sigma(I)}^{h'}$.

\end{proof}

\section{Modified twisted characters}\label{sec:char}

Note that product matrices of the form $L_I^hR_J^h$ or $R_I^hL_J^h$ encode $I$-$J$-colored paths of length 2 traveling from boson to boson or fermion to fermion, respectively. The following matrices will indicate $I$-$J$-colored diamonds in $h$ as demonstrated in Theorem~\ref{prop:diamonds}

\begin{definition}[Raised diamond matrices]\label{def:diamond}
    \begin{align*}
    M_{IJ}^h &= \frac{\sqrt{-1}}{4} \left(L_I^hR_J^h - L_J^hR_I^h \right) \\
    \widehat M_{IJ}^h &= \frac{\sqrt{-1}}{4} \left(R_I^hL_J^h - R_J^hL_I^h \right) 
    \end{align*}
\end{definition}

Since, for any $k, \ell \in \mathbb{Z}$,  the collection $\{L_1, \ldots, L_N;R_1,\ldots,R_N\}$ is equivalent to the collection 
    \[
    \left\{m_B^k m_F^\ell L_1, \ldots, m_B^k m_F^\ell L_N;\frac1{\mu_B^k} \frac1{\mu_F^\ell} R_1,\ldots,\frac1{\mu_B^k} \frac1{\mu_F^\ell} R_N\right\}, 
    \]
we have that the collection of raised diamond matrices $\{M_{IJ}, \widehat M_{IJ} \}$ is equivalent to  $\{\rho_B^k\rho_F^\ell M_{IJ}, \rho_B^k\rho_F^\ell\widehat M_{IJ} \}$, where $\rho_B = \frac{m_B}{\mu_B}$ and $\rho_F = \frac{m_F}{\mu_F}$. 

\begin{theorem}\label{prop:diamonds}
    Let $Q =  (b_i , f_k, b_j, f_\ell)$ be an $I$-$J$ 4-cycle of $H^4$ with $1\leq i<j\leq 8$, $1\leq k<\ell\leq 8$, and $1 \leq I < J \leq 4$. 
    
    Both $M_{IJ}(i,j)=0$ and $\widehat M_{IJ}(i,j)=0$ if $Q$ forms a bow-tie under $h$ ($h(b_i) = h(b_j)$, $h(f_k) = h(f_\ell)$). 
    
    Assume that $Q$ forms a diamond under $h$ and that $Q$ is labeled as in the diagram on the left in Figure~\ref{fig:diam-bt-cycles}. 
    Let $\delta =  1$ if $ A\widesim[2]{I} B$ ($I$ is red) and $\delta = -1$ if $A\widesim[2]{I} C$ ($I$ is blue). If $A = f_k$, $C = f_\ell$, then the $i,j$ principal submatrix of $M_{IJ}$ is the $2\times2$ zero matrix, and the  $k,\ell$ principal submatrix of $\widehat M_{IJ}$ is
        \[
    \widehat M_{IJ}^h[(k,\ell)] = 
    \delta
    (\rho_B\rho_F)^{\lfloor\frac n2\rfloor}\frac{\sqrt{-1}}{4}
    \begin{bmatrix}
        0 & \rho_B - 1 \\ 
        1-\rho_B & 0 
    \end{bmatrix} .
    \]
    If $A = b_i$, $C = b_j$, then  $k,\ell$ principal submatrix of $\widehat M_{IJ}$ is the $2\times2$ zero matrix, and the $i,j$ principal submatrix of $M_{IJ}$ is
    \[
    M_{IJ}^h[(i,j)] = 
    \delta
    (\rho_B\rho_F)^{\lfloor\frac n2\rfloor}\frac{\sqrt{-1}}{4}
    \begin{bmatrix}
        0 & \rho_F - 1 \\ 
        1-\rho_F & 0 
    \end{bmatrix} .
    \]
\end{theorem}
\begin{center}
\begin{figure}[h]
\centering
\begin{tikzpicture}[
    boson/.style={circle, draw, thick, fill=white, minimum size=3mm},
    fermion/.style={circle, draw, thick, fill=black, text=white, minimum size=5mm},
    edge1/.style={ultra thick, red},
    edge2/.style={ultra thick, blue},
]

\begin{small}
    
\node[boson] (b1) at (0, 1){$B$};
\node[boson] (b2) at (0, -1){$D$};

\node[fermion] (f1) at (-1, 0){$A$};
\node[fermion] (f2) at (1, 0){$C$};

\end{small}

\draw[edge1] (b1) -- (f1);
\draw[edge1] (b2) -- (f2);
\draw[edge2] (b1) -- (f2);
\draw[edge2] (b2) -- (f1);

\node[color = gray,anchor = west] at (1.5,1) {$h=n+2$};
\node[color = gray,anchor = west] at (1.5,0) {$h=n+1$};
\node[color = gray,anchor = west] at (1.5,-1) {$h=n$};
\draw[color = gray, line width = .3pt,dotted] (1.5,1)--(-1.5,1);
\draw[color = gray, line width = .3pt,dotted] (1.5,0)--(-1.5,0);
\draw[color = gray, line width = .3pt,dotted] (1.5,-1)--(-1.5,-1);

\end{tikzpicture}
\qquad \qquad
\begin{tikzpicture}[
    boson/.style={circle, draw, thick, fill=white, minimum size=3mm},
    fermion/.style={circle, draw, thick, fill=black, text=white, minimum size=5mm},
    edge1/.style={ultra thick, red},
    edge2/.style={ultra thick, blue}, scale = .8
]

\begin{small}

\node[boson] (b1) at (-1, -1){$B$};
\node[boson] (b2) at (1, -1){$D$};

\node[fermion] (f1) at (-1, 1){$A$};
\node[fermion] (f2) at (1, 1){$C$};

\end{small}

\draw[edge1] (b1) -- (f1);
\draw[edge1] (b2) -- (f2);
\draw[edge2] (b1) -- (f2);
\draw[edge2] (b2) -- (f1);

\node[color = gray,anchor = west] at (1.7,1) {$h=n+1$};
\node[color = gray,anchor = west] at (1.7,-1) {$h=n$};
\draw[color = gray, line width = .3pt,dotted] (1.7,1)--(-1.7,1);
\draw[color = gray, line width = .3pt,dotted] (1.7,-1)--(-1.7,-1);

\end{tikzpicture}
\caption{Possible configurations of any 4-cycle $Q$ in $H^4$ under a height $h$.}\label{fig:diam-bt-cycles}

\end{figure}
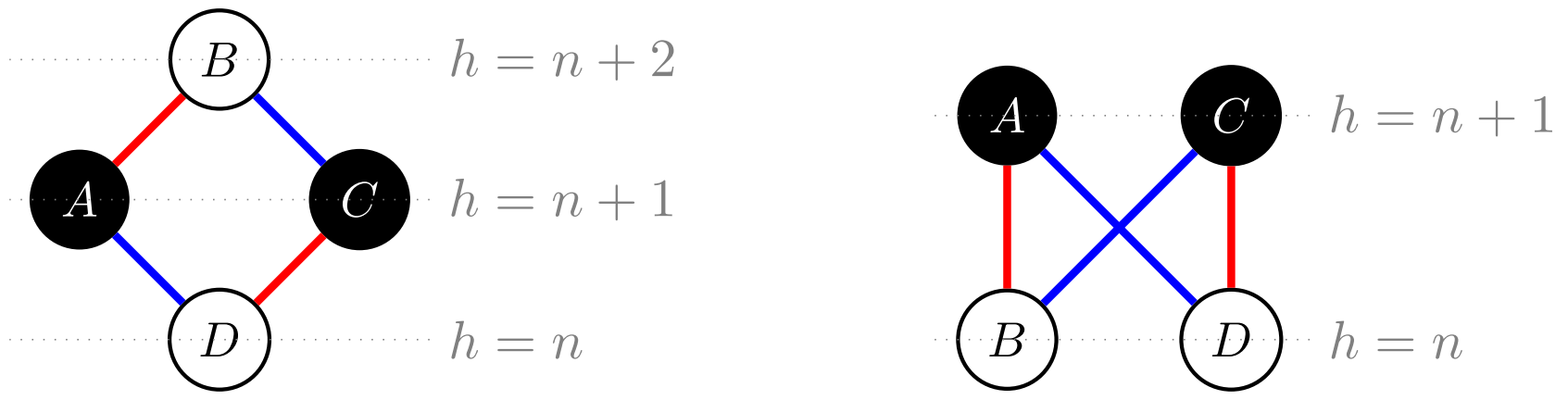
\end{center}
\begin{proof}
    Let $L_I^h$, $L_J^h$, $R_I^h$, $R_j^h$ be the relevant color-level adjacency matrices for $h$. 

    First, we show that all entries of  $M_{IJ}^h$  and $\widehat M_{IJ}^h$ that correspond to bow-tie cycles are zero. Assume that $Q$ is labeled as in the diagram on the right in Figure~\ref{fig:diam-bt-cycles}.       
Assume that $A = b_i$, $C = b_j$. If we restrict to the $i$, $j$ bosons and the $k$, $\ell$ fermions we get
    \[
    L_I^h[(i,j),(k,\ell)] = m_B^{\lfloor\frac{n+1}2\rfloor}m_F^{\lfloor\frac{n}2\rfloor}\begin{bmatrix}
1 & 0 \\
    0 & 1
    \end{bmatrix}, \quad 
    L_J^h[(i,j),(k,\ell)] =m_B^{\lfloor\frac{n+1}2\rfloor}m_F^{\lfloor\frac{n}2\rfloor}
    \begin{bmatrix}
0& 1 \\
      1 &  0
    \end{bmatrix}
    \]
    \[
    R_I^h[(i,j),(k,\ell)] =\mu_B^{-\lfloor\frac{n+1}2\rfloor}\mu_F^{-\lfloor\frac{n+1}2\rfloor}
    \begin{bmatrix}
1 & 0 \\
    0 & 1
    \end{bmatrix},\quad
    R_J^h[(i,j),(k,\ell)] =\mu_B^{-\lfloor\frac{n+1}2\rfloor} \mu_F^{\lfloor\frac{n+1}2\rfloor}
    \begin{bmatrix}0& 1 \\
     1 &  0
    \end{bmatrix}
    \]
    Applying Definition~\ref{def:diamond} gives the result. A similar proof works in the case that $A = f_k$ and $ C = f_\ell$.

    Now, we assume that $Q$ is a diamond under $h$, labeled as in the diagram on the left in Figure~\ref{fig:diam-bt-cycles}. Consider the case that $A = b_i$ and $C = b_j$. If we restrict to the $i$, $j$ bosons and the $k$, $\ell$ fermions we get
    \[
    L_I^h[(i,j),(k,\ell)] = m_B^{\lfloor\frac{n+1}2\rfloor}\begin{bmatrix}
m_F^{\lfloor\frac{n+1+\delta}2\rfloor} & 0 \\
    0 &  m_F^{\lfloor\frac{n+1-\delta}2\rfloor}
    \end{bmatrix}
    \]
    \[
    L_J^h[(i,j),(k,\ell)] =m_B^{\lfloor\frac{n+1}2\rfloor}
    \begin{bmatrix}
0& m_F^{\lfloor\frac{n+1-\delta}2\rfloor} \\
      m_F^{\lfloor\frac{n+1+\delta}2\rfloor} &  0
    \end{bmatrix}
    \]
    \[
    R_I^h[(i,j),(k,\ell)] =\mu_B^{-\lfloor\frac{n+1}2\rfloor}
    \begin{bmatrix}
\mu_F^{-\lfloor\frac{n+1+\delta}2\rfloor} & 0 \\
    0 & \mu_F^{-\lfloor\frac{n+1-\delta}2\rfloor}
    \end{bmatrix},
    \]
    \[
    R_J^h[(i,j),(k,\ell)] =\mu_B^{-\lfloor\frac{n+1}2\rfloor}
    \begin{bmatrix}
0& \mu_F^{-\lfloor\frac{n+1+\delta}2\rfloor} \\
     \mu_F^{-\lfloor\frac{n+1-\delta}2\rfloor} &  0
    \end{bmatrix}
    \]
    Applying Definition~\ref{def:diamond} gives the result. 
    
    A similar proof works in the case that $A = f_k$ and $ C = f_\ell$.
    \end{proof}

    Theorem~\ref{prop:diamonds} implies that the `unfoldings'  under $h$ can be read off from the principal submatrices of $M_{IJ}^h$ and $\widehat M_{IJ}^h$ for $1 \leq I < J \leq 4$. It is natural to suspect that these matrices would be useful finding invariants to classify Adinkra heights. The twisted and modified characters discussed below, which are related to the isomer numbers in Section~\ref{ssec:projecting}, are interesting candidates.

\begin{definition}[Twisted Character]\label{def:TChar}
A ``twisted" version of the supercharacter is defined \cite{Char09,Char13} as 
\begin{align*}
    \widetilde{\chi}(a, b) &= {\rm Tr} \left(\exp(ia{M}_{12}^h)\exp(ib{M}_{34}^h)\right)
\end{align*}
with $
\exp A = \sum_{k=0}^\infty \frac1{k!}t (A)^k$ as usual. 
\end{definition}

As an example, the twisted character for a valise chiral Adinkra is 
\begin{align*}
    \widetilde{\chi}_{\rm c}(a, b) = 4\text{cos}\left(\frac{a}{2}\right)\text{cos}\left(\frac{b}{2}\right) + 4\text{sin}\left(\frac{a}{2}\right)\text{sin}\left(\frac{b}{2}\right) 
\end{align*}
When computed for any dashed \textbf{valise} Adinkra, $\widetilde{\chi}$ can be used to recover the isomer numbers defined earlier in Section \ref{ssec:projecting}:
\begin{align*}
    \frac{\partial}{\partial a}\frac{\partial}{\partial b}\widetilde{\chi}\Big|_{a, b = 0} = 4(n_c - n_t)
\end{align*}
\begin{align*}
    \left(\frac{\partial}{\partial a}\right)^2\Big|_{a, b = 0} = \left(\frac{\partial}{\partial b}\right)^2\Big|_{a, b = 0} = -4(n_c + n_t)
\end{align*}
This hints at the possibility of using the twisted character in its raw form (like the one presented above for $\widetilde{\chi}_{\rm c}$) to classify $N=4$ heights, even though for non-valise Adinkras there will not be such a neat interpretation in terms of isomer numbers (so we do not lose any information by setting $a = b = 1$). For the purpose of such a classification, we define a modified version of the character that accounts for color permutation, as follows. 

\begin{definition}[Modified Character and Dual Character]\label{def:MChar}
\begin{align*}
\chi_h &= {\rm Tr} \left(\exp{iM}_{12}^h\exp{iM}_{34}^h + \exp{iM}_{13}^h\exp{iM}_{24}^h + \exp{iM}_{14}^h\exp{iM}_{23}^h\right),\quad \text{and}\\
\bar\chi_h &= {\rm Tr} \left(\exp{i\widehat M}_{12}^h\exp{i\widehat M}_{34}^h + \exp{i\widehat M}_{13}^h\exp{i\widehat M}_{24}^h + \exp{i\widehat M}_{14}^h\exp{i\widehat M}_{23}^h\right).
\end{align*}
\end{definition}
These values are taken modulo $\rho_B^{2k}\rho_F^{2\ell}$, for $k,\ell \in \mathbb{Z}$.

\vspace{2mm} 

\subsection{Modified characters reproduce the FWI classification}
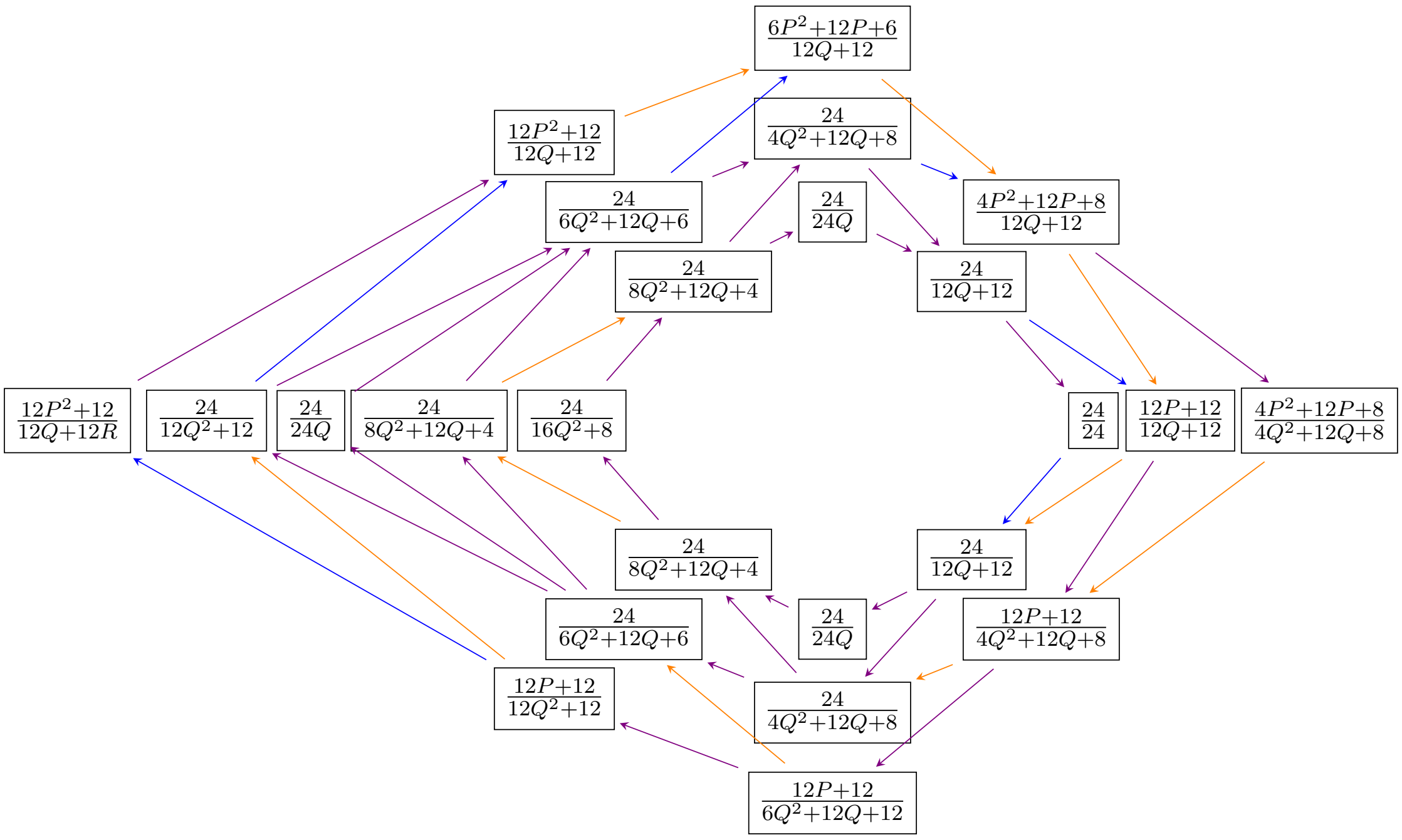
\begin{figure}[h]
\centering
\resizebox{0.98\textwidth}{!}{
       \begin{tikzpicture}[> = stealth,  shorten > = 2pt,   auto,   node distance = 1.5cm,state/.style={draw},scale=.9, outer sep = 2pt]

        \node[state]  (hfe) at (-11,0) {$\frac{12P^2 + 12}{12Q+12R}$};
        \node[state]  (h4741) at (-4,-4){$\frac{12P + 12}{12Q^2+12}$};
        \node[state]  (h1474) at (-4,4) {$\frac{12P^2+12}{12Q+12}$};
        \node[state]  (h3751) at (0,-5.5){$\frac{12P + 12}{6Q^2+12Q+12}$};
        \node[state]  (h1573) at (0,5.5){$\frac{6P^2+12P+6}{12Q+12}$};
        \node[state]  (h484a) at (-9,0){$\frac{24}{12Q^2+12}$};
        \node[state]  (h484c)  at (-7.5,0){$\frac{24}{24Q}$};
        \node[state]  (h484b)  at (-5.8,0) {$\frac{24}{8Q^2+12Q+4}$};
        \node[state]  (h484d)  at (-3.75,0){$\frac{24}{16Q^2+8}$};
        \node[state]  (h88) at (3.75,0){$\frac{24}{24}$};
        \node[state]  (h1771) at (5,0){$\frac{12P + 12}{12Q+12}$};
        \node[state]  (h2761) at(3,-3){$\frac{12P + 12}{4Q^2+12Q+8}$};
        \node[state]  (h1672) at(3,3) {$\frac{4P^2+12P+8}{12Q+12}$};
        \node[state]  (h2662)  at (7,0){$\frac{4P^2+12P+8}{4Q^2+12Q+8}$};
        \node[state]  (h781) at (2,2){$\frac{24}{12Q+12}$};
        \node[state]  (h187) at (2,-2) {$\frac{24}{12Q+12}$};
        \node[state]  (h286b) at (0,-3){$\frac{24}{24Q}$};
        \node[state]  (h286a) at (0,-4.2) {$\frac{24}{4Q^2+12Q+8}$};
        \node[state]  (h682b)  at (0,3){$\frac{24}{24Q}$};
        \node[state]  (h682a) at (0,4.2){$\frac{24}{4Q^2+12Q+8}$};
        \node[state]  (h385b) at (-2,-2){$\frac{24}{8Q^2+12Q+4}$};
        \node[state]  (h385a) at (-3,-3){$\frac{24}{6Q^2+12Q+6}$};
        \node[state]  (h583b) at (-2,2){$\frac{24}{8Q^2+12Q+4}$};
        \node[state]  (h583a)  at (-3,3){$\frac{24}{6Q^2+12Q+6}$};
       
		\path[<-, blue] (hfe)  edge node[anchor = east] {
        } (h4741);
		\path[<-, violet] (h4741)  edge node[anchor = west] {
        } (h3751);
		\path[<-, violet] (h1474)  edge node[anchor = south] {
        } (hfe);
		\path[<-, blue] (h1474)  edge node[anchor = north] {
        } (h484a);
		\path[<-, violet] (h484a)  edge node[anchor = west] {
        } (h385a);
		\path[<-, orange] (h484a)  edge node[anchor = west] {
        } (h4741);
		\path[<-, orange] (h385a)  edge node[anchor = west] {
        } (h3751);
		\path[<-, violet] (h385a)  edge node[anchor =  west] {
        } (h286a);
		\path[<-, violet] (h583a)  edge node[anchor = west] {
        } (h484a);
		\path[<-, violet] (h583a)  edge node {} (h484b);
		\path[<-, violet] (h583a)  edge node {} (h484c);
		\path[<-, violet] (h484b)  edge node {} (h385a);
		\path[<-, orange] (h484b)  edge node {} (h385b);
		\path[<-, violet] (h484c)  edge node {} (h385a);
		\path[<-, violet] (h3751)  edge node[anchor = east] {
        } (h2761);
		\path[<-, blue] (h1573)  edge node[anchor = south] {
        } (h583a);
		\path[<-, orange] (h1573)  edge node[anchor = south] {
        } (h1474);
		\path[<-, violet] (h286a)  edge node[anchor = west] {
        } (h187);
		\path[<-, orange] (h286a)  edge node[anchor = east] {
        } (h2761);
		\path[<-, violet] (h682a)  edge node[anchor = west] {
        } (h583a);
		\path[<-, violet] (h682a)  edge node {} (h583b);
		\path[<-, violet] (h385b)  edge node {} (h286a);
		\path[<-, violet] (h385b)  edge node {} (h286b);
		\path[<-, orange] (h583b)  edge node[anchor = west] {} (h484b);
		\path[<-, violet] (h583b)  edge node {} (h484d);
		\path[<-,violet] (h2761)  edge node[anchor = east] {
        } (h1771);
		\path[<-, orange] (h2761)  edge node[anchor = south] {
        } (h2662);
		\path[<-, blue] (h1672)  edge node[anchor = south] {
        } (h682a);
		\path[<-, orange] (h1672)  edge node[anchor = south] {
        } (h1573);
		\path[<-, orange] (h1771)  edge node[anchor = west] {
        } (h1672);
		\path[<-, blue] (h1771)  edge node[anchor = south] {
        } (h781);
		\path[<-, orange] (h187)  edge node[anchor = west] {
        } (h1771);
		\path[<-, blue] (h187)  edge node[anchor = east] {
        } (h88);
		\path[<-, violet] (h781)  edge node[anchor = west] {
        } (h682a);
		\path[<-, violet] (h781)  edge node {} (h682b);
		\path[<-, violet] (h286b)  edge node {} (h187);
		\path[<-, violet] (h682b)  edge node[anchor = north] {} (h583b);
		\path[<-, violet] (h484d)  edge node {} (h385b);
		\path[<-, violet] (h88)  edge node[anchor =  east] {
        } (h781);
		\path[<-, violet] (h2662)  edge node[anchor = west] {
        } (h1672);

      \end{tikzpicture}}

\caption{Characters and Dual Characters for each FWI height class arranged as on the digraph $\widetilde \Gamma_4$ in Figure~\ref{fig:Gamma_4}. Each arrow corresponds to raising a vertex. Purple arrows preserve characters, orange arrows preserve dual characters, and blue arrows substitute a character for a dual character. }\label{fig:CHARS}
\end{figure}

\begin{proposition}\label{prop:chars}
    If $h$ and $h'$ are in the same FWI class and they are either both boson-based or both fermion-based, then $\chi_h = \chi_{h'}$ and $\chi^{*}_h = \chi^{*}_{h'}$. 
    Further, if $h$ and $h'$ are in the same FWI class, but one is boson-based and the other is fermion-based, then $\chi_h = \chi^*_{h'}$ and $\chi^{*}_h = \chi_{h'}$.
\end{proposition}
\begin{proof}
This follows from Proposition~\ref{prop:FWI_L}.
\end{proof}
The characters and dual characters for all FWI-classes of $N=4$ hypercube heights (see Table~\ref{tab:hts4}) can now be calculated.  The results are summarized in Table~\ref{tab:CHARS} and Figure~\ref{fig:CHARS}. 

The modified characters are invariants of Adinkral heights which demonstrate their rich structure and may aid in their classification. Examining Figure~\ref{fig:CHARS} closely, we  see some interesting preservation of character information along adjacencies in the digraph. Recall that the FWI height classes in this picture are related by vertex raising and lowering operations; in this figure, an arrow represents raising a single vertex. Notice that in all cases, these operations preserve either the character or the dual character. In cases where raising a vertex takes a height $h$ with $k-1$ distinct height values to a height $h'$ with $k$ distinct height values, the dual character of $h$ becomes the  character of $h'$ (these are the blue arrows in Figure~\ref{fig:CHARS}). This seems natural somehow because raising that vertex takes a boson-based height to a fermion-based height (or vice versa). The instances where the dual character is preserved from $h$ to $h'$  are the orange arrows in Figure~\ref{fig:CHARS}. Note that for all height classes with 4 levels have characters and dual characters (and variables) reversed in their vertical flip height class (e.g., $h_1$ and $h_{1'}$). For heights with 3 levels, vertically flipped classes have the same characters and dual characters.  
Another observation on the modified characters and dual characters displayed here is that, adopting the notion like $\deg Q = 1$, $\deg P = 2$, and $\deg R = 3$,  there seems to be some natural way in which they ``grow'' as the classes become closer to the fully extended height and further from the valise. It would be interesting to explore the mathematical and or physical reasons for the above observations and to consider generalizations to higher dimensions and quotient Adinkras.

Table~\ref{tab:CHARS} also gives the following natural converse to Proposition~\ref{prop:chars}
\begin{proposition}
    If $\chi_h = \chi_{h'}$, $\chi^{*}_h = \chi^{*}_{h'}$, and the height signatures of $h$ and $h'$ are the same, then $h$ and $h'$ are in the same FWI class. 
\end{proposition}

 \begin{table}[h]     
            \[
            \begin{array}{|c|c|c|c|}
            \hline
            \text{FWI Label} & \text{Height Signature} & \text{Character} & \text{Dual Character}\\\hline \hline
            h_{\rm fe}&(1|4|6|4|1) & 12P^{2} + 12 & 12Q + 12R\\ \hline\hline
            h_{\rm v}&(8|8) & 24 & 24 \\\hline\hline
            h_{1}&(4|7|4|1) & 12P + 12 & 12Q^{2} + 12
\\\hline 
            h_{1'}&(1|4|7|4) & 12P^2 + 12 & 12Q + 12 \\\hline\hline 
            h_{2}&(3|7|5|1) & 12P + 12 & 6Q^2 + 12Q + 6\\\hline 
            h_{2'}&(1|5|7|3) & 6P^2 + 12P + 6 & 12Q + 12 \\\hline\hline 
            h_{3}&(2|7|6|1) & 12P + 12 & 4Q^2 + 12Q + 8\\\hline 
            h_{3'}&(1|6|7|2) & 4P^2 + 12P + 8  & 12Q + 12\\ \hline\hline 
            h_{4}&(2|6|6|2) & 4P^2 + 12P + 8 & 4Q^2 + 12Q + 8\\\hline\hline 
            h_{5}&(1|7|7|1) & 12P + 12 & 12Q + 12 \\\hline\hline 
            h_{6}&(7|8|1) & 24  & 12Q + 12 \\\hline 
            h_{6'}&(1|8|7) & 24 & 12Q + 12\\ \hline\hline 
            h_{7}&(6|8|2) & 24 & 24Q\\\hline
            h_{7'}&(2|8|6) & 24 & 24Q\\\hline\hline 
            h_{8}&(6|8|2) & 24 & 4Q^2 + 12Q + 8\\\hline 
            h_{8'}&(2|8|6) & 24 & 4Q^2 + 12Q + 8\\ \hline\hline 
            h_{9}&(5|8|3) &  24 & 8Q^2 + 12Q + 4\\\hline 
            h_{9'}&(3|8|5) & 24 & 8Q^2 + 12Q + 4 \\\hline\hline
            h_{10}&(5|8|3) & 24 & 6Q^2 + 12Q + 6\\\hline 
            h_{10'}&(3|8|5) & 24 & 6Q^2 + 12Q + 6 \\\hline\hline 
            h_{11}&(4|8|4) & 24 & 16Q^2 + 8\\ \hline\hline 
            h_{12}&(4|8|4) & 24 & 8Q^2 + 12Q + 4 \\\hline\hline 
            h_{13}&(4|8|4) & 24 & 24Q\\\hline\hline 
            h_{14}&(4|8|4) & 24 & 12Q^2 + 12 \\\hline 
            \end{array}
            \]
\caption{$\chi_h$ and $\chi_h^{*}$ corresponding to each FWI height class, assuming bosons at the lowest height. Here $P = \cos(\frac a4\rho_B(\rho_F - 1))$, $Q = \cos(\frac a4(\rho_B - 1))$, $R = \cos(\frac a4\rho_B\rho_F(\rho_B - 1))$}\label{tab:CHARS}
\end{table}

\section{Height Yielding Matrix Numbers}\label{sec:HYMNs}
In this section, we aim to describe the $N=4$ heights using the eigenvalues of Banchoff matrices $B_L^h$ and $B_R^h$, which are defined using the $L_i^h$ and $R_i^h$ matrices. In previous literature \cite{HYMN12, HYMN19, HYMN20}, for even $N$ these ``$B$-matrices" were defined as $B^h_L = L^h_NR^h_{N-1}L^h_{N-2}\cdots$ and $B^h_R = R^h_NL^h_{N-1}R^h_{N-2}\cdots$ (so $B^h_L = L^h_4R^h_3L^h_2R^h_1$ and $B^h_R = R^h_4L^h_3R^h_2L^h_1$ for $N=4$), and the eigenvalues of these matrices were called the \textit{Height Yielding Matrix Numbers} (HYMNs). Here we shall generalize this definition to account for different color-orderings, as follows. \\

\begin{definition}\label{def:HYMNs}
For  $\sigma \in S_4$, define the permutation Banchoff matrices
    \begin{align*}
        B_{L;\sigma}^h &= L_{\sigma(1)}^h R_{\sigma(2)}^h L_{\sigma(3)}^h R_{\sigma(4)}^h \\[2mm]
        B_{R;\sigma}^h &= R_{\sigma(1)}^h L_{\sigma(2)}^h R_{\sigma(3)}^h L_{\sigma(4)}^h, \\[1mm]
    \end{align*}
The eigenvalues of these matrices are the modified  HYMNs. Let \textit{left-HYMNs} refer to the eigenvalues of the $B_L$ matrices, and \textit{right-HYMNs} refer to the eigenvalues of the $B_R$ matrices.
These values are taken modulo $\rho_B^{2k}\rho_F^{2\ell}$, for $k,\ell \in \mathbb{Z}$. 
\end{definition}

\begin{proposition}
    $B_{L,\sigma}$ and $B_{R,\sigma}$, for $\sigma \in \{ {\rm id}, (2\, 3\, 4), (2\, 4\, 3)\}$ attain all possible spectra of left- or right-HYMNS. 
\end{proposition}

\begin{proof}
      
    Since  cyclic permutations of  products  of matrices are isospectral (have the same eigenvalues), the following matrices all have the same spectrum for any $\sigma \in S_4$:
    \begin{gather*}
    B_{L,\sigma}  = L_{\sigma(1)}^hR_{\sigma(2)}^hL_{\sigma(3)}^hR_{\sigma(4)}^h,\  B_{R,\sigma \cdot (1\, 2\, 3\, 4)} = R_{\sigma(2)}^hL_{\sigma(3)}^hR_{\sigma(4)}^hL_{\sigma(1)}^h, \\ B_{L,\sigma\cdot (1\, 3)(2\, 4)} =L_{\sigma(3)}^hR_{\sigma(4)}^hL_{\sigma(1)}^hR_{\sigma(2)}^h, \ B_{R,\sigma\cdot (1\, 4\, 3\, 2)}=R_{\sigma(4)}^hL_{\sigma(1)}^hR_{\sigma(2)}^hL_{\sigma(3)}^h.
    \end{gather*}
    Transposes of matrices are also isospectral. So, the matrix 
    \[
    B_{L,\sigma} = (D_BL_{\sigma(1)}^{\rm v} D_F)( \widehat D_FR_{\sigma(2)}^{\rm v}
    \widehat D_B)(  D_B L_{\sigma(3)}^{\rm v} D_F)( \widehat D_F R_{\sigma(4)}^{\rm v} \widehat D_B)
    \]
    is isospectral to the matrix 
    \[
    B_{L,\sigma}^T =( \widehat D_B (R_{\sigma(4)}^{\rm v})^T  D_F) ( \widehat D_F (L_{\sigma(3)}^{\rm v} )^T \widehat D_B) (D_B(R_{\sigma(2)}^{\rm v})^TD_F)(\widehat D_FR_{\sigma(1)}^{\rm v} D_B)
    .
    \]
      By definition, $(L_I^{\rm v})^T = R_I$ and $(R_I^{\rm v})^T = L_I$ for each $I \in \{1,2,3,4\}$. So, 
    \[
    B_{L,\sigma}^T =( \widehat D_B L_{\sigma(4)}^{\rm v} D_F) ( \widehat  D_F R_{\sigma(3)}^{\rm v}  \widehat D_B) ( D_BL_{\sigma(2)}^{\rm v}D_F)(\widehat D_F(L_{\sigma(1)}^{\rm v})^T D_B),
    \]
    which, using the fact that $ \widehat D_B^{-1} D_B = D_B \widehat D_B^{-1}$, is isospectral to $B_{L,\sigma \cdot (4\, 3\, 2\, 1)}= (D_B\widehat D_B^{-1})B_{L,\sigma}^T ( D_B\widehat D_B^{-1})^{-1}$. So, $B_{L,\sigma}$ is isospectral to each of 
    \begin{gather*}
    B_{L,\sigma \cdot (1\, 4)( 2\, 3)}  = L_{\sigma(4)}^hR_{\sigma(3)}^hL_{\sigma(2)}^hR_{\sigma(1)}^h,\  B_{R,\sigma \cdot ( 1\, 3)} = R_{\sigma(3)}^hL_{\sigma(2)}^hR_{\sigma(1)}^hL_{\sigma(4)}^h, \\ B_{L,\sigma\cdot (1\, 2)( 3\, 4)} =L_{\sigma(2)}^hR_{\sigma(1)}^hL_{\sigma(4)}^hR_{\sigma(3)}^h, \ 
    B_{R,\sigma\cdot (2\, 4)}=R_{\sigma(1)}^hL_{\sigma(4)}^hR_{\sigma(3)}^hL_{\sigma(2)}^h.
    \end{gather*}
    Set
   \begin{gather*}
   \mathcal{P}_\sigma =\{ B_{L,\sigma}, B_{L,\sigma\cdot (1\, 3)(2\, 4)}, B_{L,\sigma \cdot (1\, 4)( 2\, 3)}, B_{L,\sigma\cdot (1\, 2)( 3\, 4)}, 
   B_{R,\sigma \cdot (1\, 2\, 3\, 4)},  B_{R,\sigma\cdot (1\, 4\, 3\, 2)}, 
   B_{R,\sigma \cdot ( 1\, 3)}, B_{R,\sigma\cdot (2\, 4)}\},\\
    \mathcal{P}'_\sigma =\{ B_{R,\sigma}, B_{R,\sigma\cdot (1\, 3)(2\, 4)}, B_{R,\sigma \cdot (1\, 4)( 2\, 3)}, B_{R,\sigma\cdot (1\, 2)( 3\, 4)}, 
   B_{L,\sigma \cdot (1\, 2\, 3\, 4)},  B_{L,\sigma\cdot (1\, 4\, 3\, 2)}, 
   B_{L,\sigma \cdot ( 1\, 3)}, B_{L,\sigma\cdot (2\, 4)}\}.
   \end{gather*}
   Then for any $\sigma \in S_4$, the matrices in $\mathcal{P}_\sigma $ are all isospectral. It is easy to verify that $\mathcal{P}_{\rm id}$, $\mathcal{P}_{(2\, 3\, 4)}$,  $\mathcal{P}_{(2\, 4\, 3)}$, $\mathcal{P}'_{\rm id}$, $\mathcal{P}'_{(2\, 3\, 4)}$, and $\mathcal{P}'_{(2\, 4\, 3)}$ give all possible HYMNs values. 

\end{proof}

\begin{proposition}\label{prop:HYMNS}
    If $h$ and $h'$ are in the same FWI class, then their full set of left-HYMNs are identical (up to reordering), and so are their full set of right-HYMNs. 
\end{proposition}
\begin{proof}
This follows from Proposition~\ref{prop:FWI_L}.
\end{proof}
\begin{remark}
    A partial converse to Proposition~\ref{prop:HYMNS} is also true: If $h$ and $h'$ are two heights with the same height signature and the same left- and right-HYMNs, then $h$ and $h'$ are in the same FWI class,  except in the case of the FWI classes $\{h_{13}, h_{14}\}$, which have identical sets of HYMNs. Note that this is also the only pair of FWI classes that cannot be distinguished by counting the number of vertices at each height and  comparing the tuple of shortest distances (along edges) between all pairs of pinned vertices at the maximal height.
\end{remark}
The Right- and Left-HYMNs for heights in each FWI class have been computed, and are given in Table~\ref{tab:HYMNS} (spread across two pages) and shown visually in Figure~\ref{fig:HYMNS}.

\subsection{HYMNs for all FWI height classes}
\label{sec:A}

For each height class below, the first three sets are the left-HYMNs corresponding to the Banchoff matrices $L_1R_2L_3R_4$, $L_1R_3L_4R_2$ and $L_1R_4L_2R_3$. The next three sets are the right-HYMNs corresponding to the Banchoff matrices $R_1L_2R_3L_4$, $R_1L_3R_4L_2$ and $R_1L_4R_2L_3$.  Each eigenvalue has a superscript  denoting its multiplicity.

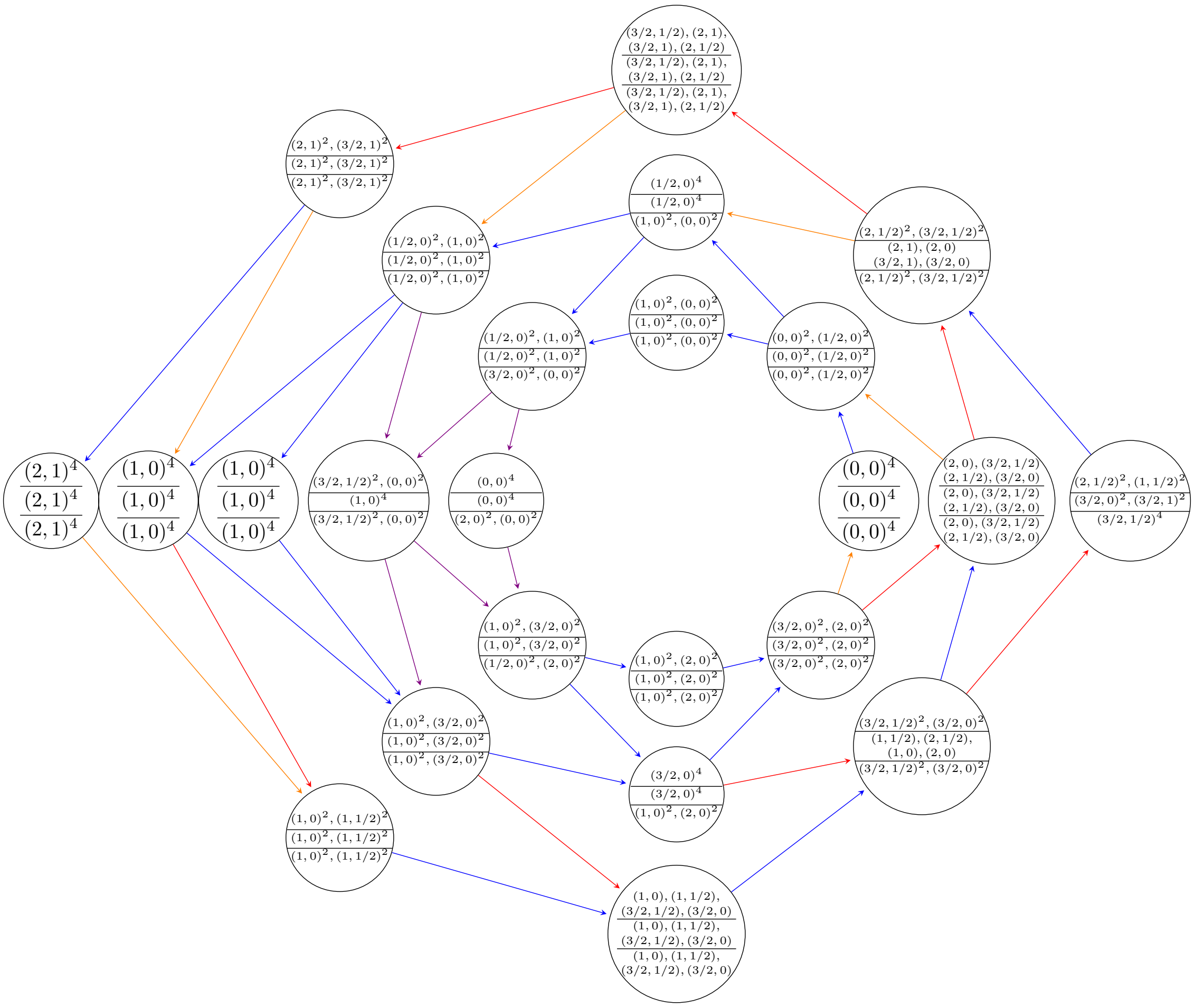
\begin{figure}[h]
\centering
\resizebox{0.99\textwidth}{!}{
       \begin{tikzpicture}[> = stealth,  shorten > = 2pt,   auto,   node distance = 1.5cm,state/.style={circle, draw, minimum size=1.1cm, inner sep = -2 pt},scale=.9]

        \node[state]  (hfe) at (-13,0) {$\begin{array}{c}(2,1)^4 \\\hline(2,1)^4 \\\hline(2,1)^4 \end{array} $};
        \node[state]  (h4741) at (-7,-7){$\begin{tiny}\begin{array}{c}
            (1,0)^2, (1,1/2)^2 \\[.5mm]\hline(1,0)^2, (1,1/2)^2 \\[.5mm]\hline (1,0)^2, (1,1/2)^2 \end{array}\end{tiny} $};
        \node[state]  (h1474) at (-7,7) {$\begin{tiny}\begin{array}{c}
            (2,1)^2, (3/2,1)^2 \\[.5mm]\hline(2,1)^2, (3/2,1)^2 \\[.5mm]\hline (2,1)^2, (3/2,1)^2 \end{array}\end{tiny} $};
        \node[state]  (h3751) at (0,-9){$\begin{tiny}\begin{array}{c}
            (1,0), (1,1/2),\\ (3/2,1/2), (3/2,0) \\\hline (1,0), (1,1/2),\\ (3/2,1/2), (3/2,0) \\\hline (1,0), (1,1/2),\\ (3/2,1/2), (3/2,0) \end{array}\end{tiny} $};
        \node[state]  (h1573) at (0,8.95){$\begin{tiny}\begin{array}{c}
            (3/2,1/2), (2,1),\\  (3/2, 1),(2,1/2) \\\hline (3/2,1/2), (2,1),\\  (3/2, 1),(2,1/2) \\\hline (3/2,1/2), (2,1),\\  (3/2, 1),(2,1/2)\end{array}\end{tiny} $};
        \node[state]  (h484a) at (-10.98,0){$\begin{array}{c}(1,0)^4 \\[.5mm]\hline(1,0)^4 \\[.5mm]\hline(1,0)^4 \end{array} $};
        \node[state]  (h484c)  at (-8.9,0){$\begin{array}{c}(1,0)^4 \\[.5mm]\hline(1,0)^4 \\[.5mm]\hline(1,0)^4 \end{array} $};
        \node[state]  (h484b)  at (-6.4,0){$\begin{tiny}\begin{array}{c}
            (3/2,1/2)^2, (0,0)^2 \\[.5mm]\hline(1,0)^4 \\[.5mm]\hline(3/2,1/2)^2, (0,0)^2 \end{array}\end{tiny} $};
        \node[state]  (h484d)  at (-3.75,0){$\begin{tiny}\begin{array}{c}(0,0)^4 \\[.5mm]\hline(0,0)^4 \\[.5mm]\hline(2,0)^2, (0,0)^2 \end{array}\end{tiny} $};
        \node[state]  (h88) at (4,0){$\begin{array}{c}
            (0,0)^4 \\[.5mm]\hline(0,0)^4 \\[.5mm]\hline(0,0)^4 \end{array} $};
        \node[state]  (h1771) at (6.55,0){$\begin{tiny}\begin{array}{c}(2,0),(3/2,1/2)\\(2,1/2),(3/2,0) \\\hline(2,0),(3/2,1/2)\\(2,1/2),(3/2,0) \\\hline(2,0),(3/2,1/2)\\(2,1/2),(3/2,0) \end{array}\end{tiny} $};
        \node[state]  (h2761) at(5.1,-5.1){$\begin{tiny}\begin{array}{c}(3/2,1/2)^2,(3/2,0)^2 \\\hline(1,1/2),(2,1/2),\\(1,0),(2,0) \\\hline (3/2,1/2)^2,(3/2,0)^2 \end{array}\end{tiny} $};
        \node[state]  (h1672) at(5.1,5.1) {$\begin{tiny}\begin{array}{c}(2,1/2)^2,(3/2,1/2)^2 \\\hline(2,1),(2,0)\\(3/2,1),(3/2,0) \\\hline (2,1/2)^2,(3/2,1/2)^2 \end{array}\end{tiny} $};
        \node[state]  (h2662)  at (9.43,0){$\begin{tiny}\begin{array}{c} (2,1/2)^2,(1,1/2)^2 \\[.5mm]\hline(3/2,0)^2,(3/2,1)^2\\[.5mm]\hline(3/2,1/2)^4 \end{array}\end{tiny} $};
        \node[state]  (h781) at (3,3){$\begin{tiny}\begin{array}{c} (0,0)^2,(1/2,0)^2 \\[.5mm]\hline(0,0)^2,(1/2,0)^2\\[.5mm]\hline(0,0)^2,(1/2,0)^2 \end{array}\end{tiny} $};
        \node[state]  (h187) at (3,-3) {$\begin{tiny}\begin{array}{c} (3/2,0)^2,(2,0)^2 \\[.5mm]\hline(3/2,0)^2,(2,0)^2\\[.5mm]\hline(3/2,0)^2,(2,0)^2 \end{array}\end{tiny} $};
        \node[state]  (h286b) at (0,-3.7){$\begin{tiny}\begin{array}{c} (1,0)^2,(2,0)^2 \\[.5mm]\hline(1,0)^2,(2,0)^2\\[.5mm]\hline(1,0)^2,(2,0)^2 \end{array}\end{tiny} $};
        \node[state]  (h286a) at (0,-6.1) {$\begin{tiny}\begin{array}{c} (3/2,0)^4 \\[.5mm]\hline(3/2,0)^4\\[.5mm]\hline(1,0)^2,(2,0)^2 \end{array}\end{tiny} $};
        \node[state]  (h682b)  at (0,3.7){$\begin{tiny}\begin{array}{c} (1,0)^2,(0,0)^2  \\[.5mm]\hline(1,0)^2,(0,0)^2\\[.5mm]\hline(1,0)^2,(0,0)^2 \end{array}\end{tiny} $};
        \node[state]  (h682a) at (0,6.2){$\begin{tiny}\begin{array}{c} (1/2,0)^4 \\[.5mm]\hline(1/2,0)^4\\[.5mm]\hline(1,0)^2,(0,0)^2 \end{array}\end{tiny} $};
        \node[state]  (h385b) at (-3,-3){$\begin{tiny}\begin{array}{c} (1,0)^2,(3/2,0)^2 \\[.5mm]\hline(1,0)^2,(3/2,0)^2\\[.5mm]\hline(1/2,0)^2,(2,0)^2 \end{array}\end{tiny} $};
        \node[state]  (h385a) at (-5,-5){$\begin{tiny}\begin{array}{c} (1,0)^2,(3/2,0)^2 \\[.5mm]\hline(1,0)^2,(3/2,0)^2\\[.5mm]\hline(1,0)^2,(3/2,0)^2 \end{array}\end{tiny} $};
        \node[state]  (h583b) at (-3,3){$\begin{tiny}\begin{array}{c} (1/2,0)^2, (1,0)^2 \\[.5mm]\hline(1/2,0)^2, (1,0)^2\\[.5mm]\hline(3/2,0)^2, (0,0)^2 \end{array}\end{tiny} $};
        \node[state]  (h583a)  at (-5,5){$\begin{tiny}\begin{array}{c} (1/2,0)^2, (1,0)^2 \\[.5mm]\hline(1/2,0)^2, (1,0)^2\\[.5mm]\hline(1/2,0)^2, (1,0)^2 \end{array}\end{tiny} $};
       
		\path[->, orange] (hfe)  edge node {} (h4741);
		\path[->, blue] (h4741)  edge node {} (h3751);
		\path[->, blue] (h1474)  edge node {} (hfe);
		\path[->, orange] (h1474)  edge node {} (h484a);
		\path[->, blue] (h484a)  edge node {} (h385a);
		\path[->, red] (h484a)  edge node {} (h4741);
		\path[->, red] (h385a)  edge node {} (h3751);
		\path[->, blue] (h385a)  edge node {} (h286a);
		\path[->, blue] (h583a)  edge node {} (h484a);
		\path[->, violet] (h583a)  edge node {} (h484b);
		\path[->, blue] (h583a)  edge node {} (h484c);
		\path[->, violet] (h484b)  edge node {} (h385a);
		\path[->, violet] (h484b)  edge node {} (h385b);
		\path[->, blue] (h484c)  edge node {} (h385a);
		\path[->, blue] (h3751)  edge node {} (h2761);
		\path[->, orange] (h1573)  edge node {} (h583a);
		\path[->, red] (h1573)  edge node {} (h1474);
		\path[->, blue] (h286a)  edge node {} (h187);
		\path[->, red] (h286a)  edge node {} (h2761);
		\path[->,blue] (h682a)  edge node {} (h583a);
		\path[->, blue] (h682a)  edge node {} (h583b);
		\path[->, blue] (h385b)  edge node {} (h286a);
		\path[->, blue] (h385b)  edge node {} (h286b);
		\path[->, violet] (h583b)  edge node {} (h484b);
		\path[->, violet] (h583b)  edge node {} (h484d);
		\path[->,blue] (h2761)  edge node {} (h1771);
		\path[->, red] (h2761)  edge node {} (h2662);
		\path[->, orange] (h1672)  edge node {} (h682a);
		\path[->, red] (h1672)  edge node {} (h1573);
		\path[->, red] (h1771)  edge node {} (h1672);
		\path[->, orange] (h1771)  edge node {} (h781);
		\path[->, red] (h187)  edge node {} (h1771);
		\path[->, orange] (h187)  edge node {} (h88);
		\path[->, blue] (h781)  edge node {} (h682a);
		\path[->, blue] (h781)  edge node {} (h682b);
		\path[->, blue] (h286b)  edge node {} (h187);
		\path[->, blue] (h682b)  edge node {} (h583b);
		\path[->, violet] (h484d)  edge node {} (h385b);
		\path[->, blue] (h88)  edge node {} (h781);
		\path[->, blue] (h2662)  edge node {} (h1672);
      \end{tikzpicture}}

\caption{HYMNs for each FWI height class shown on the digraph $\widetilde \Gamma_4$ as in Figure~\ref{fig:Gamma_4}. }\label{fig:HYMNS}
\end{figure}

\begingroup
\small
\begin{longtable}{|c|l|l|}
\caption{Descriptions of HYMNs from each height class.}\label{tab:HYMNS} \\
\hline
 & \textbf{Left HYMNs} & \textbf{Right HYMNs} \\ \hline \hline
\endhead

\hline
\endfoot

$h_{\rm fe}$ & $3 \times  (\pm \rho_B^2\rho_F)^4$ & $3 \times (\pm \rho_B^2\rho_F)^4$ \\ \hline\hline
$h_{1}$ & $3 \times[(\pm\rho_B)^2, (\pm\rho_B\rho_F^{1/2})^2]$ & $3 \times[(\pm\rho_B)^2, (\pm\rho_B\rho_F^{1/2})^2]$ \\ \hline
$h_{1'}$ & $3 \times[(\pm\rho_B^{2}\rho_F^{1})^2, (\pm\rho_B^{3/2}\rho_F)^2]$ & $3 \times[(\pm\rho_B^{2}\rho_F^{1})^2, (\pm\rho_B^{3/2})^2]$ \\ \hline\hline
$h_{2}$ & $3 \times[(\pm\rho_B),(\pm\rho_B\rho_F^{1/2}), (\pm\rho_B^{3/2}\rho_F^{1/2}), (\pm\rho_B^{3/2})]$ & $3 \times[(\pm\rho_B),(\pm\rho_B\rho_F^{1/2}), (\pm\rho_B^{3/2}\rho_F^{1/2}), (\pm\rho_B^{3/2})]$ \\ \hline
$h_{2'}$ & $3 \times[(\pm\rho_B^{2}\rho_F^{1}),(\pm\rho_B^{3/2}\rho_F), (\pm\rho_B^{3/2}\rho_F^{1/2}), (\pm\rho_B^{2}\rho_F^{1/2})]$ & $3 \times[(\pm\rho_B^{2}\rho_F^{1}),(\pm\rho_B^{3/2}\rho_F), (\pm\rho_B^{3/2}\rho_F^{1/2}), (\pm\rho_B^{2}\rho_F^{1/2})]$ \\ \hline\hline
\multirow{3}{*}{$h_{3}$} & $(\pm\rho_B^{3/2}\rho_F^{1/2})^2, (\pm\rho_B^{3/2})^2$ & $(\pm\rho_B^{3/2}\rho_F^{1/2})^2, (\pm\rho_B^{3/2})^2$ \\ \cline{2-3}
 & $(\pm\rho_B),(\pm \rho_B^{2}), (\pm\rho_B\rho_F^{1/2}), (\pm\rho_B^{2}\rho_F^{1/2})$ & $(\pm\rho_B^{3/2}\rho_F^{1/2})^2, (\pm\rho_B^{3/2})^2$ \\ \cline{2-3}
 & $(\pm\rho_B^{3/2}\rho_F^{1/2})^2, (\pm\rho_B^{3/2})^2$ & $(\pm\rho_B),(\pm \rho_B^{2}), (\pm\rho_B\rho_F^{1/2}), (\pm\rho_B^{2}\rho_F^{1/2})$ \\ \hline\hline
\multirow{3}{*}{$h_{3'}$} & $(\pm\rho_B^{2}\rho_F^{1/2})^2, (\pm\rho_B^{3/2}\rho_F^{1/2})^2$ & $(\pm\rho_B^{2}\rho_F^{1}),(\pm \rho_B^{2}), (\pm\rho_B^{3/2}\rho_F), (\pm\rho_B^{3/2})$ \\ \cline{2-3}
 & $(\pm\rho_B^{2}\rho_F^{1}),(\pm \rho_B^{2}), (\pm\rho_B^{3/2}\rho_F), (\pm\rho_B^{3/2})$ & $(\pm\rho_B^{2}\rho_F^{1/2})^2, (\pm\rho_B^{3/2}\rho_F^{1/2})^2$ \\ \cline{2-3}
 & $(\pm\rho_B^{2}\rho_F^{1/2})^2, (\pm\rho_B^{3/2}\rho_F^{1/2})^2$ & $(\pm\rho_B^{2}\rho_F^{1/2})^2, (\pm\rho_B^{3/2}\rho_F^{1/2})^2$ \\ \hline\hline
\multirow{3}{*}{$h_{4}$} & $(\pm\rho_B^{2}\rho_F^{1/2})^2, (\pm\rho_B\rho_F^{1/2})^2$ & $(\pm\rho_B^{3/2})^2, (\pm\rho_B^{3/2}\rho_F)^2$ \\ \cline{2-3}
 & $(\pm\rho_B^{3/2})^2, (\pm\rho_B^{3/2}\rho_F)^2$ & $(\pm\rho_B^{2}\rho_F^{1/2})^2, (\pm\rho_B\rho_F^{1/2})^2$ \\ \cline{2-3}
 & $(\pm\rho_B^{3/2}\rho_F^{1/2})^4$ & $\pm(3/2,1/2)^4$ \\ \hline\hline
$h_{5}$ & $3\times[(\pm \rho_B^{2}), (\pm\rho_B^{3/2}\rho_F^{1/2}), (\pm\rho_B^{2}\rho_F^{1/2}), (\pm\rho_B^{3/2})]$ & $3\times[(\pm \rho_B^{2}), (\pm\rho_B^{3/2}\rho_F^{1/2}), (\pm\rho_B^{2}\rho_F^{1/2}), (\pm\rho_B^{3/2})]$ \\ \hline\hline
$h_{6}$ & $3 \times[(\pm1)^2, (\pm\rho_B^{1/2})^2]$ & $3 \times[(\pm1)^2, (\pm\rho_B^{1/2})^2]$ \\ \hline
$h_{6'}$ & $3 \times[(\pm\rho_B^{3/2})^2, (\pm\rho_B^2)^2]$ & $3 \times[(\pm\rho_B^{3/2})^2, (\pm\rho_B^2)^2]$ \\ \hline\hline
$h_{7}$ & $3\times[ (\pm\rho_B)^2, (\pm1)^2]$ & $3\times[ (\pm\rho_B)^2,(\pm1)^2]$ \\ \hline
$h_{7'}$ & $3\times[(\pm\rho_B)^2,(\pm\rho_B^2)^2]$ & $3\times[ (\pm\rho_B)^2, (\pm\rho_B^2)^2]$ \\ \hline\hline
\multirow{3}{*}{$h_{8}$} & $(\pm\rho_B^{1/2})^4$ & $(\pm\rho_B)^2, (\pm1)^2$ \\ \cline{2-3}
 & $(\pm\rho_B^{1/2})^4$ & $(\pm\rho_B^{1/2})^4$ \\ \cline{2-3}
 & $(\pm\rho_B)^2, (\pm1)^2$ & $(\pm\rho_B^{1/2})^4$ \\ \hline\hline
\multirow{3}{*}{$h_{8'}$} & $(\pm\rho_B^{3/2})^4$ & $(\pm\rho_B)^2, (\pm\rho_B^2)^2$ \\ \cline{2-3}
 & $(\pm\rho_B^{3/2})^4$ & $(\pm\rho_B^{3/2})^4$ \\ \cline{2-3}
 & $(\pm\rho_B)^2,(\pm\rho_B^2)^2$ & $(\pm\rho_B^{3/2})^4$ \\ \hline\hline
\multirow{3}{*}{$h_{9}$} & $(\pm\rho_B)^2, (\pm\rho_B^{1/2})^2$ & $(\pm\rho_B^{3/2})^2, (\pm1)^2$ \\ \cline{2-3}
 & $(\pm\rho_B)^2, (\pm\rho_B^{1/2})^2$ & $(\pm\rho_B)^2, (\pm\rho_B^{1/2})^2$ \\ \cline{2-3}
 & $(\pm\rho_B^{3/2})^2, (\pm1)^2$ & $(\pm\rho_B)^2, (\pm\rho_B^{1/2})^2$ \\ \hline\hline
\multirow{3}{*}{$h_{9'}$} & $(\pm\rho_B)^2,(\pm\rho_B^{3/2})^2$ & $(\pm\rho_B^{1/2})^2, (\pm\rho_B^2)^2$ \\ \cline{2-3}
 & $(\pm\rho_B)^2, (\pm\rho_B^{3/2})^2$ & $(\pm\rho_B)^2, (\pm\rho_B^{3/2})^2$ \\ \cline{2-3}
 & $(\pm\rho_B^{1/2})^2, (\pm\rho_B^2)^2$ & $(\pm\rho_B)^2, (\pm\rho_B^{3/2})^2$ \\ \hline\hline
$h_{10}$ & $3 \times[(\pm\rho_B^{1/2})^2, (\pm\rho_B)^2]$ & $3 \times[(\pm \rho_B^{1/2})^2, (\pm\rho_B)^2]$ \\ \hline
$h_{10'}$ & $3 \times[(\pm\rho_B^{3/2})^2, (\pm\rho_B)^2]$ & $3 \times[(\pm\rho_B^{3/2})^2, (\pm\rho_B)^2]$ \\ \hline\hline
\multirow{3}{*}{$h_{11}$} & $(\pm\rho_B)^4$ & $(\pm\rho_B^2)^2,(\pm1)^2$ \\ \cline{2-3}
 & $(\pm\rho_B)^4$ & $(\pm\rho_B)^4$ \\ \cline{2-3}
 & $(\pm \rho_B^2)^2, (\pm1)^2$ & $(\pm\rho_B)^4$ \\ \hline\hline
\multirow{3}{*}{$h_{12}$} & $(\pm\rho_B^{3/2}\rho_F^{1/2})^2, (\pm1)^2$ & $(\pm\rho_B^{3/2}\rho_F^{1/2})^2, (\pm1)^2$ \\ \cline{2-3}
 & $(\pm\rho_B)^4$ & $(\pm\rho_B^{3/2}\rho_F^{1/2})^2, (\pm1)^2$ \\ \cline{2-3}
 & $(\pm\rho_B^{3/2}\rho_F^{1/2})^2, (\pm 1)^2$ & $(\pm\rho_B)^4$ \\ \hline\hline
$h_{13}$ & $3 \times(\pm\rho_B)^4$ & $3 \times(\pm\rho_B)^4$ \\ \hline\hline
$h_{14}$ & $3 \times(\pm\rho_B)^4$ & $3 \times(\pm \rho_B)^4$ \\ \hline\hline
$h_{\rm v}$ & $3 \times (\pm 1)^4$ & $3 \times(\pm 1)^4$ \\ \hline
\end{longtable}
\endgroup

Examining Figure~\ref{fig:HYMNS} closely, we see some interesting structure among these spectra. We display them in place of the corresponding FWI class, mimicking the structure in Figure~\ref{fig:Gamma_4}. Note that there is a kind of vertical symmetry: Among heights with $k$ levels, certain eigenvalues along the equator (i.e., $(1,0)$ for $k =3$) are ``neutral'', and the spectra above and below the equator related by a vertical flip (i.e., for $h_3$ and $h_{3'}$) are dual to each other (i.e., $(1/2,0)$ and $(3/2,0)$).  

 The structure among the arrows (vertex  lowering) and the HYMNs values is also interesting to note. For some ordering among the $(a_i,b_i)$, we have:
 \begin{itemize}
     \item The blue arrows send 
     \[
     \{(a_i,b_i)\}_{i=1, 2, 3, 4} \mapsto \{(a_i,b_i)\}_{i=1, 2}\cup \{(a_i+1/2,b_i)\}_{i= 3, 4}
     \]
     \item The red arrows send  
     \[
     \{(a_i,b_i)\}_{i=1, 2, 3, 4}  \mapsto \{(a_i,b_i)\}_{i=1, 2}\cup \{(a_i,b_i+1/2)\}_{i= 3, 4} .
     \]
     \item The orange arrows send 
     \[
     \{(a_i,b_i)\}_{i=1, 2, 3, 4} \mapsto \{(b_i,a_i-2)\}_{i=1, 2}\cup \{(b_i,a_i-2+1/2)\}_{i= 3, 4} .
     \]
 \end{itemize}

Making these observations explicit might help us understand the distribution of and relationship among Adinkral heights. Again, further investigation of the rich structure demonstrated by these invariants for this and more general cases would be interesting. 

\section{Outlook}

The present work classifies Adinkra heights in the physically significant case of the $N=4$ hypercube using both graph-theoretic and matrix-theoretic methods. A coarser classification scheme for the $N=5$ case is given in \cite{FW}; this classification applies the geometry and number theory of complex curves, building on the work of \cite{geom2}. Classification of heights for other Adinkra chromotopologies is of obvious interest. We now describe some avenues for future exploration of particular physical importance.

The first of these avenues includes studies of 4D, $N = 4$ ($\mathcal{N}=1$) theories that arise
from examining the sets of graphs related to additional supermultiplets.  The class of graphs investigated in this study only relates to 4D supermultiplets with 8 bosonic and 8 fermionic degrees of freedom that describe irreducible off-shell representations. Combinatorially, the natural next step entails classifying all possible heights on an $N=4$ Adinkra obtained as the quotient of the $N=4$ hypercube by a doubly even code. This would yield a class of graphs with 4 bosonic and 4 fermionic degrees of freedom. 
However, these are only minimal representations.  The analog in conventional compact Lie algebras is to study only representations that are described by Young diagrams with a single box.  It is well known that in these algebras, higher dimensional representations require Young diagrams with greater than a single box. 

Studying reducible representations is of great interest in our context as well. More complex Adinkras are also of clear interest in their own right. In the realm of Adinkras relevant for four-dimensional supersymmetrical theories, there are more complicated graphs associated with supermultiplets with (4 + 8$p$) bosonic and (4 + 8$p$) fermionic degrees of freedom (where $p$ is any integer). 
For
the case where $p$ = 1, there are minimal supergravity and
complex linear supermultiplets \cite{G-2,Char13, anatomy}. For the case where $p$ = 2, there are the non-minimal supergravity 
\cite{Char13} and gravitino supermultiplets \cite{MS}.

As we have already observed, the Adinkra height invariants reveal extremely rich structure. Further investigation of the underlying physical and mathematical sources of this structure is a clear priority for future work.
An obvious path involves further examining the relationship between
the combinatorial approach to characterizing heights using FWI and rainbow classes  and the physically motivated
HYMNs and twisted character approaches. In particular, it would be interesting to find an explanation for the preservation and inversion of characters and dual characters seen along arrows in Figure~\ref{fig:CHARS}, as well as the variation of spectra seen in the arrows in Figure~\ref{fig:HYMNS}. All of these approaches to classifying Adinkra heights could potentially be extended to hypercubes with larger $N$ and to quotient Adinkras; the structure of heights in such cases is largely unexplored.

\section*{Data Availability Statement}

Not applicable.

\section*{Code Availability Statement}
The python scripts that we wrote to calculate the modified characters and HYMNs of all the 4-color Adinkra heights can be accessed at \url{https://github.com/digvijayroyvarier/AdinkraHeights}. The tabulated results are outputted by these scripts as .txt files. This repository also contains a .txt file with a raw list of all 990 heights, which is required as input when running the code. 

\section*{Acknowledgments}

We thank Charles Doran, Kevin Iga, and Tristan H\"ubsch for interesting and useful conversations.

The research of S.J.G. is currently supported in part by
the Clark Leadership Chair in Science endowment at the University of Maryland--College Park and the Army Research Office under Contract WNW911NF2520117. S.J.G. thanks ChatGPT (OpenAI) for assistance in literature search and in checking the details of the derivation described in the section on sigma models. 

U.A.W. was supported by ICERM, the Institute for Computational and Experimental Research in Mathematics, Brown University, during the preparation of this manuscript, and thanks ICERM for an inspiring and productive visit.

We further thank the humans whose contributions to large language models such as ChatGPT often go unrecognized, including quality control workers, content moderators, and our fellow scholars.

\end{document}